\documentclass[structabstract]{aa}
\usepackage{graphicx}
\usepackage{stfloats}    
\usepackage{txfonts}
\usepackage{color}
\newcommand{\tablenotea}[1]{\parbox{8.7cm}{ \indent \footnotesize{\textsc{}~#1}}}
\newcommand{\tablenoteb}[1]{\parbox{8.7cm}{ \indent \footnotesize{\textsc{Notes.--}~#1}}}
\newcommand{\tablenotec}[1]{\parbox{18.5cm}{ \indent \footnotesize{\textsc{}~#1}}}
\newcommand{\tablenoted}[1]{\parbox{14.2cm}{ \indent \footnotesize{\textsc{}~#1}}}
\newcommand{\tablerefs}[1]{\parbox{8.7cm}{ \indent \footnotesize{\textsc{References.--}~#1}}}
\newcommand{\nature}{Nature}
\newcommand{\science}{Science}
\newcommand{\chemrev}{Chem. Rev.}
\newcommand{\acp}{Atm. Chem. Phys.}
\newcommand{\jpcrd}{J. Phys. Chem. Ref. Data}
\newcommand{\fdis}{Faraday Discuss.}
\newcommand{\ijms}{Int. J. Mass Spectrom.}
\newcommand{\molphys}{Mol. Physics}
\newcommand{\cp}{Chem. Phys.}
\newcommand{\cpl}{Chem. Phys. Lett.}
\newcommand{\jpb}{J. Phys. B: At. Mol. Opt. Phys.}
\newcommand{\pccp}{Phys. Chem. Chem. Phys.}
\newcommand{\rgsp}{Rev. Geophys. Space Phys.}
\newcommand{\jesrp}{J. Electron Spectrosc. Relat. Phenom.}
\newcommand{\jms}{J. Mol. Spectr.}
\newcommand{\jmst}{J. Mol. Struct.}
\newcommand{\jpc}{J. Phys. Chem.}
\newcommand{\jpca}{J. Phys. Chem. A}
\newcommand{\jphotchem}{J. Photochem.}
\newcommand{\appopt}{Appl. Opt.}
\newcommand{\jams}{J. Mol. At. Sci.}

\newcommand{\prsla}{Proc. Roy. Soc. London Ser. A}
\newcommand{\jcsft}{J. Chem. Soc., Faraday Trans. 2}
\newcommand{\ijck}{Int. J. Chem. Kinet.}
\newcommand{\natastro}{Nature Astron.}
\begin{document}
\title{The chemistry of disks around T\,Tauri and Herbig\,Ae/Be stars}
\titlerunning{The chemistry of disks\\ around T\,Tauri and Herbig\,Ae/Be stars}
\authorrunning{Ag\'undez et al.}

\author{Marcelino Ag\'undez\inst{1}, Evelyne Roueff\inst{2}, Franck Le Petit\inst{2}, and Jacques Le Bourlot\inst{2,3}}

\institute{Instituto de F\'isica Fundamental, CSIC, C/ Serrano 123, E-28006 Madrid, Spain \\
\email{marcelino.agundez@csic.es} \and 
Sorbonne Universit\'e, Observatoire de Paris, Universit\'e PSL, CNRS, LERMA, F-92190 Meudon, France \and
Universit\'e Paris-Diderot, Sorbonne Paris-Cit\'e, F-75013 Paris, France}

\date{Received; accepted}


\abstract
{Infrared and (sub-)mm observations of disks around T\,Tauri and Herbig\,Ae/Be stars point to a chemical differentiation between both types of disks, with a lower detection rate of molecules in disks around hotter stars.}
{To investigate the underlying causes of the chemical differentiation indicated by observations we perform a comparative study of the chemistry of T\,Tauri and Herbig\,Ae/Be disks. This is one of the first studies to compare chemistry in the outer regions of these two types of disks.}
{We developed a model to compute the chemical composition of a generic protoplanetary disk, with particular attention to the photochemistry, and applied it to a T\,Tauri and a Herbig\,Ae/Be disk. We compiled cross sections and computed photodissociation and photoionization rates at each location in the disk by solving the FUV radiative transfer in a 1+1D approach using the Meudon PDR code and adopting observed stellar spectra.}
{The warmer disk temperatures and higher ultraviolet flux of Herbig stars compared to T\,Tauri stars induce some differences in the disk chemistry. In the hot inner regions, H$_2$O, and simple organic molecules like C$_2$H$_2$, HCN, and CH$_4$ are predicted to be very abundant in T\,Tauri disks and even more in Herbig\,Ae/Be disks, in contrast with infrared observations that find a much lower detection rate of water and simple organics toward disks around hotter stars. In the outer regions, the model indicates that the molecules typically observed in disks, like HCN, CN, C$_2$H, H$_2$CO, CS, SO, and HCO$^+$, do not have drastic abundance differences between T\,Tauri and Herbig\,Ae disks. Some species produced under the action of photochemistry, like C$_2$H and CN, are predicted to have slightly lower abundances around Herbig\,Ae stars due to a narrowing of the photochemically active layer. Observations indeed suggest that these radicals are somewhat less abundant in Herbig\,Ae disks, although in any case the inferred abundance differences are small, of a factor of a few at most. A clear chemical differentiation between both types of disks concerns ices. Owing to the warmer temperatures of Herbig\,Ae disks, one expects snowlines lying farther away from the star and a lower mass of ices compared to T\,Tauri disks.}
{The global chemical behavior of T\,Tauri and Herbig\,Ae/Be disks is quite similar. The main differences are driven by the warmer temperatures of the latter, which result in a larger reservoir or water and simple organics in the inner regions and a lower mass of ices in the outer disk.}


\maketitle

\section{Introduction}

Circumstellar disks around young stars, the so-called protoplanetary disks, are an important link in the evolution from molecular clouds to planetary systems. These disks allow to feed with matter the young star and provide the scenario in which planets form. The study of the physical and chemical conditions of these objects is thus of paramount importance to understand how and from which type of material do planets form. Protoplanetary disks are mainly composed of molecular gas and dust. The last two decades have seen significant progress in the study of their chemical composition thanks to astronomical observations at wavelengths from the millimeter to the ultraviolet domains.

Observations with ground-based mm and sub-mm telescopes such as the 30m antenna of the Institut de Radioastronomie Millim\'etrique (IRAM), the James Clerk Maxwell Telescope (JCMT), and the APEX 12m telescope, which are sensitive to the outer cool regions of disks, have provided information on the presence of various gaseous molecules such as CO, HCO$^+$, H$_2$CO, C$_2$H, HCN, HNC, CN, CS, and SO (\cite{dut1997} 1997; \cite{kas1997} 1997, 2014; \cite{van2001} 2001; \cite{thi2004} 2004; \cite{fue2010} 2010; \cite{gui2013} 2013, 2016; \cite{pac2015} 2015). Interferometers that operate at (sub-)millimeter wavelengths such as the OVRO millimeter array, the IRAM array at Plateau de Bure (PdBI, now known as NOEMA), and the Submillimeter Array (SMA) have also allowed to perform sensitive observations leading to the detection of new molecules such as N$_2$H$^+$ and HC$_3$N, and to obtain maps of the emission distribution of some molecules with angular resolutions down to a few arcsec (\cite{qi2003} 2003, 2008, 2013a; \cite{dut2007} 2007, 2011; \cite{pie2007} 2007; \cite{sch2008} 2008; \cite{hen2010} 2010; \cite{obe2010} 2010, 2011; \cite{cha2012a} 2012a,b; \cite{fue2012} 2012; \cite{gra2015} 2015; \cite{tea2015} 2015; \cite{pac2016} 2016). In recent years, the advent of the Atacama Large Millimeter Array (ALMA) is making possible to characterize the molecular content of protoplanetary disks with an unprecedented sensitivity and angular resolution (down to sub-arcsec scales). For example, thanks to ALMA it has been possible to detect new molecules such as cyclic C$_3$H$_2$ (\cite{qi2013b} 2013b; \cite{bergin2016} 2016), CH$_3$CN (\cite{obe2015} 2015), and CH$_3$OH (\cite{wal2016} 2016), and to image the CO snowline in a few disks (\cite{mat2013} 2013; \cite{qi2013c} 2013c, 2015; \cite{sch2016} 2016; \cite{zha2017} 2017). Infrared (IR) observations using space telescopes such as \emph{Spitzer} and ground-based facilities such as the Very Large Telescope (VLT) and the Keck Observatory telescopes have provided a view of the molecular content of the very inner regions of protoplanetary disks, where absorption and emission lines from molecules such as CO, CO$_2$, H$_2$O, OH, HCN, C$_2$H$_2$, and CH$_4$ have been routinely observed (\cite{lah2006} 2006; \cite{gib2007} 2007; \cite{sal2007} 2007, 2008, 2011; \cite{car2008} 2008, 2011, 2014; \cite{pon2010a} 2010a,b; \cite{naj2010} 2010, 2013; \cite{kru2011} 2011; \cite{dop2011} 2011; \cite{fed2011} 2011; \cite{man2012} 2012; \cite{bas2013} 2013; \cite{gib2013} 2013; \cite{sar2014} 2014; \cite{ban2017} 2017). The launch of the \emph{Herschel Space Observatory} was also very helpful to investigate the chemical content at far-IR wavelengths, with the detection of molecules such as CH$^+$ (\cite{thi2011} 2011) and NH$_3$ (\cite{sal2016} 2016), and the exhaustive characterization of H$_2$O and OH from the inner to the outer regions (\cite{hog2011} 2011; \cite{riv2012} 2012; \cite{mee2012} 2012; \cite{fed2012} 2012, 2013; \cite{pod2013} 2013). Probing the molecular gas in the inner regions of disks through the most abundant molecule, H$_2$, has also been possible thanks to observations at ultraviolet (UV) wavelengths using the \emph{Hubble Space Telescope} (\cite{ing2009} 2009; \cite{fra2012} 2012).

On the theoretical side, the chemical structure of protoplanetary disks has been also widely studied during the last two decades. Early models focused on the one dimensional radial structure of disks along the midplane (\cite{aik1996} 1996, 1997, 1999; \cite{wil1998} 1998; \cite{aik1999a} 1999a), although it was later on recognized that the chemical composition presents also an important stratification along the vertical direction, with a structure consisting of three main layers: the cold midplane where molecules are mostly condensed as ices on dust grains, a warm upper layer where a rich chemistry takes place, and the uppermost surface layer where photochemistry driven by stellar and interstellar far ultraviolet (FUV) photons regulates the chemical composition (\cite{aik1999b} 1999b, 2001; \cite{wil2000} 2000; \cite{aik2002} 2002). In recent years there has been an interest in identifying the main processes that affect the abundance and distribution of molecules in protoplanetary disks. An important number of studies have addressed in detail the role of processes such as the interaction with FUV and X-ray radiation (\cite{wil2000} 2000; \cite{mar2002} 2002; \cite{ber2003} 2003; \cite{ilg2006a} 2006a; \cite{agu2008} 2008; \cite{are2011} 2011; \cite{fog2011} 2011; \cite{wal2010} 2010, 2012), the interplay between the thermal and chemical structure (\cite{gla2004} 2004; \cite{woi2009} 2009), turbulent mixing and other transport processes (\cite{ilg2004} 2004; \cite{sem2006} 2006; \cite{ilg2006b} 2006b; \cite{wil2006} 2006; \cite{aik2007} 2007; \cite{hei2011} 2011; \cite{sem2011} 2011), and the evolution of dust grains as they grow by coagulation and settle onto the midplane regions (\cite{aik2006} 2006; \cite{fog2011} 2011; \cite{aki2013} 2013). Some studies have investigated the impact of using different chemical networks (\cite{sem2004} 2004; \cite{ilg2006c} 2006c; \cite{kam2017} 2017) and the sensitivity to uncertainties in the rate constants of chemical reactions (\cite{vas2008} 2008). Various specific aspects of the chemistry of protoplanetary disks have also been addressed in detail as, for example, deuterium fractionation (\cite{wil2007} 2007; \cite{wil2009} 2009; \cite{thi2010} 2010; \cite{fur2013} 2013; \cite{yan2013} 2013; \cite{alb2014} 2014), the formation and survival of water vapour (\cite{dom2005} 2005; \cite{gla2009} 2009; \cite{bet2009} 2009; \cite{du2014} 2014), and the formation of particular species such as benzene (\cite{woo2007} 2007) and complex organic molecules (\cite{wal2014} 2014).

Overall, protoplanetary disks are complex systems where many different processes such as gas phase chemistry, interaction with stellar and interstellar FUV photons, transport processes, adsorption and desorption from dust grains, chemical reactions on grain surfaces, and grain evolution are all together governing the chemical composition (see reviews by \cite{ber2007} 2007, \cite{hen2013} 2013, \cite{dut2014} 2014, and \cite{pon2014} 2014).

Disks are commonly found around young low-mass (T\,Tauri) and intermediate-mass (Herbig\,Ae/Be) stars, which have quite different masses and effective temperatures, and thus may affect differently the disk chemical composition. For example, Herbig\,Ae/Be stars have a higher ultraviolet flux and disks around them are warmer than around T\,Tauri stars. Indeed, T\,Tauri and Herbig\,Ae disks have been extensively observed from millimeter to IR wavelengths and it has been found that the detection rate of molecules (for example, H$_2$O, C$_2$H$_2$, HCN, CH$_4$, CO$_2$, H$_2$CO, C$_2$H, and N$_2$H$^+$) is strikingly lower toward Herbig\,Ae disks than toward disks around T\,Tauri stars (\cite{man2008} 2008; \cite{sch2008} 2008; \cite{pon2010a} 2010a; \cite{obe2010} 2010, 2011; \cite{fed2011} 2011, 2012, 2013; \cite{sal2011} 2011; \cite{riv2012} 2012; \cite{mee2012} 2012; \cite{gui2016} 2016; \cite{ban2017} 2017). This fact may indicate that there is a marked chemical differentiation between both types of disks. Most theoretical studies on the chemistry of protoplanetary disks have focused on disks around T\,Tauri-like stars, and only a few have studied Herbig\,Ae/Be disks (e.g., \cite{jon2007} 2007). Here we present a comparative study in which we investigate the two dimensional distribution of molecules in disks around T\,Tauri and Herbig\,Ae/Be stars. In a recent study, \cite{wal2015} (2015) have investigated the differences in the chemical composition between disks around stars of different spectral type, focusing on the inner disk regions. In this study, we make a thorough investigation of the main chemical differences and similarities between T\,Tauri and Herbig\,Ae/Be disks from the inner to the outer disk regions, which to our knowledge has not been investigated in detail. We are particularly concerned with a detailed treatment of the photochemistry and with the impact of the FUV illumination from these two types of stars on the chemical composition of the disk. To this purpose we have implemented the Meudon PDR code in the disk model to compute photodestruction rates at each location in the disk in a 1+1D approach. We adopted FUV stellar spectra coming from observations of representative T\,Tauri and Herbig\,Ae/Be stars. In Sec.~\ref{sec:model} we present in detail our physical and chemical disk model, with a particular emphasis on the photochemistry (further described in Appendices~\ref{sec:app_sections} and \ref{sec:app_rates}), in Sec.~\ref{sec:results} we present the resulting abundance distributions of important families of molecules for our T\,Tauri and Herbig\,Ae/Be disk models and compare them with results from observations, in Sec.~\ref{sec:influence} we analyse the influence on the chemistry of the stellar spectra and the method used to compute photodestruction rates, and we summarize the main conclusions found in this work in Sec.~\ref{sec:conclusions}.

\section{The disk model} \label{sec:model}

We consider a passively irradiated disk in steady state around a T\,Tauri or Herbig\,Ae/Be star. We solve the thermal and chemical structure of the disk using a procedure which can be summarized as follows. We first solve the dust temperature distribution in the disk using the {\footnotesize RADMC} code (\cite{dul2004} 2004). We assume that gas and dust are thermally coupled except for the surface layers of the disk where we estimate the gas kinetic temperature following \cite{kam2004} (2004). We then solve the radiative transfer of interstellar FUV photons along the vertical direction and of stellar FUV photons along the direction from the star using the Meudon PDR code (\cite{lep2006} 2006) to get the photodissociation and photoionization rates of the different species at each disk location. We finally solve the chemical composition at each location in the disk as a function of time including gas phase chemical reactions, processes induced by FUV photons and cosmic rays, and interactions of gas particles with dust grains (adsorption and desorption processes).

\begin{table}
\caption{Model parameters.} \label{table:parameters}
\centering
\begin{tabular}{l@{\hspace{1.4cm}}r}
\hline \hline
Parameter                                               & Value / Source \\
\hline \multicolumn{2}{c}{Disk} \\
\hline
Disk mass ($M_{\rm disk}$)                              & 0.01 M$_{\odot}$ \\
Inner disk radius ($R_{\rm in}$)                        & 0.5 au \\
Outer disk radius ($R_{\rm out}$)                       & 500 au \\
Radial surface density power index ($\epsilon$)         & 1.0 $\dag$ \\
Gas-to-dust mass ratio ($\rho_{\rm g}/\rho_{\rm d}$)    & 100 \\
Dust composition                                        & 70 \% silicate \\
                                                        & 30 \% graphite \\
Minimum dust grain radius ($a_{\rm min}$)               & 0.001 $\mu$m \\
Maximum dust grain radius ($a_{\rm max}$)               & 1 $\mu$m \\
Dust size distribution power index ($\beta$)            & 3.5 \\
Interstellar radiation field                            & \cite{dra1978} (1978) \\
Cosmic-ray ionization rate of H$_2$ ($\zeta$)           & $5\times10^{-17}$ s$^{-1}$ \\
\hline \multicolumn{2}{c}{T\,Tauri star} \\
\hline
Stellar mass ($M_*$)                                    & 0.5 M$_{\odot}$ \\
Stellar radius ($R_*$)                                  & 2 R$_{\odot}$ \\
Stellar effective temperature ($T_*$)                   & 4000 K \\
\hline \multicolumn{2}{c}{Herbig\,Ae/Be star} \\
\hline
Stellar mass ($M_*$)                                    & 2.5 M$_{\odot}$ \\
Stellar radius ($R_*$)                                  & 2.5 R$_{\odot}$ \\
Stellar effective temperature ($T_*$)                   & 10,000 K \\
\hline
\end{tabular}
\tablenotea{$\dag$ $\epsilon = -12$ at $r < R_{\rm in}$.}
\end{table}

\subsection{Physical model} \label{sec:physical_model}

We adopt a fiducial model of disk representative of objects commonly found around T\,Tauri and Herbig\,Ae/Be stars (see parameters in Table~\ref{table:parameters}). The disk extends between an inner radius $R_{\rm in}$ of 0.5 au and an outer radius $R_{\rm out}$ of 500 au from the star and has a mass $M_{\rm disk}$ of 0.01 M$_\odot$. We consider that the radial distribution of the surface density $\Sigma$ is given by a power law of the type
\begin{equation}
\Sigma = \Sigma_0 (r/r_0)^{-\epsilon} \label{eq-surface-density}, \label{eq:surface_density}
\end{equation}
where $\Sigma_0$ is the surface density at a reference radius $r_0$ and the exponent $\epsilon$ is chosen to be 1.0 between $R_{\rm in}$ and $R_{\rm out}$. To avoid the abrupt disappearance of the disk at the inner radius we set $\epsilon$ to $-$12 at $r < R_{\rm in}$, which results in a soft continuation of the disk at the inner regions.

We consider that dust is present in the disk with a uniform abundance and size distribution, i.e., which does not vary with radius nor height over midplane. We adopt a gas-to-dust mass ratio of 100 and consider spherical grains typically present in the interstellar medium, i.e., with the composition being a mixture of 70 \% of silicate and 30 \% of graphite (with optical properties taken from \cite{dra1984} 1984, \cite{lao1993} 1993, and \cite{wei2001} 2001), and the size distribution being given by the power law
\begin{equation}
n(a) \propto a^{-\beta}, \label{eq:grain_size_distribution}
\end{equation}
where $n(a)$ is the number of grains of radius $a$, the minimum and maximum grain radius $a_{\rm min}$ and $a_{\rm max}$ are 0.001 and 1 $\mu$m, and the exponent $\beta$ takes a value of 3.5 according to \cite{mat1977} (1977).

As stellar parameters we adopt typical values of T\,Tauri and Herbig\,Ae/Be stars. The T\,Tauri star is assumed to have a mass $M_*$ of 0.5 M$_{\odot}$, a radius $R_*$ of 2 R$_{\odot}$, and an effective temperature $T_*$ of $\sim$4000 K, typical values of T\,Tauri stars of spectral type M0-K7 in Taurus (\cite{ken1995} 1995). In the case of the star of type Herbig\,Ae/Be we adopt a mass of 2.5 M$_{\odot}$, a radius of 2.5 R$_{\odot}$, and an effective temperature of $\sim$10,000 K, typical values of early Ae and late Be stars (e.g., \cite{mar2008} 2008; \cite{mon2009} 2009).

\subsubsection{Stellar and interstellar FUV spectra} \label{subsubsec:uv_field}

The irradiation from the central star is of great importance for the disk because it dominates the heating of gas and dust. The stellar and interstellar spectra at FUV wavelengths are also of prime importance because they control the photochemistry that takes place in the disk surface.

The FUV component of the interstellar radiation field (ISRF) adopted here is given by
\begin{equation}
I_{\lambda} = \frac{1}{4 \pi} \Big(\frac{6.3622 \times 10^7}{\lambda^4} - \frac{1.0238 \times 10^{11}}{\lambda^5} + \frac{4.0813 \times 10^{13}}{\lambda^6} \Big), \label{eq:draine}
\end{equation}
where $\lambda$ is the wavelength in $\AA$, $I_{\lambda}$ is the specific intensity in units of erg s$^{-1}$ cm$^{-2}$ $\AA^{-1}$ sr$^{-1}$, and the expression has been obtained by fitting to the radiation field given by \cite{dra1978} (1978). The expression in Eq.~(\ref{eq:draine}) is similar to that given by \cite{lep2006} (2006), except for an erratum in their first term. The ISRF given by Eq.~(\ref{eq:draine}) is complemented with another component that accounts for the emission at longer wavelengths (from $\sim$2000 $\AA$ to the near IR), for which we adopt the radiation field given by \cite{mat1983} (1983) in the form of a combination of three diluted black bodies
\begin{eqnarray}
I_{\lambda} = 1.05 \times 10^{-14} B_\lambda (7127 {\rm K}) & + 1.28 \times 10^{-13} B_{\lambda}(4043 {\rm K}) \nonumber \\
                      & + 3.30 \times 10^{-13} B_{\lambda}(2930 {\rm K}), \label{eq:mathis}
\end{eqnarray}
where $B_\lambda (T)$ is the Planck law at temperature $T$.

\begin{figure}
\centering
\includegraphics[angle=0,width=\columnwidth]{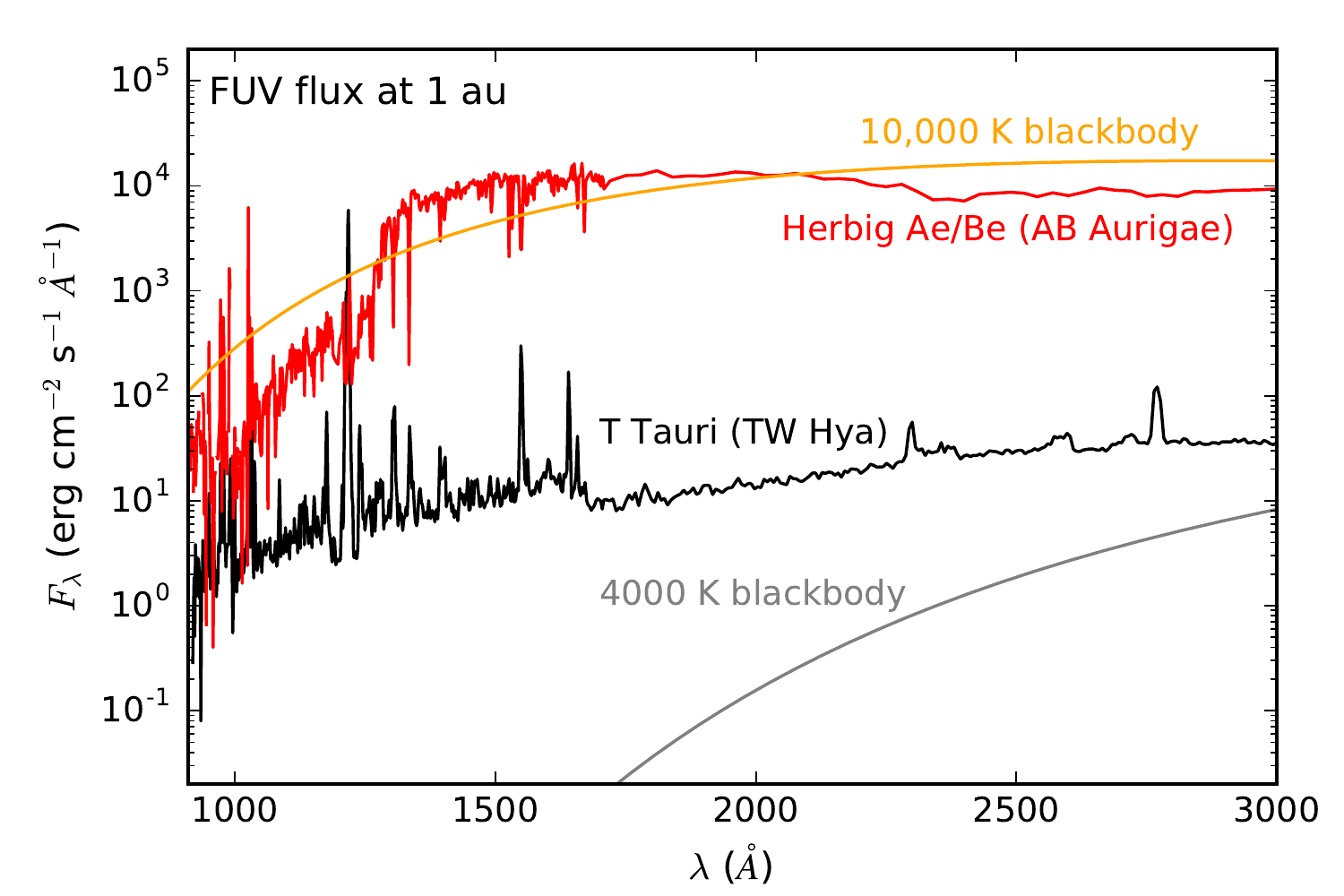}
\caption{Stellar FUV flux at 1 au of the T\,Tauri star (emitting as TW\,Hya and as a 4000 K blackbody) and of the Herbig\,Ae/Be star (emitting as AB\,Aurigae and as a 10,000 K blackbody).} \label{fig:uv_spectra}
\end{figure}

Concerning stellar spectra, in the case of the T\,Tauri star we adopt as a proxy of the stellar spectrum that of TW\,Hya, which consists of observations with the \emph{Far Ultraviolet Spectroscopic Explorer} (FUSE) in the 900-1150 $\AA$ wavelength range (data from program C0670102) and \emph{Hubble} STIS observations in the 1150-3150 $\AA$ range (\cite{her2002} 2002; \cite{ber2003} 2003). The intrinsic stellar brightness is calculated adopting a distance to the star of 51 pc (\cite{mam2005} 2005), a stellar radius of 1 R$_{\odot}$ (\cite{web1999} 1999), and a negligible interstellar reddening (\cite{ber2003} 2003). At wavelengths longer than 3150 $\AA$ we use a Kurucz model spectrum (\cite{cas2004} 2004)\footnote{\texttt{http://wwwuser.oats.inaf.it/castelli}} with an effective temperature of 4000 K, a surface gravity of 10$^{1.5}$ cm s$^{-2}$, and solar metallicity, scaled to match the flux of TW\,Hya around 3150 $\AA$. The resulting spectrum is similar to that presented by \cite{fra2014} (2014) in the 1150-1750 $\AA$ wavelength range, although their continuum level is about twice below our adopted spectrum. In the case of the star of type Herbig\,Ae/Be we use as a proxy the FUV spectrum of AB\,Aurigae, which consists of observations taken with FUSE in the 900-1190 $\AA$ wavelength range (data from program P2190301) and with \emph{Hubble} STIS in the 1190-1710 $\AA$ range (\cite{rob2001} 2001; \cite{ayr2010} 2010). The intrinsic brightness of AB\,Aurigae is calculated adopting a distance to the star of 144 pc, a stellar radius of 2.41 R$_{\odot}$, and an interstellar extinction of 0.48 mag (\cite{mar2008} 2008). For the correction due to extinction we use the method and coefficients of \cite{fit2007} (2007). Longwards of 1710 $\AA$ we use a Kurucz spectrum (\cite{cas2004} 2004) with an effective temperature of 9750 K (close to that of AB\,Aurigae), a surface gravity of 10$^{2.0}$ cm s$^{-2}$, and solar metallicity.

The FUV spectra adopted for the T\,Tauri and Herbig\,Ae/Be stars are shown in Fig.~\ref{fig:uv_spectra}, where we also compare with spectra corresponding to blackbodies at the temperatures of the T\,Tauri and Herbig\,Ae/Be stars, 4000 K and 10,000 K, respectively. It is seen that AB\,Aurigae outshines by 2-3 orders of magnitude the FUV flux of TW\,Hya because of the much higher effective temperature. However, T\,Tauri stars usually have an important FUV excess and may become very bright in lines such as Ly$\alpha$ (at 1215.67 $\AA$). It is worth noting that TW\,Hya is brighter than AB\,Aurigae in the Ly$\alpha$ line and that in TW\,Hya the fraction of flux emitted in Ly$\alpha$ is about 30 \% of the total flux emitted in the 910-2400 $\AA$ wavelength range. We also note that while the FUV spectra of AB\,Aurigae is similar to that of a 10,000 K blackbody, a 4000 K blackbody is a bad approximation for a T\,Tauri star as it completely misses the FUV excess. As will be discussed in Sec.~\ref{sec:influence}, this has important consequences for the chemistry of the disk.

\subsubsection{Dust and gas temperature}

Given the input parameters characteristic of the star ($M_*$, $R_*$, and stellar spectrum), the radial distribution of surface density in the disk given by Eq.~(\ref{eq:surface_density}), the dust-to-gas mass ratio, and the dust properties, we solve for the two dimensional distribution of the dust temperature in the disk using the {\footnotesize RADMC} code (\cite{dul2004} 2004)\footnote{\texttt{http://www.mpia-hd.mpg.de/$\sim$dullemon/radtrans/radmc}}. {\footnotesize RADMC} is a two dimensional Monte Carlo code that solves the dust continuum radiative transfer in circumstellar disks and yields the dust temperature as a function of radius $r$ and height $z$ over the midplane. The vertical distribution of the volume density of particles $n(z)$ is assumed to be given by hydrostatic equilibrium as
\begin{equation}
\frac{d n(z)}{n(z)} = - \frac{\mu G M_*}{k T_d(z)} \frac{z}{(z^2 + r^2)^{3/2}} dz, \label{eq:hydrostatic_equilibrium}
\end{equation}
where $\mu$ is the mean mass of particles, $G$ the gravitational constant, $k$ the Boltzmann constant, and $T_d(z)$ is the vertical distribution of dust temperature. We proceed iteratively to find $n(z)$ at each radius $r$ according to Eq.~(\ref{eq:hydrostatic_equilibrium}) and consistently with the vertical distribution of dust temperature $T_d(z)$ computed at each radius $r$.

\begin{figure*}
\centering
\includegraphics[angle=0,width=\textwidth]{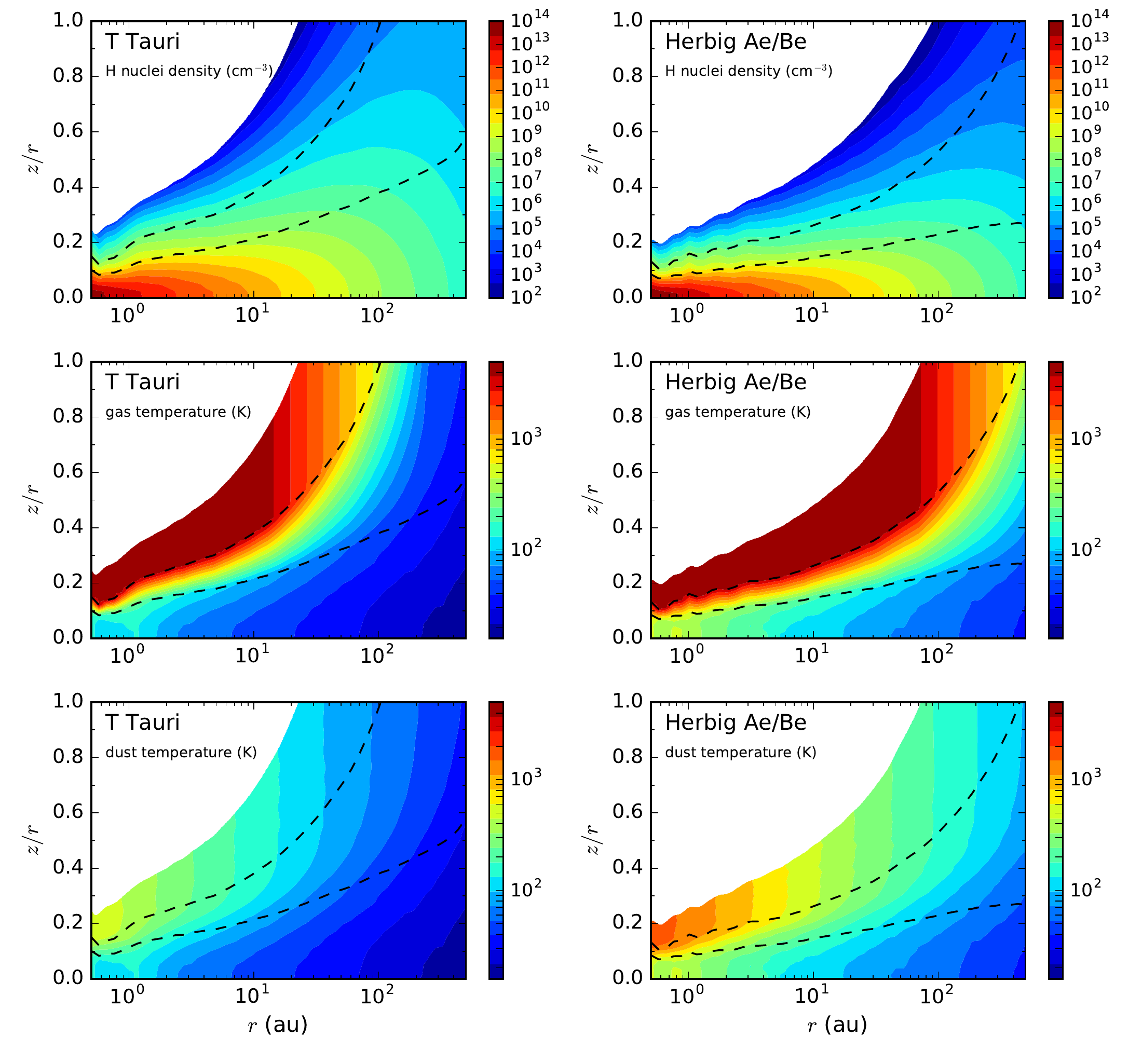}
\caption{Calculated volume density of H nuclei (top), gas temperature (middle), and dust temperature (bottom) as a function of radius $r$ and height over radius $z/r$ for the T\,Tauri (left) and Herbig\,Ae/Be (right) disks. The dashed lines indicate the location where $A_V$ in the outward vertical direction takes values of 0.01 and 1.} \label{fig:map_phys}
\end{figure*}

We assume that gas and dust are thermally coupled, i.e., gas and dust temperatures are equal, except for the disk surface. Models dealing with the computation of the gas temperature in protoplanetary disks find that the thermal coupling of gas and dust is a good approximation over most of the disk and that this assumption breaks down at the surface layers of the disk, where the visual extinction $A_V$ in the vertical outward direction becomes lower than $\sim$1 (\cite{kam2004} 2004; \cite{woi2009} 2009; \cite{wal2010} 2010). In these surface layers the gas can be much warmer than dust grains. In order to take this into account we use the following approximation for the gas temperature. We assume that the gas temperature is equal to the dust temperature in those regions where $A_V$ in the vertical outward direction is higher than 1. Following the study by \cite{kam2004} (2004), we assume that in the uppermost regions of the disk, where $A_V < 0.01$, the gas temperature is equal to the evaporation temperature of a hydrogen atom, calculated as the temperature at which the most probable speed of particles equals the escape velocity from the disk, i.e.,
\begin{equation}
T_{\rm evap} = \frac{G M_* m_{\rm H}}{k r},
\end{equation}
where $m_{\rm H}$ is the mass of a hydrogen atom. Finally, at regions intermediate between $A_V=1$ and $A_V=0.01$ we approximate the gas temperature through a linear interpolation with height. In Fig.~\ref{fig:map_phys} we show the two dimensional distributions of the volume density of particles and of the gas and dust temperatures calculated for the disks around the T\,Tauri and Herbig\,Ae/Be stars. The flared shape, which is apparent in both disks, is more prominent in the T\,Tauri disk owing to the lower gravity of the star. The most significant difference between the physical structure of both disks is that the disk around the Herbig\,Ae/Be star is significantly warmer than the disk around the T\,Tauri star because of the higher stellar irradiation of the former.

\subsection{Chemical model} \label{subsec:chemical_model}

Once the physical structure of the disk (temperature of gas and dust and volume density of particles) is calculated at steady state, we solve for the temporal evolution of the chemical composition at each location in the disk. Since transport processes are neglected, each location in the disk evolve independently of other disk regions. We solve the chemical composition as a function of time up to 1 Myr, which is of the order of the typical ages of protoplanetary disks, for a grid consisting of 50 radii (logarithmically spaced from $R_{\rm in}$ to $R_{\rm out}$) and 200 heights (generated specifically for each radius to properly sample the different regimes of $A_V$ in the vertical direction). The chemical network includes 252 species (97 neutral species, 133 positive ions plus the negative ion H$^-$ and free electrons, and 20 ice molecules) involving the elements H, He, C, N, O, S, Cl, and F. Atoms of Si, P, Fe, Na, and Mg are also included because their ionized forms are important in controlling the degree of ionization in certain disk regions. We adopt as initial chemical composition that calculated with a pseudo-time-dependent chemical model (where the chemical evolution is solved under fixed physical conditions) of a cold dense cloud with standard parameters (density of H nuclei of $2\times10^4$ cm$^{-3}$, temperature of 10 K, visual extinction of 10 mag, and cosmic-ray ionization rate of H$_2$ of $5\times10^{-17}$ s$^{-1}$) at a time of 0.1 Myr. The elemental abundances adopted, based in the so-called "low metal" case (e.g., \cite{gra1982} 1982; \cite{lee1998} 1998), are listed in Table~3 of \cite{agu2013} (2013). The chemical network comprises 5533 processes including gas phase chemical reactions, cosmic-ray induced processes, photodissociations and photoionizations due to stellar and interstellar FUV photons, and exchange processes between the gas and ice mantle phases (adsorption and desorption). At this stage, the model does not include X-ray induced processes and grain-surface reactions, except for the formation of H$_2$. X rays may be an important source of chemical differentiation between disks around T\,Tauri and Herbig\,Ae/Be stars because the former are more important X-ray emitters (see, e.g., \cite{tel2007} 2007). Moreover, winds and magnetic fields in actively accreting T\,Tauri systems may lead to cosmic-ray exclusion so that ionization in the disk can be dominated by X rays rather than cosmic rays (\cite{cle2015} 2015). Therefore, it will be interesting to study the impact of X rays on the chemistry of the two types of disks in the future. In any case, previous chemical models of T\,Tauri disks have found that the gas-phase chemistry is not greatly affected by X rays. The species whose abundance is most affected are, according to \cite{are2011} (2011), the ions present in the disk surface OH$^+$, H$_2$O$^+$, H$_3$O$^+$, and N$^+$, while \cite{wal2012} (2012) find that N$_2$H$^+$ is the most sensitive species to X rays. We however note that the models of \cite{are2011} (2011) and \cite{wal2012} (2012) did not consider cosmic-ray exclusion, unlike the study of \cite{cle2015} (2015). Chemical reactions occurring on the surface of dust grains are likely to have an effect on the chemical composition of cool midplane regions, especially regarding complex organic molecules (\cite{wal2014} 2014), although the distribution of abundant molecules and the main chemical patterns in the disk are probably not very much affected by such processes. We also plan to investigate this particular aspect in the future.

\subsubsection{Gas phase chemical reactions}

The vast majority of gas phase chemical reactions included can be grouped into two main categories: ion-neutral reactions and neutral-neutral reactions. The subset of ion-neutral reactions has been constructed based on databases originally devoted to the study of the chemistry of cold interstellar clouds, such as the {\footnotesize UMIST} database for astrochemistry (\cite{woodall2007} 2007; \cite{mce2013} 2013)\footnote{\texttt{http://udfa.ajmarkwick.net}} and the Ohio State University ({\footnotesize OSU}) database, formerly maintained by E. Herbst and currently integrated into the Kinetic Database for Astrochemistry ({\footnotesize KIDA}; \cite{wak2012} 2012, 2015)\footnote{\texttt{http://kida.obs.u-bordeaux1.fr}}. Rate constants of ion-neutral reactions have been taken from the previous databases and from the literature on chemical kinetics. In particular, a large part of the rate constants has been revised according to the compilation by \cite{ani2003} (2003)\footnote{\texttt{http://trs.jpl.nasa.gov/handle/2014/7981}}. The chemical kinetics of exothermic ion-neutral reactions is rather simple because in most cases the kinetics is dominated by long range electrostatic forces. In the case of reactions in which the neutral species is non polar the theory indicates that the rate constant is independent of temperature and is given by the Langevin value. If the neutral species has an electric dipole moment the expression found by \cite{su1982} (1982) can be used to evaluate the rate constant and its dependence with temperature (\cite{mae2009} 2009; see more details in \cite{wak2010} 2010). For ion-non polar reactions for which there is no experimental data, the rate constant has been approximated as the Langevin value. In the case of ion-polar reactions, we have used the Su-Chesnavich approach to evaluate the rate constant of reactions not studied experimentally and to obtain the temperature dependence of the rate constant of reactions which have been only characterized at one single temperature, usually around 300 K.

The part of the chemical network involving ions includes also dissociative recombinations of positive ions with electrons and radiative recombinations between cations and electrons. The set of reactions and associated rate constants have been mainly taken from databases such as {\footnotesize UMIST} (\cite{woodall2007} 2007; \cite{mce2013} 2013) and {\footnotesize KIDA} (\cite{wak2012} 2012, 2015). Information on the chemical kinetics of dissociative recombinations has largely benefited from experiments carried out with ion storage rings (\cite{flo2006} 2006; \cite{gep2008} 2008).

The subset of neutral-neutral reactions has been constructed from chemical kinetics databases, such as the one by NIST (\cite{man2013} 2013)\footnote{\texttt{http://kinetics.nist.gov}}, databases devoted to the study of interstellar chemistry, such as {\footnotesize UMIST} (\cite{woodall2007} 2007; \cite{mce2013} 2013) and {\footnotesize KIDA} (\cite{wak2012} 2012, 2015), compilations for application in atmospheric chemistry, such as the evaluations by {\footnotesize IUPAC} (\cite{atk2004} 2004, 2006)\footnote{\texttt{http://iupac.pole-ether.fr/}} and JPL (\cite{san2011} 2011)\footnote{\texttt{http://jpldataeval.jpl.nasa.gov/}}, and compilations for use in combustion chemistry, such as the {\footnotesize IUPAC} evaluation by \cite{bau2005} (2005) and the Leeds methane oxidation mechanism (\cite{hug2001} 2001) or the mechanism by \cite{kon2000} (2000). A good number of reaction rate constants have been taken from specific experimental and theoretical studies found in the literature on chemical kinetics. For example, we have included the numerous measurements at low and ultra low temperatures carried out with the {\footnotesize CRESU} apparatus (\cite{smi2006} 2006). It is important to note that some regions of protoplanetary disks may have temperatures up to some thousands of degrees Kelvin and therefore it is necessary to include reactions that become fast at high temperatures, i.e., reactions which are endothermic and/or have activation barriers. Chemical kinetics data for such reactions are to a large extent based on chemical networks used in previous chemical models of warm gas in protoplanetary nebulae (\cite{cer2004} 2004) and inner regions of protoplanetary disks (\cite{agu2008} 2008), whose original sources of data are mainly the combustion chemistry databases listed above. A similar high temperature chemical network has also been used by \cite{har2010} (2010) to model the chemistry of active galactic nuclei of galaxies. We include also three body reactions and their reverse process (thermal dissociation) with H$_2$, He, and H acting as third body. Three body reactions become important at densities above $\sim$10$^{10}$ cm$^{-3}$, values that are reached in the innermost midplane regions of protoplanetary disks, while thermal dissociations become important at high temperatures. We use an expanded version of the set of three body reactions and thermal dissociations compiled by \cite{agu2006} (2006). An important aspect of the neutral-neutral subset of reactions is that for many of the endothermic reactions for which chemical kinetics data are not available the rate constants have been calculated through detailed balance from the rate constant of the reverse exothermic reaction and the thermochemical properties of the species involved. Thermochemical data in the form of NASA polynomial coefficients (\cite{mcb2002} 2002) have been obtained from compilations like those by \cite{kon2000} (2000) and \cite{bur2005} (2005)\footnote{\texttt{http://burcat.technion.ac.il/dir}}.

\subsubsection{Cosmic-ray induced processes}

The processes induced by cosmic rays play also an important role in the chemistry of protoplanetary disks. We include the direct ionization of H$_2$ and atoms by cosmic-ray impact, together with photoprocesses induced by secondary electrons produced in the direct ionization of H$_2$, the so-called Prasad-Tarafdar mechanism (\cite{pra1983} 1983; \cite{gre1989} 1989). The rates of these processes are expressed in terms of the cosmic-ray ionization rate of H$_2$ ($\zeta$), for which we adopt a value of $5\times10^{-17}$ s$^{-1}$ (see Table~\ref{table:parameters}), and are taken from astrochemical databases such as {\footnotesize UMIST} (\cite{woodall2007} 2007; \cite{mce2013} 2013) and {\footnotesize KIDA} (\cite{wak2012} 2012, 2015).

\subsubsection{Photoprocesses} \label{sec:photochemistry}

Photodissociation and photoionization processes caused by stellar and interstellar FUV photons are a key aspect of the chemistry of protoplanetary disks as they control the chemical composition of the surface layers, from where much of the molecular emission arises. In order to treat in detail these processes we have used the Meudon PDR code (\cite{lep2006} 2006; \cite{goi2007} 2007; \cite{gon2008} 2008)\footnote{\texttt{http://ism.obspm.fr}}, where PDR stands for photodissociation region, to compute the photodissociation and photoionization rates of various important species at each location in the disk. We use the version 1.4 of the Meudon PDR code, with some practical modifications to make it more versatile and integrate it into the protoplanetary disk model. The Meudon PDR code solves the FUV radiative transfer in one dimension for a plane-parallel cloud illuminated on one side by a FUV source. In protoplanetary disks the geometry involves two dimensions (radial and vertical) and there are two different FUV sources, the star illuminating from the central position and the interstellar radiation field illuminating isotropically from outside the disk. We thus adopt a 1+1D approach. On the one hand we solve the radiative transfer of FUV interstellar radiation as it propagates from outside the disk to the midplane along a series of vertical directions located at the different radii of the grid described in Sec.~\ref{subsec:chemical_model}. On the other hand we solve the radiative transfer of FUV stellar radiation as it travels from the star through the disk along a series of directions given by a grid of 200 angles from the midplane (covering the range from 0$^{\circ}$ to almost 90$^{\circ}$). The Meudon PDR code assumes that stellar photons arrive from a direction perpendicular to the plane-parallel surface of the cloud. In disks, the star may illuminate the disk with small grazing angles, in particular in the inner regions where the flaring shape of the disk is less marked. It is thus likely that when computing the FUV energy density along the different directions from the star, at low penetration depths the Meudon PDR code underestimates the contribution of FUV photons scattered by dust from nearby regions around the disk surface. However, it is not straightforward to properly correct by this geometrical effect without moving to 2D and thus we do not apply any specific correction for it here.

\begin{figure}
\centering
\includegraphics[angle=0,width=\columnwidth]{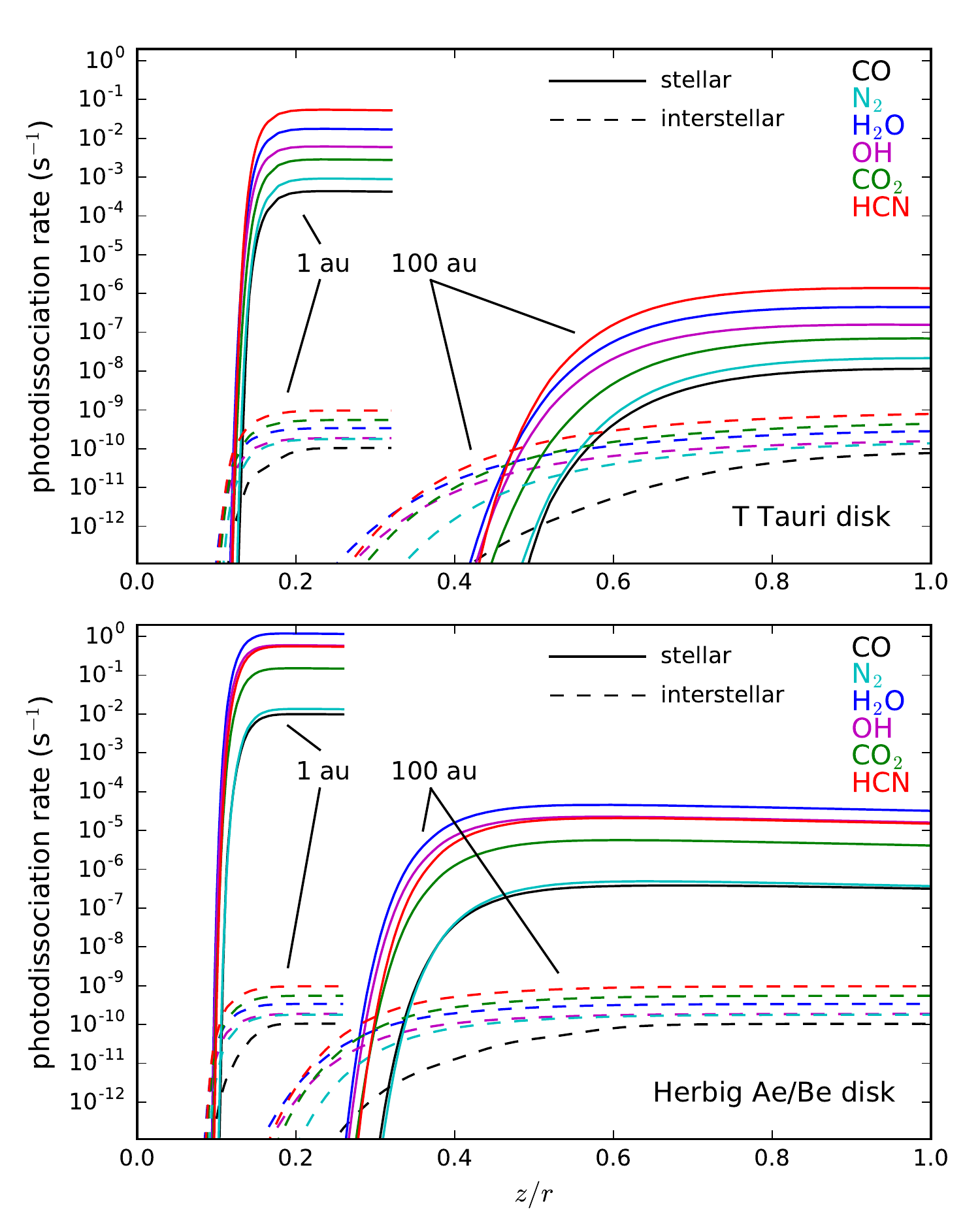}
\caption{Contribution of the stellar (solid lines) and interstellar (dashed lines) FUV fields to the photodissociation rates of various molecules as a function of the height over radius ($z/r$) at 1 au and 100 au in the T\,Tauri (upper panel) and Herbig\,Ae/Be (lower panel) disks.} \label{fig:photorates}
\end{figure}

In summary, once the physical structure of the disk is calculated at steady state (using the {\footnotesize RADMC} code as described in Sec.~\ref{sec:physical_model}) and prior to the computation of the temporal evolution of the chemical composition, we use the Meudon PDR code to evaluate the photodissociation and photoionization rates at each location in the disk by calculating the FUV flux as a function of wavelength due to stellar and interstellar radiation and using the relevant wavelength-dependent cross sections. Our approach to treat photochemistry is thus different from other state-of-the-art chemical models of protoplanetary disks in which the FUV radiative transfer is solved in 2D but only in a few broad spectral bands (e.g., \cite{woi2016} 2016) and it is in essence more similar to the series of models by \cite{wal2012} (2012, 2014, 2015).

We have compiled cross sections for 29 molecules and 8 atoms (see Appendix~\ref{sec:app_sections}). The photodissociation rate of H$_2$ and CO are computed by solving the excitation and the line-by-line radiative transfer taking into account self and mutual shielding effects. In the case of photoprocesses for which cross section data is not available we have approximated the rate using a parametric expression in which the rate $\Gamma$, in units of s$^{-1}$, is expressed as a function of the visual extinction $A_V$ as
\begin{equation}
\Gamma = \chi \alpha \exp (-\gamma A_V), \label{eq:photorate}
\end{equation}
where $\chi$ is the FUV\footnote{The FUV is chosen to cover the wavelength range from the Lyman cutoff at 911.776~$\AA$ to the limit of the Habing field at 2400~$\AA$.} energy density with respect to that of the ISRF of \cite{dra1978} (1978), $\alpha$ is the rate under a given unattenuated radiation field with $\chi=1$, and the coefficient $\gamma$ controls the decrease in the rate with increasing visual extinction. The coefficients $\alpha$ and $\gamma$ are specific of each photoprocess and depend also on the spectral shape of the FUV field. Values of $\alpha$ and $\gamma$ corresponding to the ISRF have been taken from the databases such as the {\footnotesize UMIST} database for astrochemistry (\cite{woodall2007} 2007; \cite{mce2013} 2013) and the OSU and KIDA databases (\cite{wak2012} 2012, 2015), as well as from the compilation by \cite{van2006} (2006), recently revised by \cite{hea2017} (2017).

In Fig.~\ref{fig:photorates} we show the contribution of the stellar and interstellar radiation fields to the photodissociation rate of various molecules as a function of height over the disk midplane. This is shown at two radial distances from the star (1 au and 100 au) for the T\,Tauri and Herbig\,Ae/Be disks. We see that in the uppermost regions of the disk, photoprocesses are clearly dominated by the stellar radiation field. However, as we go deeper into the disk, at some point the contribution of the ISRF becomes more important than the stellar one because of the more marked increase of the visual extinction against stellar light than against interstellar photons. Therefore, depending on the location in the disk, photodestruction can be driven by the ISRF or by the star.

\subsubsection{Reactions with vibrationally excited H$_2$}

In the disk surface the gas is strongly illuminated by FUV photons and vibrationally excited states of molecular hydrogen are easily populated through FUV fluorescence. Since the reactivity of H$_2$ can be quite different when it is in the ground or in excited vibrational states --the internal energy of H$_2$ can be used to overcome or diminish endothermicities or activation barriers which are present when H$_2$ is in its ground vibrational state-- we have included a few reactions of H$_2$ with specific rate constants for each vibrational state. Concretely, we have included the reactions of H$_2$ with C$^+$ (important in the formation of CH$^+$), He$^+$, O, OH, and CN, with the rate constant expressions compiled by \cite{agu2010} (2010), and the reaction of H$_2$ and S$^+$, which may be important in the synthesis of the ion SH$^+$, with the rate constant expressions calculated by \cite{zan2013} (2013). The populations of the different vibrational states of H$_2$ are computed at each location in the disk with the Meudon PDR code.

\subsubsection{Adsorption processes} \label{subsubsec:adsorption}

The adsorption of gas species onto the surface of dust grains is treated in a rather simple and standard way. The adsorption rate of a gas species $i$, in units of s$^{-1}$, is given by
\begin{equation}
R_i^{ads} = \alpha_i v_i \langle \sigma_d n_d \rangle, \label{eq:adsorption_rate}
\end{equation}
where $\alpha_i$ and $v_i$ are the sticking coefficient and thermal velocity of each gas species $i$. The sticking coefficient is assumed to be 1 for all species and the thermal velocity is evalued as $\sqrt{3 k T_k / m_i}$, where $T_k$ is the gas kinetic temperature and $m_i$ is the mass of each gas species $i$. The term $\langle \sigma_d n_d \rangle$ is the product of the geometric cross section and the volume density of dust particles averaged over the grain size distribution given by Eq.~(\ref{eq:grain_size_distribution}) and is evaluated following the formalism of \cite{leb1995} (1995). To simplify, we only consider adsorption of a limited number of stable neutral species, among which there are the most typically abundant ice constituents (see Table~\ref{table:ices}).

\subsubsection{Desorption processes} \label{subsubsec:desorption}

Species which have been adsorbed on dust grains forming ice mantles can return to the gas phase through a variety of desorption mechanisms. In protoplanetary disks, the most important desorption processes are thermal desorption and photodesorption by FUV photons. We also include desorption induced by cosmic rays. Based on the results of experimental work using isotopic markers (\cite{ber2012} 2012, 2013) and molecular dynamics calculations (\cite{and2008} 2008), we consider that only molecules from the top two monolayers can desorb efficiently ($N_l$ = 2). The term due to desorption can therefore be written in the kinetic rate equations as
\begin{equation}
\frac{dn_i}{dt} = - \frac{dn_i^{ice}}{dt} = (R_i^{thd} + R_i^{pd} + R_i^{crd}) n_i^{ice, {\rm desorbable}}, \label{eq:desorption_rate}
\end{equation}
where $n_i$ and $n_i^{ice}$ are the volume densities of species $i$ in the gas and ice phases, respectively, $R_i^{thd}$, $R_i^{pd}$, $R_i^{crd}$ are the desorption rates, in units of s$^{-1}$, of thermal desorption, photodesorption, and desorption induced by cosmic rays, respectively (see below). The quantity $n_i^{ice, {\rm desorbable}}$ is the volume density of species $i$ in the top desorbable ice layers, which is simply equal to $n_i^{ice}$ when these layers are not fully occupied while it is given by $n_i^{ice}$ multiplied by the factor ($n_{top}$/$n_{tot}^{ice}$) otherwise. The volume density of sites in the top desorbable layers $n_{top}$ is given by $n_s 4 \langle \sigma_d n_d \rangle N_l$ (where $n_s$ is the surface density of sites, typically $\sim1.5\times10^{15}$~cm$^{-2}$; see \cite{has1992} 1992) and $n_{tot}^{ice}$ is the sum of the volume densities of all ice species, i.e., $\sum_i n_i^{ice}$. This implementation of desorption in the kinetic rate equations is similar to that of, e.g., \cite{aik1999} (1996) and \cite{woi2009} (2009); see also \cite{cup2017} (2017).

\begin{table}
\caption{Ice species, binding energies, and photodesorption yields.} \label{table:ices}
\centering
\begin{tabular}{lr@{\hspace{0.3cm}}l@{\hspace{0.5cm}}r@{\hspace{0.3cm}}l}
\hline \hline
Species         & $E_D$ (K) & Ref. & \multicolumn{1}{c}{$Y$ (molecule~photon$^{-1}$)} & Ref. \\
\hline
CH$_4$         & 1000 & \tiny{(1) $\diamondsuit$} & $10^{-3}$ & \tiny{(11)} \\
C$_2$H$_2$ & 2587 & \tiny{(2) $\ddag$}            & $10^{-3}$ & \tiny{(11)} \\
C$_2$H$_4$ & 3487 & \tiny{(2) $\ddag$}            & $10^{-3}$ & \tiny{(11)} \\
C$_2$H$_6$ & 4387 & \tiny{(2) $\ddag$}            & $10^{-3}$ & \tiny{(11)} \\
H$_2$O         & 5773 & \tiny{(3) $\diamondsuit$} & $(1.3+0.032~T_d)\times10^{-3}$ & \tiny{(12) $*$} \\
O$_2$           & 1161 & \tiny{(4) $\dag~\natural$} & $(2.8^a, 2.3^b, 2.4^c)\times10^{-3}$ & \tiny{(13) $\S$} \\
CO                & 1575 & \tiny{(5) $\dag~\natural$} & $(9.5^a, 4.8^b, 10.3^c)\times10^{-3}$ & \tiny{(14) $\S$} \\
CO$_2$        & 2346 & \tiny{(4) $\dag~\natural$} & $(8.2^a, 2.0^b, 0.98^c)\times10^{-4}$ & \tiny{(15) $\S~*$} \\
H$_2$CO      & 3260 & \tiny{(6) $\dag$}              & $10^{-3}$ & \tiny{(11)} \\
CH$_3$OH   & 4990 & \tiny{(7) $\diamondsuit$} & $(1.5^a, 1.1^b, 1.3^c)\times10^{-4}$ & \tiny{(16) $\S~*$} \\
HCOOH        & 5570 & \tiny{(2) $\ddag$}            & $10^{-3}$ & \tiny{(11)} \\
NH$_3$        & 3830 & \tiny{(1) $\diamondsuit$} & $10^{-3}$ & \tiny{(11)} \\
N$_2$           & 1435 & \tiny{(5) $\dag~\natural$} & $(2.0^a, 1.6^b, 0.89^c)\times10^{-3}$ & \tiny{(13) $\S$} \\
HCN              & 2050 & \tiny{(2) $\ddag$}            & $10^{-3}$ & \tiny{(11)} \\
H$_2$S        & 1945  & \tiny{(8) $\sharp$}           & $10^{-3}$ & \tiny{(11)} \\
CS                & 1900 & \tiny{(2) $\ddag$}             & $10^{-3}$ & \tiny{(11)} \\
H$_2$CS      & 2700 & \tiny{(2) $\ddag$}            & $10^{-3}$ & \tiny{(11)} \\
SO                & 2600 & \tiny{(2) $\ddag$}            & $10^{-3}$ & \tiny{(11)} \\
SO$_2$        & 3900 & \tiny{(9) $\dag$}              & $10^{-3}$ & \tiny{(11)} \\
OCS             & 3440 & \tiny{(10) $\diamondsuit$} & $10^{-3}$ & \tiny{(11)} \\
\hline
\end{tabular}
\tablenoteb{
$\diamondsuit$ pure ice; $\dagger$ amorphous water ice substrate ($\natural$ submonolayer regime); $\sharp$ solid SO$_2$ substrate; $\ddag$ estimated for water ice substrate; $\S$ wavelength-dependent photodesorption yield is convolved over the 7-13.6 eV range with the interstellar$^a$, T\,Tauri$^b$, and Herbig\,Ae/Be$^c$ FUV radiation fields described in section~\ref{subsubsec:uv_field}; $*$ yields of direct photodesorption and fragmentation have been measured.
}
\tablerefs{
(1) \cite{lun2014} (2014); (2) \cite{gar2006} (2006); (3) \cite{fra2001} (2001); (4) \cite{nob2012a} (2012a); (5) \cite{fay2016} (2016); (6) \cite{nob2012b} (2012b); (7) \cite{dor2015} (2015); (8) \cite{san1993} (1993); (9) \cite{sch2003} (2003), assuming that $E_D$ is directly proportional to the desorption temperature (see, e.g., \cite{mar2014} 2014); (10) \cite{bur2010} (2010); (11) assumed; (12) pure H$_2$O ice ($>8$~ML) in the 18-100~K range; the fraction of H$_2$O molecules photodesorbed is $(0.42 + 0.002~T_d)$, the remaining results in fragmentation into OH + H (\cite{obe2009} 2009); (13) pure O$_2$ (30~ML) and $^{15}$N$_2$ (60~ML) ices at 15~K (\cite{fay2013} 2013); (14) pure CO ice (10 ML) at 18~K (\cite{fay2011} 2011); (15) pure $^{13}$CO$_2$ (10~ML) ice at 10~K; the fraction of CO$_2$ molecules desorbed is 0.37, 0.42, and 0.22 under the interstellar, T\,Tauri, and Herbig\,Ae/Be FUV fields, respectively, the remaining results in fragmentation into CO + O (\cite{fil2014} 2014); (16) pure CH$_3$OH (20~ML) ice at 10~K; the fraction of CH$_3$OH molecules desorbed is 0.08, while the fragmentation channels CH$_3$ + OH, H$_2$CO + H$_2$, CO + H$_2$ + H$_2$ occur with fractions of 0.05, 0.07, and 0.80, almost independently of the radiation field (\cite{ber2016} 2016; see also \cite{cru2016} 2016).
}
\end{table}

\emph{-- Thermal desorption}. This process, which depends on the dust temperature $T_d$ and the binding energy of adsorption of each species $E_D$, controls to a large extent the distribution of ices in protoplanetary disks and the location of the different snow lines of each molecule. The thermal desorption rate, in units of s$^{-1}$, of an adsorbed species $i$ is given by
\begin{equation}
R_i^{thd} = \nu_{0,i} \exp{(-E_{D,i}/T_d)}, \label{eq:thermal_desorption}
\end{equation}
where $\nu_{0,i}$ is the characteristic vibration frequency of the adsorbed species $i$ (evaluated as $\sqrt{2 n_s k E_{D,i}/(\pi^2 m_i)}$, where the binding energy $E_{D,i}$ is expressed in units of K. Binding energies have been measured in the laboratory by depositing volatile species on different cold substrates and using temperature programmed desorption methods (see \cite{bur2010} 2010 and references therein). In general, binding energies show little dependence with the chemical or morphological nature of the substrate as long as the ice under study consists of various monolayers (ML). If desorption occurs at submonolayer coverage, desorption energies can be quite different depending on the substrate, e.g., they tend to be higher if employing water ice as substrate than using silicates (\cite{nob2012a} 2012a). The translation of these laboratory experiments into a realistic model of thermal desorption in protoplanetary disks is complicated because ices are heterogeneous mixtures, thought to be dominated by water ice but whose composition probably varies between cold and warmer regions. Also, molecules may selectively co-desorb with other trapped species, experience volcano desorption following the crystallization of water ice, or co-desorb with some of the major ice constituents (\cite{col2004} 2004). Here we have adopted the simple and usual approach in which the thermal desorption of each species is controlled by a specific binding energy. We have collected values of $E_D$ experimentally measured, when possible using a water ice substrate and under a submonolayer regime. For those molecules for which experimental binding energies are not available, we have adopted the values estimated by \cite{gar2006} (2006) for a water ice substrate based on the previous compilation by \cite{has1993} (1993) and the experimental study of \cite{col2004} (2004). The binding energies of the ice molecules considered and the corresponding references are given in Table~\ref{table:ices}.

\emph{-- Photodesorption}. The absorption of FUV photons of stellar or interstellar origin (or generated through the Prasad-Tarafdar mechanism) by icy dust grains can induce the desorption of molecules on the ice surface. In regions where the dust temperature is too cold to allow for thermal desorption, photodesorption can provide an efficient means to bring ice molecules to the gas phase. The photodesorption rate, in units of s$^{-1}$, of an adsorbed species $i$ is given by
\begin{equation}
R_i^{pd} = Y_i F_{\rm FUV} \frac{ \langle \sigma_d n_d \rangle}{4 \langle \sigma_d n_d \rangle n_s N_l}, \label{eq:photodesorption}
\end{equation}
where $Y_i$ is the yield of molecules desorbed per incident photon, $F_{\rm FUV}$ is the FUV photon flux (in units of photon cm$^{-2}$ s$^{-1}$), and the expression of $R_i^{pd}$, when inserted in Eq.~(\ref{eq:desorption_rate}), naturally accounts for the fact that ices are not pure but consist of multiple constituents and that desorption is only effective from the top desorbable layers (see, e.g., \cite{cup2017} 2017). FUV photons may have an stellar or interstellar origin (in which case $F_{\rm FUV}$ is evaluated with the Meudon PDR code at each position in the disk) or can be generated through the Prasad-Tarafdar mechanism, in which case we adopt $F_{\rm FUV}$ = $2\times10^3$ photon cm$^{-2}$ s$^{-1}$ (values between 750 and 10$^4$ photon cm$^{-2}$ s$^{-1}$ have been reported in the literature; e.g., \cite{har1990} 1990; \cite{she2004} 2004). Note that this latter value scales with the cosmic-ray ionization rate. Experiments carried out to study the photodesorption of pure ices or binary ice mixtures (\cite{obe2009} 2009; \cite{mun2010} 2010; \cite{fay2011} 2011, 2013; \cite{ber2013} 2013; \cite{fil2014} 2014; \cite{mar2015} 2015) suggest that the main underlying mechanism, called desorption induced by electronic transition (DIET), involves absorption of FUV photons in the $\sim5$ top monolayers and electronic excitation of the absorbing molecules, followed by energy redistribution to neighbouring molecules, which may break their intermolecular bonds and be ejected into the gas phase. The efficiency of photodesorption is thus regulated by the ability of the molecules present in the ice surface and sub-surface to absorb FUV photons through electronic transitions. If these transitions are dissociative the situation becomes more complex because the fragments may desorb directly, recombine in the ice and then desorb, or diffuse through the ice forming new molecules that may also desorb (e.g., \cite{and2008} 2008). Here we adopt a simple approach in which ice molecules may desorb directly or as fragments upon FUV irradiation, with yields based on experimental data for CO, N$_2$, O$_2$, H$_2$O, and CO$_2$ (see Table~\ref{table:ices}). Desorption of fragments has been observed upon irradiation of pure ices of H$_2$O, CO$_2$, and CH$_3$OH (\cite{obe2009} 2009; \cite{fil2014} 2014; \cite{mar2015} 2015; \cite{ber2016} 2016; \cite{cru2016} 2016), and it is likely that desorption of dissociation fragments and new species formed in situ in the ice dominates over direct desorption for other polyatomic molecules. However, in the absence of experimental photodesorption yields for molecules other than CO, N$_2$, O$_2$, H$_2$O, CO$_2$, and CH$_3$OH we have assumed that direct desorption dominates with assumed values for $Y_i$. It is interesting to note that photodesorption yields of CO, O$_2$, N$_2$, CO$_2$, and CH$_3$OH have been measured as a function of wavelength using synchrotron techniques (\cite{fay2011} 2011, 2013; \cite{fil2014} 2014; \cite{ber2016} 2016), which permits to compute the yield $Y_i$ under different FUV fields (see values for the ISRF and the T\,Tauri and Herbig\,Ae/Be stellar radiation fields in Table~\ref{table:ices}). Note for example that the photodesorption yield of CO$_2$ is almost one order of magnitude higher under the ISRF than under a Herbig\,Ae/Be stellar field.

\emph{-- Cosmic-ray induced desorption}. This mechanism is driven by the impact of cosmic rays on dust grains. The energy deposited on dust grains upon impact of relativistic heavy nuclei of iron results in a local heating that induces the thermal desorption of the ice molecules present in the heated region. According to \cite{has1993} (1993), the desorption rate induced by cosmic rays, in units of s$^{-1}$, of a species $i$ is given by
\begin{equation}
R_i^{crd} = 3.16\times10^{-19} R_i^{thd}({\rm 70 K}), \label{eq:cosmic_ray_desorption}
\end{equation}
where the numerical factor stands for the fraction of the time spent by grains in the vicinity of a temperature of 70 K, at which much of the desorption is assumed to occur in the formalism of \cite{has1993} (1993), and is derived adopting the Fe cosmic ray flux estimated by \cite{leg1985} (1985) for the local interstellar medium and dust grains with a radius of 0.1~$\mu$m. The term $R_i^{thd}$(70 K) is the rate of thermal desorption of species $i$, given by Eq.~(\ref{eq:thermal_desorption}), evaluated at 70~K. The desorption rate $R_i^{crd}$ scales with the cosmic-ray ionization rate. 

\begin{table*}[ht]
\caption{Summary of molecules (other than H$_2$ and CO) observed in disks and abundances derived.} \label{table:observations}
\centering
\begin{tabular}{llllll}
\hline \hline
Species & Spectral signature, region probed & \multicolumn{2}{c}{Detection rate, abundance, and references} \\
 & & \multicolumn{1}{c}{T\,Tauri} & \multicolumn{1}{c}{Herbig\,Ae/Be} \\
\hline
H$_2$O         & IR emission, inner atmosphere                         & many disks \hfill $^{(1, 2, 3, 4, 5, 6, 7, 8, 9, 10, 11)}$                & one disk: HD\,163296 \hfill $^{(4, 7, 11, 12, 13, 14, 15)}$ \\
                      &                                                                           & $(0.4-800)\times10^{18}$ cm$^{-2}$ \hfill $^{(6, 7)}$                & $10^{14}-10^{15}$ cm$^{-2}$ \hfill $^{(14)}$ \\
                      & sub-mm emission, outer disk                            & two disks: TW\,Hya, DG\,Tau \hfill $^{(16, 17, 18)}$                  & one disk: HD\,100546 \hfill $^{(18, 19)}$ \\
                      &                                                                           & $\sim10^{-7}$ relative to H$_2$ \hfill $^{(16)}$                          & \\

\hline
OH                 & IR emission, inner atmosphere                         & many disks \hfill $^{(2, 3, 4, 6, 7, 8, 11, 15, 20)}$                      & a dozen of disks \hfill $^{(4, 7, 11, 12, 21)}$ \\
                      &                                                                           & $(0.4-200)\times10^{15}$ cm$^{-2}$ \hfill $^{(6, 7)}$                 & $(0.01-200)\times10^{15}$ cm$^{-2}$ \hfill $^{(12, 15, 21)}$ \\
\hline
C$_2$H$_2$ & IR emission/absorption, inner disk                    & many disks \hfill $^{(2, 6, 7, 8, 22, 23, 24)}$                               & none \\
                      &                                                                           & $(0.5-70)\times10^{15}$ cm$^{-2}$ \hfill $^{(6, 7, 22, 23, 24)}$ & \\
\hline
HCN              & IR emission/absorption, inner disk                    & many disks \hfill $^{(2, 6, 7, 8, 22, 23, 24)}$                               & none \\
                      &                                                                           & $(0.5-65)\times10^{15}$ cm$^{-2}$ \hfill $^{(6, 7, 22, 23, 24)}$ & \\
                      & mm emission, outer disk                                   & many disks \hfill $^{(27, 28, 29, 30, 31, 32, 33, 35)}$                 & many disks \hfill $^{(28, 29, 30, 31, 32, 33, 34)}$ \\
                      &                                                                           & $(0.2-10.6)\times10^{12}$ cm$^{-2}$ \hfill $^{(28, 31, 32, 33)}$ & $(0.1-1)\times10^{12}$ cm$^{-2}$ \hfill $^{(31, 33, 34)}$ \\
\hline
HNC              & mm emission, outer disk                                   & two disks: DM\,Tau, TW\,Hya \hfill $^{(27, 36)}$                         & one disk: HD\,163296 \hfill $^{(36)}$ \\
                      &                                                                           & HNC/HCN = 0.3-0.4 \hfill $^{(27)}$                                            & HNC/HCN = $0.1-0.2$ \hfill $^{(36) a}$ \\
\hline
CH$_4$         & IR absorption, inner disk                                   & one disk: GV\,Tau \hfill $^{(25)}$                                                & none \\
                      &                                                                           & $2.8\times10^{17}$ cm$^{-2}$ \hfill $^{(25)}$                             & \\
\hline
CO$_2$         & IR emission/absorption, inner disk                    & many disks \hfill $^{(6, 7, 22, 24, 26)}$                                      & one disk: HD\,101412 \hfill $^{(7)}$ \\
                      &                                                                           & $(0.04-10)\times10^{16}$ cm$^{-2}$ \hfill $^{(7, 22, 24, 26)}$  & $10^{16}$ cm$^{-2}$ \hfill $^{(7)}$ \\
\hline
C$_2$H         & mm emission, outer disk                                   & many disks \hfill $^{(27, 32, 33, 37, 38)}$                                  & two disks: AB\,Aur, MWC\,480 \hfill $^{(33, 34, 37, 39, 40)}$ \\
                      &                                                                           & $(0.4-8.5)\times10^{13}$ cm$^{-2}$ \hfill $^{(33, 37) b}$          & $(0.6-1)\times10^{13}$ cm$^{-2}$ \hfill $^{(33)}$ \\
\hline
CN                 & mm emission, outer disk                                   & many disks \hfill $^{(27, 28, 29, 30, 31, 32, 33)}$                       & a handful of disks \hfill $^{(28, 29, 30, 31, 33, 34)}$ \\
                      &                                                                           & $(0.7-11.5)\times10^{13}$ cm$^{-2}$ \hfill $^{(28, 31, 32, 33)}$ & $(0.1-1.5)\times10^{13}$ cm$^{-2}$ \hfill $^{(28, 31, 33, 34)}$ \\
\hline
H$_2$CO      & mm emission, outer disk                                   & a dozen of disks \hfill $^{(27, 28, 29, 30, 33, 41, 43, 44)}$         & a handful of disks \hfill $^{(30, 33, 34, 42, 45, 46)}$ \\
                      &                                                                           & $(0.9-4.4)\times10^{12}$ cm$^{-2}$ \hfill $^{(33) c}$                   & $(1.6-3.3)\times10^{12}$ cm$^{-2}$ \hfill $^{(33) d}$  \\
\hline
CH$_3$OH   & mm emission (ALMA), outer disk                       & one disk: TW\,Hya; $(3-6)\times10^{12}$ cm$^{-2}$ \hfill $^{(47)}$ & none \\
\hline
HC$_3$N      & mm emission, outer disk                                   & two disks: GO\,Tau, LkCa\,15; $\sim10^{12}$ cm$^{-2}$ \hfill $^{(48)}$ & one disk: MWC\,480; $\sim10^{12}$ cm$^{-2}$ \hfill $^{(48, 49)}$ \\
\hline
CH$_3$CN    & mm emission (ALMA), outer disk                      & none                                                                                            & one disk: MWC\,480; $\sim10^{13}$ cm$^{-2}$ \hfill $^{(49)}$ \\
\hline
$c$-C$_3$H$_2$ & mm emission (ALMA), outer disk               & one disk: TW\,Hya \hfill $^{(38)}$                                                & one disk: HD\,163296; $10^{12}-10^{13}$ cm$^{-2}$ \hfill $^{(50)}$ \\
\hline
NH$_3$         & sub-mm emission, outer disk                            & one disk: TW\,Hya \hfill $^{(51)}$                                                & none \\
                      &                                                                           & $(0.2-17)\times10^{-11}$ relative to H$_2$ \hfill $^{(51)}$          & \\
\hline
CS                 & mm emission, outer disk                                   & a dozen of disks \hfill $^{(27, 32, 33, 52)}$                                 & two disks: AB\,Aur, MWC\,480 \hfill $^{(33, 34, 40)}$ \\
                      &                                                                           & $(1-20.9)\times10^{12}$ cm$^{-2}$ \hfill $^{(32, 33, 52)}$          & $(0.4-6.3)\times10^{12}$ cm$^{-2}$ \hfill $^{(33, 34)}$ \\
\hline
SO                 & mm emission, outer disk                                   & two disks: CI\,Tau, GM\,Aur \hfill $^{(33)}$                                  & one disk: AB\,Aur \hfill $^{(33, 34, 40, 42)}$ \\
                      &                                                                           & $(7.4-9)\times10^{12}$ cm$^{-2}$ \hfill $^{(33)}$                        & $(0.5-10)\times10^{12}$ cm$^{-2}$ \hfill $^{(33, 34, 42)}$ \\
\hline
HCO$^+$      & mm emission, outer disk                                   & many disks \hfill $^{(28, 29, 30, 33, 35, 53, 54, 55)}$                  & many disks \hfill $^{(28, 29, 30, 33, 34, 54, 56)}$  \\
                      &                                                                           & $(0.3-20)\times10^{12}$ cm$^{-2}$ \hfill $^{(28, 33, 53, 54, 55)}$ & $(0.2-5)\times10^{12}$ cm$^{-2}$ \hfill $^{(28, 33, 34, 54) e}$ \\
\hline
N$_2$H$^+$ & mm emission, outer disk                                   & a handful of disks \hfill $^{(29, 30, 44, 53, 54, 57)}$                   & one disk: HD\,163296 \hfill $^{(45)}$  \\
                      &                                                                           & $(0.1-30)\times10^{12}$ cm$^{-2}$ \hfill $^{(53, 54) f}$             & $1.7\times10^{11}$ cm$^{-2}$ \hfill $^{(45)}$  \\
\hline
CH$^+$         & sub-mm emission, mid disk                              & none                                                                                            & two disks: HD\,100546, HD\,97048 \hfill $^{(15, 58)}$  \\
                      &                                                                           &                                                                                                    & $4.3\times10^{12}$ cm$^{-2}$ \hfill $^{(58) g}$ \\
\hline
\end{tabular}
\tablenotec{References: (1) \cite{car2004} (2004); (2) \cite{car2008} (2008); (3) \cite{sal2008} (2008); (4) \cite{pon2010a} (2010a); (5) \cite{pon2010b} (2010b); (6) \cite{car2011} (2011); (7) \cite{sal2011} (2011); (8) \cite{man2012} (2012); (9) \cite{riv2012} (2012); (10) \cite{sar2014} (2014); (11) \cite{ban2017} (2017); (12) \cite{fed2011} (2011); (13) \cite{mee2012} (2012); (14) \cite{fed2012} (2012); (15) \cite{fed2013} (2013); (16) \cite{hog2011} (2011); (17) \cite{pod2013} (2013); (18) \cite{du2017} (2017); (19) \cite{van2014} (2014); (20) \cite{car2014} (2014); (21) \cite{man2008} (2008); (22) \cite{lah2006} (2006); (23) \cite{gib2007} (2007); (24) \cite{bas2013} (2013); (25) \cite{gib2013} (2013); (26) \cite{kru2011} (2011); (27) \cite{dut1997} (1997); (28) \cite{thi2004} (2004); (29) \cite{obe2010} (2010); (30) \cite{obe2011} (2011); (31) \cite{cha2012a} (2012a); (32) \cite{kas2014} (2014); (33) \cite{gui2016} (2016); (34) \cite{fue2010} (2010); (35) \cite{fue2012} (2012); (36) \cite{gra2015} (2015); (37) \cite{hen2010} (2010); (38) \cite{bergin2016} (2016); (39) \cite{sch2008} (2008); (40) \cite{pac2015} (2015); (41) \cite{aik2003} (2003); (42) \cite{pac2016} (2016); (43) \cite{loo2015} (2015); (44) \cite{obe2017} (2017); (45) \cite{qi2013a} (2013a); (46) \cite{car2017} (2017); (47) \cite{wal2016} (2016); (48) \cite{cha2012b} (2012b); (49) \cite{obe2015} (2015); (50) \cite{qi2013b} (2013b); (51) \cite{sal2016} (2016); (52) \cite{dut2011} (2011); (53) \cite{qi2003} (2003); (54) \cite{dut2007} (2007); (55) \cite{tea2015} (2015); (56) \cite{mat2013} (2013); (57) \cite{qi2013c} (2013c); (58) \cite{thi2011} (2011).\\
Notes: $^a$ Values are line intensity ratios rather than abundance ratios. $^b$ Out of this range, \cite{kas2014} (2014) derive $N$(C$_2$H) = $5.1\times10^{15}$ cm$^{-2}$ in TW\,Hya. $^c$ Out of this range, \cite{aik2003} (2003) derive $N$(H$_2$CO) = $(7.2-19)\times10^{12}$ cm$^{-2}$ in LkCa\,15. $^d$ \cite{car2017} (2017) derive a H$_2$CO abundance of $(2-5)\times10^{-12}$ relative to H$_2$ in HD\,163296. $^e$ Out of this range, \cite{mat2013} (2013) derive $N$(HCO$^+$) = $1.5\times10^{14}$ cm$^{-2}$ in HD\,163296. $^f$ Out of this range, \cite{qi2013c} (2013c) derive $N$(N$_2$H$^+$) = $10^{14}-10^{15}$ cm$^{-2}$ in TW\,Hya. $^g$ In the same object, HD\,100546, \cite{fed2013} (2013) derive $N$(CH$^+$) = $10^{16}-10^{17}$ cm$^{-2}$ for an emitting area inner to 50-70 au.}
\end{table*}

\subsubsection{Formation of H$_2$ on grain surfaces}

The kinetics of H$_2$ formation on grain surfaces in interstellar space is usually described as
\begin{equation}
\frac{d n({\rm H_2})}{dt} = R_f n_{\rm H} n({\rm H}), \label{eq:h2_formation}
\end{equation}
where $n_{\rm H}$ is the volume density of H nuclei, $n$(H) and $n$(H$_2$) are the volume densities of neutral H atoms and H$_2$ molecules, and $R_f$ is the formation rate parameter for which usually the canonical value of $3\times10^{-17}$~cm$^3$~s$^{-1}$ derived by \cite{jur1975} (1975) for diffuse interstellar clouds is adopted. Here, we evaluate $R_f$ as
\begin{equation}
R_f = \frac{1}{2} S_{\rm H} \epsilon_{\rm H_2} v_{\rm H} \langle \sigma_d n_d \rangle \frac{1}{n_{\rm H}}, \label{eq:rf}
\end{equation}
where $v_{\rm H}$ is the thermal velocity of H atoms, evaluated as $\sqrt{3 k T_k / m_{\rm H}}$. The sticking coefficient of H atoms $S_{\rm H}$ depends on the gas kinetic temperature $T_k$ and is evaluated through the expression
\begin{equation}
S_{\rm H} (T_k)= S_0 \frac{(1+\beta T_k/T_0)}{(1+T_k/T_0)^\beta}, \label{eq:stick_h}
\end{equation}
where we have adopted $S_0$ = 1, $T_0$ = 25 K, and $\beta$ = 2.5, based on the experimental study carried out by \cite{cha2012} (2012) for a silicate surface. The recombination efficiency $\epsilon_{\rm H_2}$ in Eq.~(\ref{eq:rf}) depends on the dust temperature $T_d$ according to the expression derived by \cite{caz2002a} (2002a,b) and is evaluated with the parameters provided by \cite{caz2002a} (2002a) in their Table 1, with an updated value of 12,200 K for the desorption energy of chemisorbed H, as calculated by \cite{gou2009} (2009) for an olivine surface. Currently, the kinetics of grain-surface H$_2$ formation considered in the model accounts for the Langmuir-Hinshelwood mechanism. In the future it will be worth to consider also the Eley-Rideal mechanism, which is expected to increase the H$_2$ formation efficiency at high gas temperatures (e.g., \cite{leb2012} 2012; \cite{bro2014} 2014).

\section{Results} \label{sec:results}







In this section we present the calculated abundance distributions of various molecules in our fiducial T\,Tauri and Herbig\,Ae/Be disk models, and compare them with available constraints from observations, with the stress put on the similarities and differences between both types of disks. We focus on molecules that have been observed in disks at IR or (sub-)mm wavelengths. Detected species in disks have been summarized by \cite{dut2014} (2014). In Table~\ref{table:observations} we provide an updated an comprehensive summary of the molecules observed in T\,Tauri and Herbig\,Ae/Be disks. We first concentrate on molecules observed through IR observations, which are sensitive to the hot inner disk: H$_2$O and OH (Sec.~\ref{subsec:h2o}), and simple organics such as C$_2$H$_2$, HCN, CH$_4$, and CO$_2$ (Sec.~\ref{subsec:c2h2}). We then focus on molecules observed at (sub-)mm wavelengths, which trace the outer disk: the radicals C$_2$H and CN (Sec.~\ref{subsec:c2h}) and other organic molecules with a certain complexity, such as H$_2$CO (Sec.~\ref{subsec:complex}), the sulfur-bearing molecules CS and SO (Sec.~\ref{subsec:sulfur}), and molecular ions (Sec.~\ref{subsec:ions}). We finally discuss the abundance distributions of ices, for which most observational constraints consist of determining the location of the CO snowline (Sec.~\ref{subsec:ices}). Abundances are mostly expressed as column densities because this is the quantity provided by most observational studies. Nonetheless, sometimes we use the term fractional abundance, which hereafter refers to the abundance relative to the total number of H nuclei.

\subsection{Water and hydroxyl radical} \label{subsec:h2o}

\begin{figure*}
\centering
\includegraphics[angle=0,width=0.88\textwidth]{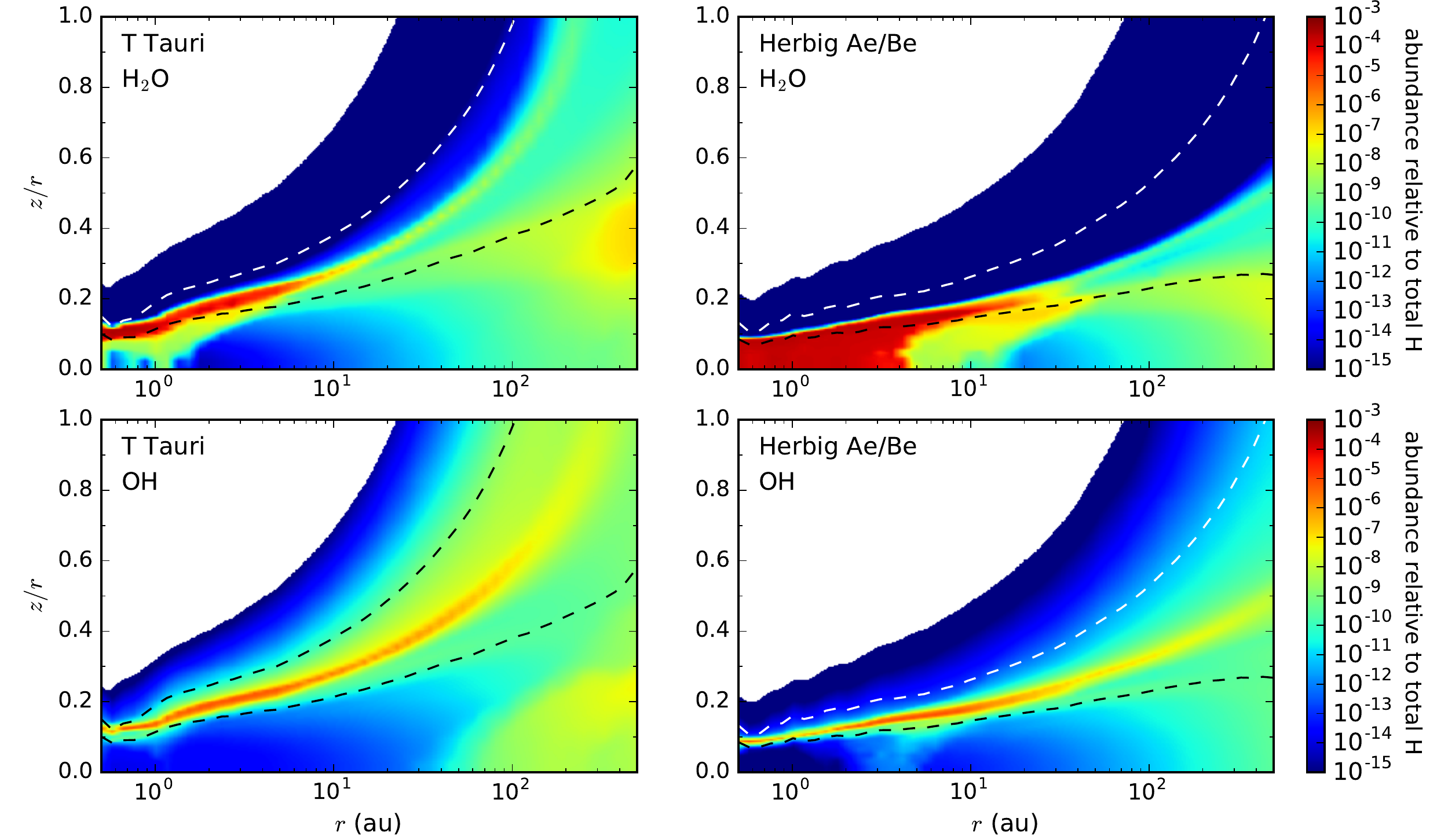}
\caption{Calculated distributions of H$_2$O and OH as a function of radius $r$ and height over radius $z/r$ for the T\,Tauri (left) and Herbig\,Ae/Be (right) disks. The dashed lines indicate the location where $A_V$ in the outward vertical direction takes values of 0.01 and 1.} \label{fig:map_h2o}
\end{figure*}

In the last years, near-IR to sub-mm observations have provided important constraints on the presence of water and its related radical (OH) in protoplanetary disks. At near- and mid-IR wavelengths, the spectra of disks around T\,Tauri stars show emission of hot H$_2$O and OH arising from the inner disk ($<$ a few au) atmosphere (\cite{car2004} 2004; \cite{car2008} 2008, 2011, 2014; \cite{sal2008} 2008, 2011; \cite{pon2010a} 2010a,b; \cite{man2012} 2012; \cite{sar2014} 2014; \cite{ban2017} 2017), while in disks around Herbig\,Ae/Be stars, emission by OH is relatively common but there is a striking lack of H$_2$O emission (\cite{man2008} 2008; \cite{pon2010a} 2010a; \cite{fed2011} 2011; \cite{sal2011} 2011; \cite{ban2017} 2017). Far-IR observations with \emph{Herschel}/PACS have essentially confirmed that the water detection rate is much higher in T\,Tauri disks than in disks around Herbig\,Ae/Be stars (\cite{riv2012} 2012; \cite{mee2012} 2012; \cite{fed2012} 2012, 2013). Therefore, infrared observations suggest that water could be intrinsically less abundant in disks around Herbig\,Ae/Be stars than around T\,Tauri stars. Such a trend is however not corroborated by our models.

As can be seen in Fig.~\ref{fig:map_h2o}, in the T\,Tauri disk model, water is present with fractional abundances of $\sim$10$^{-4}$ in a surface layer at $A_V\sim$1 in the inner $\sim$10 au from the star, while in the Herbig\,Ae/Be disk, the warmer temperatures make the region of high H$_2$O abundance to extend radially beyond 10 au and vertically down to the midplane. In the inner disk, water becomes very abundant in regions warmer than $\sim$200 K, where the reaction OH + H$_2$ is activated, and sufficiently shielded from FUV photons, while OH is present in a thin layer on top of H$_2$O resulting from its photodissociation. In our generic T\,Tauri disk model, the inner midplane regions are not warm enough to sustain a high water abundance, and thus most H$_2$O (and all OH) are present in upper layers. We however note that the presence of water vapor in the inner midplane is very sensitive to the temperature and that other T\,Tauri disk models (e.g., \cite{wal2015} 2015) find high H$_2$O abundances in these regions. Therefore, the model predicts that as the stellar luminosity increases and the disk becomes warmer water vapor is more abundant.

%

\begin{figure*}
\centering
\includegraphics[angle=0,width=\textwidth]{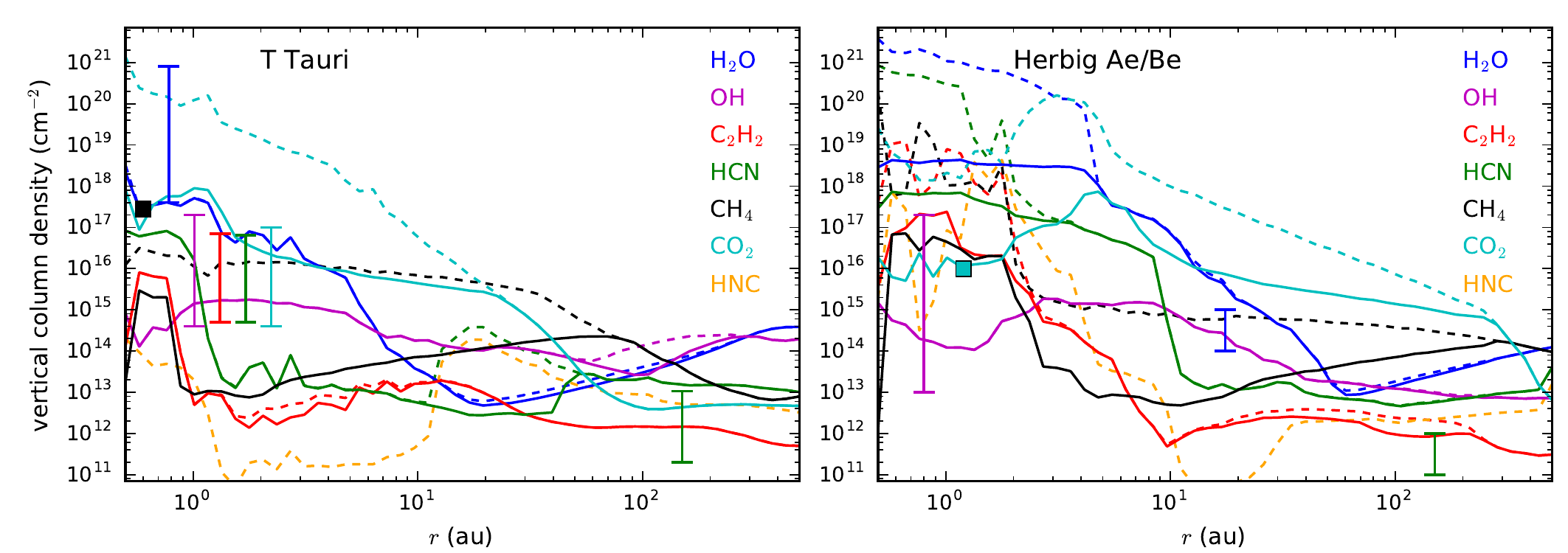}
\caption{Calculated vertical column densities down to the $A_V$=10 surface (solid lines) and down to the midplane (dashed lines) of H$_2$O, OH, and simple organics as a function of radius $r$ for the T\,Tauri (left panel) and Herbig\,Ae/Be (right panel) disks. For HNC, only the total column density down to the midplane is shown. Column densities derived from observations are indicated by vertical lines (when there are ranges of values) or by squares (when only one value is available), with their radial locations corresponding to the approximate region probed by observations (see Table~\ref{table:observations}).} \label{fig:cdmol_h2o}
\end{figure*}

The vertical column densities calculated for H$_2$O and OH in the IR-observable atmosphere\footnote{We consider the IR-observable atmosphere to extend down to the $A_V$ = 10 layer, where the optical depth at 10~$\mu$m is of the order of unity.} of the inner T\,Tauri disk are 10$^{17}$-10$^{18}$ cm$^{-2}$ and $\sim$10$^{15}$ cm$^{-2}$, respectively, which are in the low range of observed values (see left panel in Fig.~\ref{fig:cdmol_h2o}). The calculated OH/H$_2$O column density ratio in the inner disk, 10$^{-2}$-10$^{-3}$, is between the values derived in AA Tau, DR Tau, and AS 205A (0.1-0.3; \cite{car2008} 2008; \cite{sal2008} 2008) and that found from a study of a much larger sample of T\,Tauri disks ($\sim$10$^{-3}$;  \cite{sal2011} 2011). Previous chemical models of inner T\,Tauri disks (\cite{agu2008} 2008; \cite{wal2015} 2015) find H$_2$O and OH column densities and ratios of the same order of magnitude than the ones calculated by us. In summary, chemical models of T\,Tauri disks predict the existence of an important reservoir of hot water in the inner regions (formed by warm gas-phase chemistry) and smaller amounts of OH (formed by FUV photodissociation of water), with numbers that are roughly in agreement with those derived from IR observations.

In disks around Herbig\,Ae/Be stars, OH has been detected in about a dozen of objects with a broad range of column densities (see Table~\ref{table:observations}). In our Herbig\,Ae/Be disk model, the calculated vertical column density of OH is $\sim$10$^{15}$ cm$^{-2}$ across the first 10 au, decreasing down to $\sim$10$^{13}$ cm$^{-2}$ in the outer disk, in the low range of values derived from observations (see right panel in Fig.~\ref{fig:cdmol_h2o}). \cite{wal2015} (2015) calculate somewhat higher values for the inner 10 au, 10$^{16}$-10$^{17}$ cm$^{-2}$, more in line with the high range of observed values. Water has been convincingly detected at IR wavelengths only around one Herbig star, HD\,163296 (\cite{mee2012} 2012; \cite{fed2012} 2012). The H$_2$O and OH column densities derived in this disk are similar, in the range 10$^{14}$-10$^{15}$ cm$^{-2}$, and the emitting region for both species is constrained to be 15-20 au from the star. In our Herbig\,Ae/Be disk model, the column density of water is very large in the inner disk (out to $\sim$4 au), where it is very abundant in the midplane (see right panels in Figs.~\ref{fig:map_h2o} and \ref{fig:cdmol_h2o}), and experiences a sharp abundance decline with increasing radius. At 15-20 au from the star, where water is no longer present in the midplane but in upper disk layers, the model yields $N$(H$_2$O) = 10$^{15}$-10$^{16}$ cm$^{-2}$ and $N$(OH) = 10$^{14}$-10$^{15}$ cm$^{-2}$, with a OH/H$_2$O ratio of $\sim$0.1, values which are not far from those derived in the HD\,163296 disk. In the Herbig\,Ae disk model of \cite{wal2015} (2015), at 10 au (the farthest radius studied by these authors) the column densities of H$_2$O and OH are in the range 10$^{16}$-10$^{17}$, with the OH/H$_2$O ratio approaching unity. The model of \cite{wal2015} (2015) and ours do a reasonable job at explaining the order of magnitude of the water and OH observations of HD\,163296. It however remains puzzling to explain the extremely low detection rate of water in Herbig\,Ae/Be disks, as compared with T\,Tauri disks, taking into account that chemical models (\cite{wal2015} 2015 and this work) predict that in disks around Herbig stars, water should be even more abundant than in T\,Tauri disks. Several explanations have been proposed (see \cite{ant2016} 2016 and references therein), most of which are related to observational aspects (e.g., the higher level of infrared continuum in Herbig\,Ae/Be disks and the lower sensitivity reached for detection of emission lines above the continuum) than with substantive differences in the chemistry between disks around low- and intermediate-mass pre-main sequence stars.

Our model predicts that the reservoir of hot water present in the inner regions of disks around T\,Tauri and Herbig\,Ae/Be stars vanishes typically beyond 10 au from the star, owing to thermal deactivation of the water-forming reaction OH + H$_2$ and to freeze-out onto dust grains. This drastic decline of various orders of magnitude in the abundance and column density of water (see Figs.~\ref{fig:map_h2o} and \ref{fig:cdmol_h2o}) has been observationally probed by mid- to far-IR observations in a few protoplanetary disks (\cite{ble2016} 2016). The model however predicts that there exists an additional reservoir of cold water in the outer parts of T\,Tauri and Herbig\,Ae/Be disks, typically beyond 100 au and at intermediate heights (see Fig.~\ref{fig:map_h2o}). Water in these regions arises from the FUV photodesorption of water ice and reaches peak fractional abundances of $\sim$10$^{-7}$, with typical vertical column densities of the order of 10$^{13}$-10$^{14}$ cm$^{-2}$ (see Fig.~\ref{fig:cdmol_h2o}). This outer reservoir of water is also predicted by previous chemical models of T\,Tauri disks which include photodesorption (e.g., \cite{wil2000} 2000; \cite{woi2009} 2009;  \cite{sem2011} 2011; \cite{wal2012} 2012) and has been detected with \emph{Herschel}/HIFI in the T\,Tauri disks TW\,Hya and DG\,Tau (\cite{hog2011} 2011; \cite{pod2013} 2013) and in the Herbig\,Be disk HD\,100546 (\cite{van2014} 2014; \cite{du2017} 2017), although it remains elusive to detection in many other protoplanetary disks (\cite{ber2010} 2010; \cite{du2017} 2017). It is remarkable that the detection rate of water at sub-mm wavelengths is not so different between T\,Tauri and Herbig\,Ae/Be disks as it is at IR wavelengths (see Table~\ref{table:observations}), which suggests that the low detection rate of H$_2$O IR emission in Herbig objects may not be due to an intrinsic deficit of water with respect to T\,Tauri systems.

\subsection{Simple organics: C$_2$H$_2$, HCN, CH$_4$, and CO$_2$} \label{subsec:c2h2}

\begin{figure*}
\centering
\includegraphics[angle=0,width=0.88\textwidth]{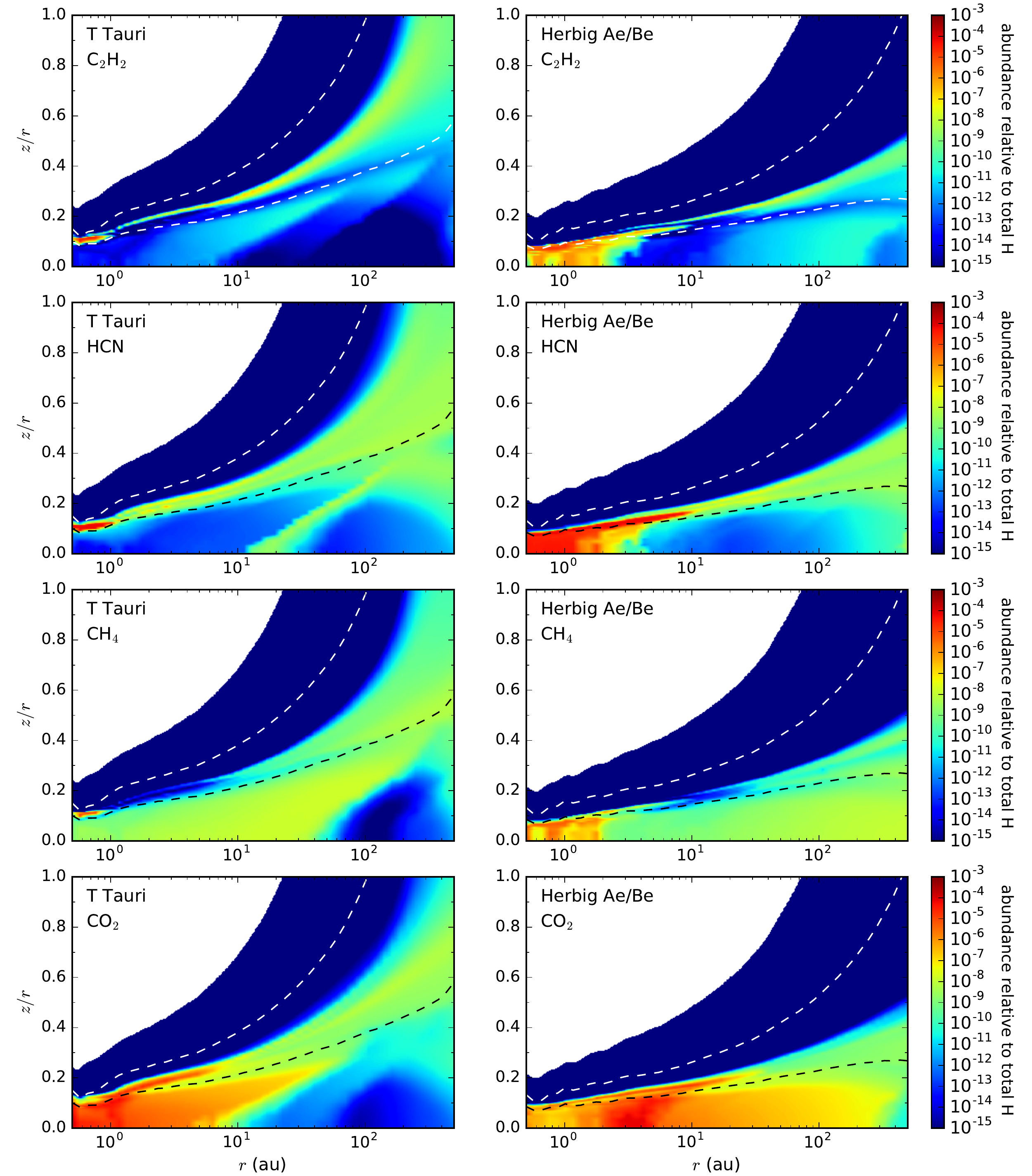}
\caption{Same as Fig.~\ref{fig:map_h2o} but for C$_2$H$_2$, HCN, CH$_4$, and CO$_2$.} \label{fig:map_c2h2}
\end{figure*}

It is known that there exists an important reservoir of simple organic molecules in protoplanetary disks. Thanks to observations at near- and mid-IR wavelengths, lines of molecules such as acetylene, hydrogen cyanide, methane, and carbon dioxide have been detected in absorption (\cite{lah2006} 2006; \cite{gib2007} 2007; \cite{gib2013} 2013; \cite{bas2013} 2013) and in emission (\cite{car2008} 2008, 2011; \cite{sal2011} 2011; \cite{kru2011} 2011; \cite{man2012} 2012; \cite{naj2013} 2013; \cite{pas2013} 2013). These observations probe hot gas located in the inner (a few au) disk, where these molecules are found with large abundances. It is noteworthy that the vast majority of infrared detections of simple organics correspond to disks around T\,Tauri stars rather than to disks around Herbig\,Ae/Be stars, where neither C$_2$H$_2$, HCN, nor CH$_4$ are detected, and only CO$_2$ has been detected in one disk, HD\,101412 (\cite{sal2011} 2011). It is therefore tempting to think that Herbig\,Ae/Be disks are less rich in simple organics than disks around T\,Tauri stars. However, similarly to the case of H$_2$O, we do not find such a trend in our models.

In our T\,Tauri disk model, C$_2$H$_2$, HCN, and CH$_4$ are formed with high fractional abundances (a few $\times$10$^{-5}$) in the atmosphere ($A_V$$\sim$1) of the inner (within a few au) regions of the disk (see Fig.~\ref{fig:map_c2h2}). Their synthesis is driven by FUV photochemistry in a warm gas (see \cite{agu2008} 2008; \cite{bas2013} 2013; \cite{wal2015} 2015). In the Herbig\,Ae/Be disk model, the region over which these molecules have large fractional abundances extends to larger radii compared to the T\,Tauri disk, as a consequence of the higher temperatures. Moreover, in the Herbig\,Ae/Be disk, C$_2$H$_2$, HCN, and CH$_4$ are formed abundantly in the inner midplane, where the synthesis is not related to photochemistry but to the fact that the chemical composition tends toward thermochemical equilibrium in these hot, dense, and FUV-shielded regions. Both the T\,Tauri and Herbig\,Ae/Be disk models show a progressive disappearance of simple organics as one moves radially from the star, in this order: CH$_4$, C$_2$H$_2$, and HCN. This behavior, already predicted by \cite{agu2008} (2008), is a consequence of the requirements of temperature that each molecule has to activate its corresponding formation routes, with CH$_4$ being the most demanding. Such a trend is also found in the disk models of \cite{wal2015} (2015) for C$_2$H$_2$ and HCN. Note that CH$_4$, unlike C$_2$H$_2$ and HCN, is also predicted to be moderately abundant in cool midplane regions, where the synthesis is driven by ion-molecule routes. In these regions however the calculated abundance of CH$_4$ could be especially uncertain if grain-surface chemistry (not included in our models) plays an important role.

The calculated vertical column densities of C$_2$H$_2$ and HCN in the IR-observable atmosphere of the inner T\,Tauri disk (within 1 au from the star) are large, with maxima in the range 10$^{15}$-10$^{17}$ cm$^{-2}$, in good agreement with observed values (see left panel in Fig.~\ref{fig:cdmol_h2o}). Observations of T\,Tauri disks indicate that C$_2$H$_2$ and HCN have similar abundances, although there is a significant dispersion, and that they are somewhat less abundant than water vapor (C$_2$H$_2$/HCN = 0.04-20; HCN/H$_2$O = 10$^{-3}$-10$^{-1}$;\cite{lah2006} 2006; \cite{gib2007} 2007; \cite{car2011} 2011; \cite{sal2011} 2011; \cite{man2012} 2012; \cite{bas2013} 2013). These observed ratios are in line with the values found in the T\,Tauri disk model. Methane has only been detected in one T\,Tauri disk, GV\,Tau, in absorption (\cite{gib2013} 2013). These authors derive a column density of $2.8\times10^{17}$ cm$^{-2}$ and a rotational temperature of 750 K, which implies that the detected CH$_4$ is distributed in the inner disk. In GV\,Tau, CH$_4$ is somewhat more abundant than C$_2$H$_2$ and HCN. Note however that in DR\,Tau, the non detection of CH$_4$ in emission implies that it has an abundance similar to or smaller than C$_2$H$_2$ and HCN (\cite{man2012} 2012). In our T\,Tauri disk model, CH$_4$ reaches a column density of the order of those of C$_2$H$_2$ and HCN in the IR-observable atmosphere of the inner disk (see left panel in Fig.~\ref{fig:cdmol_h2o}). Neither C$_2$H$_2$, HCN, or CH$_4$ have been observed in Herbig\,Ae/Be disks, although our model predicts that they should be even more abundant than in T\,Tauri disks (see Figs.~\ref{fig:cdmol_h2o} and \ref{fig:map_c2h2}). Similar conclusions are found in the models by \cite{wal2015} (2015). The lack of simple organics in the spectra of Herbig\,Ae/Be disks has not been investigated to the extent of the lack of water, but it is likely that observational rather than chemical effects are at the origin of it.

Carbon dioxide has been extensively observed in T\,Tauri disks but only in one Herbig\,Ae/Be disk, where the derived column density is within the range of values found in T\,Tauri disks. In our models, CO$_2$ is formed abundantly, mostly in the inner disk ($<10$ au in the T\,Tauri disk and $<100$ au in the Herbig disk) and over most of the vertical structure (see Fig.~\ref{fig:map_c2h2}). The formation of CO$_2$ occurs in the gas-phase, mainly through the reaction OH + CO, and is less demanding in terms of temperature than the formation of C$_2$H$_2$, HCN, and CH$_4$. Therefore, CO$_2$ extends over larger radii than the other simple organics. The calculated column density of CO$_2$ in the IR-observable atmosphere of the inner ($<10$ au) T\,Tauri disk is in the range 10$^{16}$-10$^{18}$ cm$^{-2}$ (see left panel in Fig.~\ref{fig:cdmol_h2o}), in the high range of observed values. As with the other simple organics, according to the model there is no apparent chemical reason for a lower amount of CO$_2$ in disks around Herbig\,Ae/Be stars than in T\,Tauri disks.

The simple organic molecules discussed here experience a drastic decline in their column densities with increasing radius, especially for C$_2$H$_2$ and HCN (see Fig.~\ref{fig:cdmol_h2o}). This extended and cooler reservoir of simple organics can be probed at mm wavelengths in the case of polar molecules like HCN. In fact, HCN has been extensively characterized this way in protoplanetary disks (\cite{dut1997} 1997; \cite{thi2004} 2004; \cite{fue2010} 2010, 2012; \cite{obe2010} 2010, 2011; \cite{cha2012a} 2012a; \cite{kas2014} 2014; \cite{gui2016} 2016). Although the statistics of Herbig\,Ae disks is low, these studies suggest that T\,Tauri disks can retain in their outer parts somewhat larger HCN abundances than Herbig\,Ae disks (see Table~\ref{table:observations}). The column densities calculated for HCN beyond 100 au are $\sim$10$^{13}$ cm$^{-2}$ in both the T\,Tauri and the Herbig\,Ae/Be disks, in the high range of observed values (see Fig.~\ref{fig:cdmol_h2o}). The model predicts a slightly higher amount of HCN in the outer regions of the T\,Tauri disk compared to the Herbig disk. It is interesting to note that the contrary is found in the inner regions. The reason is that the chemical synthesis of HCN is different in nature in the hot inner disk than in the cool outer regions. In the outer disk ($>$100 au), HCN is mainly present at intermediate heights with fractional abundances of $\sim$10$^{-8}$ relative to H$_2$ (see Fig.~\ref{fig:map_c2h2}). In these regions, HCN is formed by the same gas-phase chemical routes that operate in cold interstellar clouds, i.e., through ion-molecule reactions that lead to the precursor ion HCNH$^+$, which by dissociative recombination yields HCN as well as its isomer HNC. Thus, both observations and our model suggest that HCN is somewhat more abundant in the outer regions of T\,Tauri disks compared to Herbig\,Ae/Be disks. We however caution that on the observational side, the statistics of Herbig\,Ae disks is low, and on the theoretical one, the difference is small and could result from the particular set of parameters adopted in the models.

Hydrogen isocyanide, closely related to HCN from a chemical point of view, has been only detected around the T\,Tauri stars DM\,Tau and TW\,Hya and the Herbig\,Ae star HD\,163296 (\cite{dut1997} 1997; \cite{gra2015} 2015). The lower detection rate of HNC in disks compared to its most stable isomer HCN suggests that HNC is less abundant than HCN, although there is also a selection effect as lines of HNC have not been targeted as often as those of HCN (e.g., \cite{obe2011} 2011; \cite{gui2016} 2016). Observations indicate that HNC is indeed somewhat less abundant than HCN, with HNC/HCN ratios in the range 0.1-0.4 (see Table~\ref{table:observations}). These values may need to be revised down due to differences in the collisional excitation rate coefficients of HCN and HNC (\cite{sar2010} 2010). The calculated distribution of HNC approximately follows that of HCN, but at a lower level of abundance (see Fig.~\ref{fig:cdmol_h2o}). The model shows that there is a gap in the column density of HNC which occurs at 1-10 au in the T\,Tauri disk (shifted to larger radii in the Herbig\,Ae/Be disk). There is observational evidence of such a gap in the distribution of HNC from SMA interferometric observations of TW\,Hya (\cite{gra2015} 2015). While the presence of HNC in the hot inner disk is related to the existence of large amounts of HCN, a fraction of which isomerizes to HNC in the hot gas phase, in the cool outer disk, HNC has an abundance of the order of that of HCN because the two isomers share the same chemical formation routes, with the molecular ion HCNH$^+$ as main precursor. The calculated HNC/HCN ratio in the outer disk ($>$100 au) is 0.3-0.4 in both the T\,Tauri and the Herbig\,Ae/Be disks, in line with the values derived from observations. Interferometric observations indicate that HNC does not extend out to radii as large as HCN in TW\,Hya (\cite{gra2015} 2015), a feature that is not predicted by our T\,Tauri disk model. A possible explanation could be related to an enhanced photodissociation cross section of HNC compared to that of HCN, something that is suggested by theoretical calculations (\cite{che2016} 2016; \cite{agu2017} 2017).

\subsection{Radicals C$_2$H and CN} \label{subsec:c2h}

\begin{figure*}
\centering
\includegraphics[angle=0,width=0.88\textwidth]{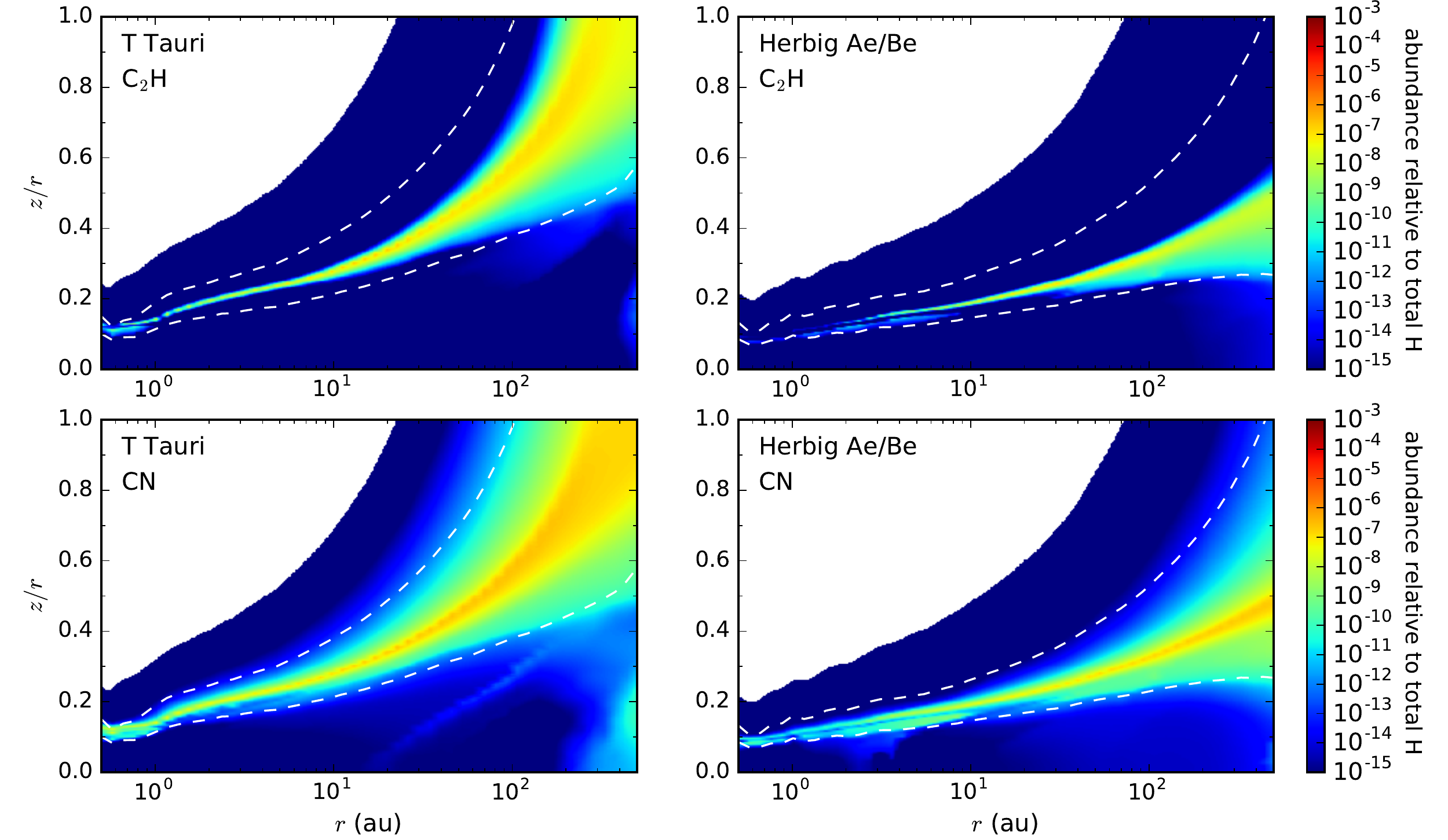}
\caption{Same as Fig.~\ref{fig:map_h2o} but for C$_2$H and CN.} \label{fig:map_c2h}
\end{figure*}

The radicals C$_2$H and CN, usually considered as good tracers of regions affected by FUV radiation such as PDRs, are among the most conspicuous molecules detected in protoplanetary disks at mm wavelengths. These observations probe the emission from the outer disk, out to some hundreds of au from the star. Constraints on the abundances of C$_2$H and CN mainly come from observations of a few protoplanetary disks, the widely studied T\,Tauri disks DM\,Tau, LkCa\,15, and TW\,Hya, and the Herbig\,Ae disks HD\,163296, MWC\,480, and AB\,Aur (\cite{thi2004} 2004; \cite{sch2008} 2008; \cite{hen2010} 2010; \cite{fue2010} 2010; \cite{cha2012a} 2012a; \cite{kas2014} 2014) and from the more recent study of \cite{gui2016} (2016), in which a larger sample of disks was observed. These studies suggest that C$_2$H and CN can reach higher abundances in T\,Tauri disks than in Herbig\,Ae disks (see Table~\ref{table:observations}). This would be in line with the general observational finding of a higher detection rate of molecules in T\,Tauri disks compared to Herbig\,Ae disks (e.g., \cite{obe2011} 2011; \cite{gui2016} 2016). However, the confirmation of such hypothesis is still hampered by the low statistics of Herbig\,Ae disks observed to date. Observations indicate that C$_2$H is present with an abundance similar to that of CN in most disks, with C$_2$H/CN ratios in the range 0.2-3.5 (\cite{gui2016} 2016), and no significant difference between T\,Tauri and Herbig\,Ae disks. It is also found that the CN radical is significantly more abundant than HCN in all observed disks, with CN/HCN abundance ratios in the range 4-30 (\cite{gui2016} 2016), and again no substantive difference between T\,Tauri and Herbig\,Ae disks. In summary, observations at mm wavelengths tell us that T\,Tauri disks can retain in the outer regions larger abundances of C$_2$H and CN than Herbig\,Ae disks, and that in round numbers the C$_2$H/CN ratio is $\sim1$ and the CN/HCN ratio is $\sim10$ in both types of disks. It is also interesting to note that recent ALMA observations have found that C$_2$H emission shows a ring-like distribution in the TW\,Hya and DM\,Tau disks (\cite{bergin2016} 2016).

\begin{figure*}
\centering
\includegraphics[angle=0,width=0.88\textwidth]{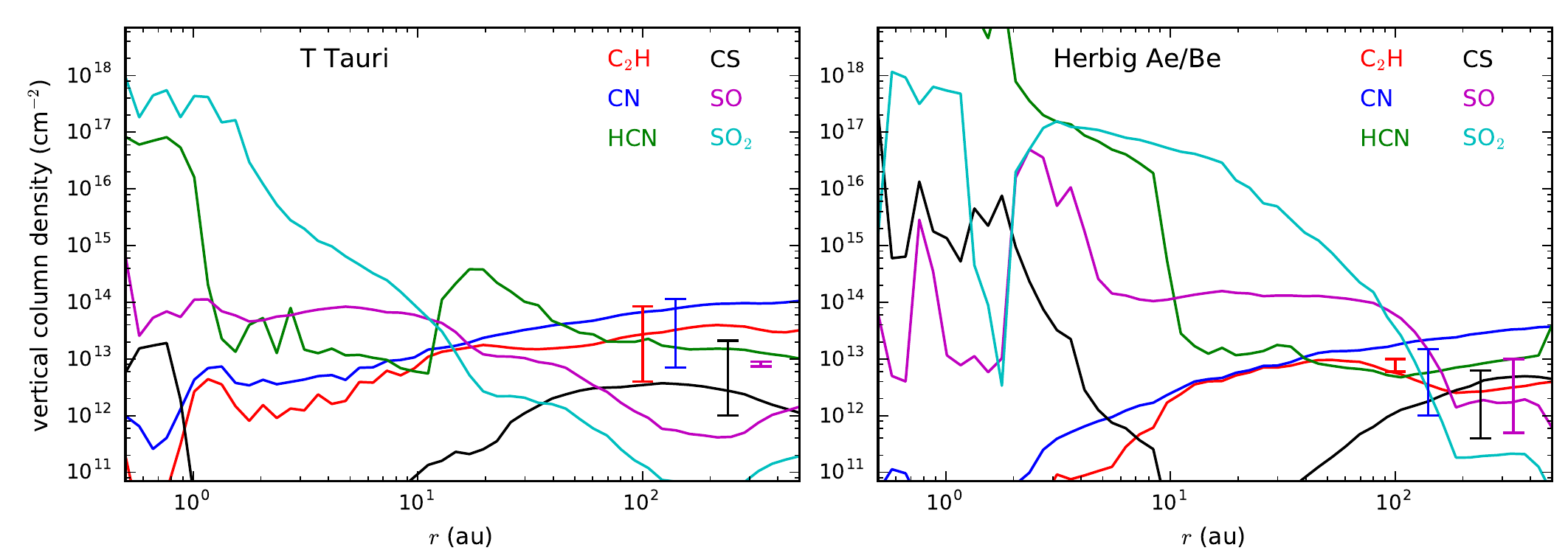}
\caption{Calculated vertical column densities (down to the midplane) of the radicals C$_2$H and CN and some sulfur-bearing molecules as a function of radius $r$ for the T\,Tauri (left panel) and Herbig\,Ae/Be (right panel) disks. HCN is also shown to allow for the visualization of the CN/HCN ratio. The ranges of column densities derived from observations are indicated by the vertical lines plotted in the outer disk (see Table~\ref{table:observations}).} \label{fig:cdmol_c2h}
\end{figure*}

In the T\,Tauri and Herbig\,Ae/Be disk models, C$_2$H and CN are essentially located in a relatively thin layer in the disk surface, between the $A_V$ = 1 and $A_V$ = 0.01 layers (see Fig.~\ref{fig:map_c2h}), with vertical column densities of the order of 10$^{13}$ cm$^{-2}$ in the outer disk (see Fig.~\ref{fig:cdmol_c2h}). Protoplanetary disks can be seen as a PDR with the typical layered structure CO/C/C$^+$ along the vertical direction, and C$_2$H and CN become abundant in the layer where neutral atomic carbon reaches its maximum abundance. The formation of C$_2$H and CN is therefore associated to the disk PDR and to the availability of atomic C in a FUV illuminated gas. Our results regarding the T\,Tauri disk are similar to those of previous models of disks around low-mass young stars (e.g., \cite{aik1999b} 1999b, 2002; \cite{wil2000} 2000; \cite{van2003} 2003; \cite{wil2006} 2006; \cite{wal2010} 2010; \cite{sem2011} 2011). There are some differences between the T\,Tauri and the Herbig\,Ae/Be disks. The higher gravity of the Herbig\,Ae/Be star makes the layer containing C$_2$H and CN to be more compressed toward lower heights (see Fig.~\ref{fig:map_c2h}), although the most significant difference is that the calculated column densities of C$_2$H and CN in the outer disk are somewhat lower in the Herbig\,Ae/Be disk (see Fig.~\ref{fig:cdmol_c2h}). The reason is that the Herbig disk is illuminated by a more intense FUV field from the star than the T\,Tauri disk, and this narrows the photochemically active layer and limits the ability of photochemistry to build molecules, resulting in smaller amounts of molecules specifically formed by the action of photochemistry. This is a general difference between disks around T\,Tauri and Herbig stars that is better appreciated in the cool outer disk ($>$100 au), where warm gas-phase chemistry is inhibited and thus cannot counterbalance the effect of FUV photons. For example, beyond 100 au, the radical OH (a species typically formed under the action of photochemistry) is significantly more abundant in the T\,Tauri disk than in the Herbig disk (see Fig.~\ref{fig:cdmol_h2o}). In the case of C$_2$H and CN, the fact that the abundances are lower in the Herbig disk is linked to a slight defficiency in the abundance of neutral atomic carbon, which is a key starting point to form both C$_2$H and CN. The calculated column densities of C$_2$H and CN are in line with the values derived from observations. Moreover, the slight overabundance (a factor of a few) of these radicals in the T\,Tauri disk compared to the Herbig\,Ae/Be disk is in line the observational suggestion that T\,Tauri disks can retain higher abundances of these two radicals. Another salient feature of the model is that the column densities of both C$_2$H and CN increase with increasing radius in both the T\,Tauri and the Herbig\,Ae/Be disks. The predicted inner gap could be consistent with the ring-like distribution found for C$_2$H in the TW\,Hya and DM\,Tau disks (\cite{bergin2016} 2016). These authors however propose an scenario in which the ring morphology observed for C$_2$H, and also for cyclic C$_3$H$_2$, is related to the evolution of ice-coated dust due to the combined effect of coagulation, gravitational settling, and drift.


The model indicates that HCN is much more abundant than CN in the inner regions of both types of disks, although as one moves away from the star, the CN/HCN ratio increases, reaching values above unity at radii $>$50 au (see Fig.~\ref{fig:cdmol_c2h}). The overabundance of CN with respect to HCN in the outer disk is in agreement with observations. According to the model, in the outer disk C$_2$H and CN do not spatially coexist with HCN, the radicals being exclusively present in a relatively thin layer at the disk surface while HCN is located at lower heights. There is some controversy regarding the region where C$_2$H and CN are present in T\,Tauri disks because mm observations derive low excitation temperatures ($\lesssim$10 K) for these two radicals (\cite{hen2010} 2010; \cite{cha2012a} 2012a; \cite{hil2017} 2017), which suggest that the emission could arise from cold midplane regions rather than from the warm disk surface. If true, this would be in strong disagreement with the predictions of our model and previous chemical models of T\,Tauri disks (e.g., \cite{aik1999b} 1999b, 2002; \cite{wil2000} 2000; \cite{van2003} 2003; \cite{wil2006} 2006; \cite{wal2010} 2010; \cite{sem2011} 2011), which also locate the radicals C$_2$H and CN well above the midplane. Observations and chemical models of T\,Tauri disks could still be reconciled if in the outer regions ($>$100 au) of the surface layer containing C$_2$H and CN (where densities are in the range 10$^{5}$-10$^{6}$ cm$^{-3}$ and gas kinetic temperatures are between 30 K and 100 K) these two radicals are subthermally excited. Note however that this could have implications for the column densities derived from observations under the assumption of local thermal equilibrium (LTE). Dedicated non-LTE excitation and radiative transfer calculations (e.g., \cite{aik2002} 2002) and comparison with mm observations are needed to shed light on this issue.

\subsection{Organic molecules of a certain complexity} \label{subsec:complex}

In this section we briefly comment on various organic molecules observed in disks that are more complex than those treated in Sec.~\ref{subsec:c2h2}. We refer to H$_2$CO and CH$_3$OH, the hydrogenation descendants of carbon monoxide, the cyanides HC$_3$N and CH$_3$CN, and the cyclic isomer of the hydrocarbon C$_3$H$_2$.

Formaldehyde has been quite commonly observed in protoplanetary disks, with a higher detection rate in T\,Tauri disks than in Herbig\,Ae disks (\cite{obe2010} 2010, 2011; \cite{gui2016} 2016), but very similar column densities, a few times $10^{12}$ cm$^{-2}$, in both types of disks (see Table~\ref{table:observations}). Interferometric observations of the disks DM\,Tau, TW\,Hya, and HD\,163296 have provided interesting constraints on the distribution and origin of H$_2$CO (\cite{qi2013a} 2013a; \cite{loo2015} 2015; \cite{obe2017} 2017; \cite{car2017} 2017). These observations point to the presence of an inner component, consistent with gas-phase formation, and an abundance enhancement in the outer disk (beyond the CO snowline) resulting from formation on grain surfaces by hydrogenation of CO ice followed by desorption. The recent detection of methanol in the TW\,Hya disk (\cite{wal2016} 2016) also points to a similar outer disk origin driven by the hydrogenation of CO ice. In our model, which does not include grain-surface chemistry, calculated column densities in the outer regions of the T\,Tauri and Herbig\,Ae/Be disks are around 10$^{12}$ cm$^{-2}$ for H$_2$CO (i.e., not far from observed values), but vanishingly small for CH$_3$OH. We note that if most of the H$_2$CO and CH$_3$OH in the outer disk indeed come from CO ice through grain-surface chemistry, one should expect a significant differentiation between T\,Tauri and Herbig disks because the latter are warmer and should have a lower reservoir of the precursor CO ice.

Other relatively large organic molecules observed in disks are HC$_3$N, CH$_3$CN, and cyclic C$_3$H$_2$ (see Table~\ref{table:observations}). Observations indicate that these molecules are present in the outer disk, out to a few hundreds of au. In these regions, our model predict column densities below the observed values by 1-2 orders of magnitude. Calculated column densities for these molecules are quite different among chemical models in the literature. For example, for HC$_3$N, \cite{cha2012b} (2012b) calculate column densities 1-2 orders of magnitude above the observed ones, while \cite{wal2014} (2014) find values ten times lower than observed. The dispersion of calculated column densities between different chemical models and the poor agreement with observations suggests that the chemistry of these moderately complex molecules is not yet as robust as for smaller species.The probable role of grain-surface reactions in regulating their abundances makes it worth to revisit their chemistry in T\,Tauri and Herbig\,Ae disks with an expanded chemical network including grain-surface chemistry.

We note that our model predicts that organic molecules like H$_2$CO, CH$_3$OH, HC$_3$N, CH$_3$CN, and $c$-C$_3$H$_2$ are enhanced in the warm inner regions (within a few au from the star), in particular in the Herbig\,Ae/Be disk, following the abundance enhancement of simple organics such as C$_2$H$_2$ and HCN (see Fig.~\ref{fig:cdmol_h2o}). Such inner reservoir of complex organics, which is the result of hot gas-phase chemistry, could be detectable at millimeter wavelengths with ALMA provided the angular resolution and sensitivity are high enough.

\subsection{Sulfur-bearing molecules: CS and SO} \label{subsec:sulfur}

\begin{figure*}
\centering
\includegraphics[angle=0,width=0.88\textwidth]{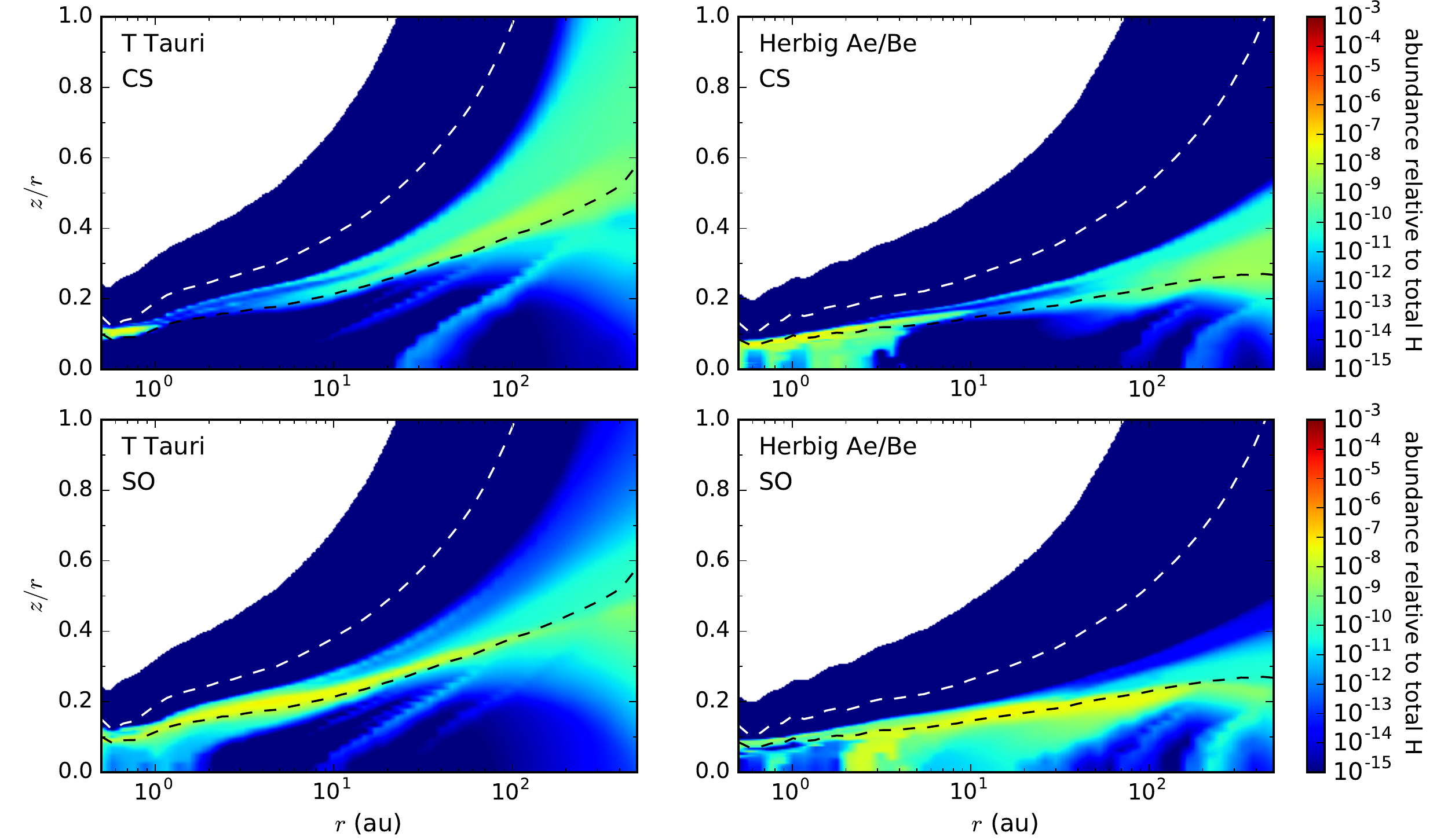}
\caption{Same as Fig.~\ref{fig:map_h2o} but for CS and SO.} \label{fig:map_sulfur}
\end{figure*}

A couple of sulfur-bearing molecules, CS and SO, have been detected in protoplanetary disks. The observation of these species is interesting because it provides information on the degree of depletion of sulfur and because they can be used to probe phenomena such as turbulence and dust traps (\cite{gui2012} 2012; \cite{pac2016} 2016). CS was among the first molecules detected in disks (\cite{dut1997} 1997) and since then has been observed in various T\,Tauri disks and in two Herbig\,Ae disks, while SO was first observed in the Herbig\,Ae disk AB\,Aur and has later on been observed in a couple of T\,Tauri disks (see references in Table~\ref{table:observations}). These observations were carried out at mm wavelengths and thus trace the outer regions of disks. Although the statistics of observed disks is small, it seems that both CS and SO have column densities of the order of 10$^{12}$-10$^{13}$ cm$^{-2}$ and that there is no substantial difference between T\,Tauri and Herbig\,Ae disks (see Table~\ref{table:observations}).













In our model, both CS and SO are mostly distributed in a layer located at intermediate heights (around $A_V\sim1$), both in the T\,Tauri and the Herbig\,Ae/Be disks (see Fig.~\ref{fig:map_sulfur}). In the inner regions (within a few au from the star) of the Herbig\,Ae/Be disk, these two molecules extend down to lower heights. The main formation pathways to both molecules involve various fast neutral-neutral reactions and thermal desorption from dust grains. The calculated column densities in the outer disk (beyond 100 au) are around 10$^{12}$ cm$^{-2}$ in both the T\,Tauri and the Herbig\,Ae/Be disks, in line with the values derived from observations (see Fig.~\ref{fig:cdmol_c2h}). The S-bearing molecule SO$_2$ has not been detected in disks but it is predicted to have column densities of the same order or slightly lower than SO in the outer disk (see Fig.~\ref{fig:cdmol_c2h}). In the inner regions, the column densities of CS, SO, and SO$_2$ are enhanced by various orders of magnitude, especially in the case of the Herbig\,Ae/Be disk, although these regions are likely to be strongly spatially diluted in the mm observations.

The column densities of CS and SO calculated here for the Herbig\,Ae/Be disk are similar to those obtained in the chemical model of the AB\,Aur disk presented by \cite{fue2010} (2010) and \cite{pac2015} (2015). The main difference is that those models predict an enhancement in the column densities of CS and SO in the 100-200 au region, something that is not seen in the Herbig\,Ae/Be disk model presented here. The main reason of such difference is that those models assumed a more simplistic initial composition, with all the sulfur being initially in the form of CS, while in the present model we adopt a more realistic initial composition in which sulfur is initially distributed in various forms, mostly as S, S$^+$, CS, and CS ice (see Sec.~\ref{subsec:chemical_model}). The choice of the initial composition can have non-negligible effects on the calculated abundances at ages typical of protoplanetary disks, something that we plan to investigate in detail in the future.

\subsection{Molecular ions} \label{subsec:ions}

\begin{figure*}
\centering
\includegraphics[angle=0,width=0.88\textwidth]{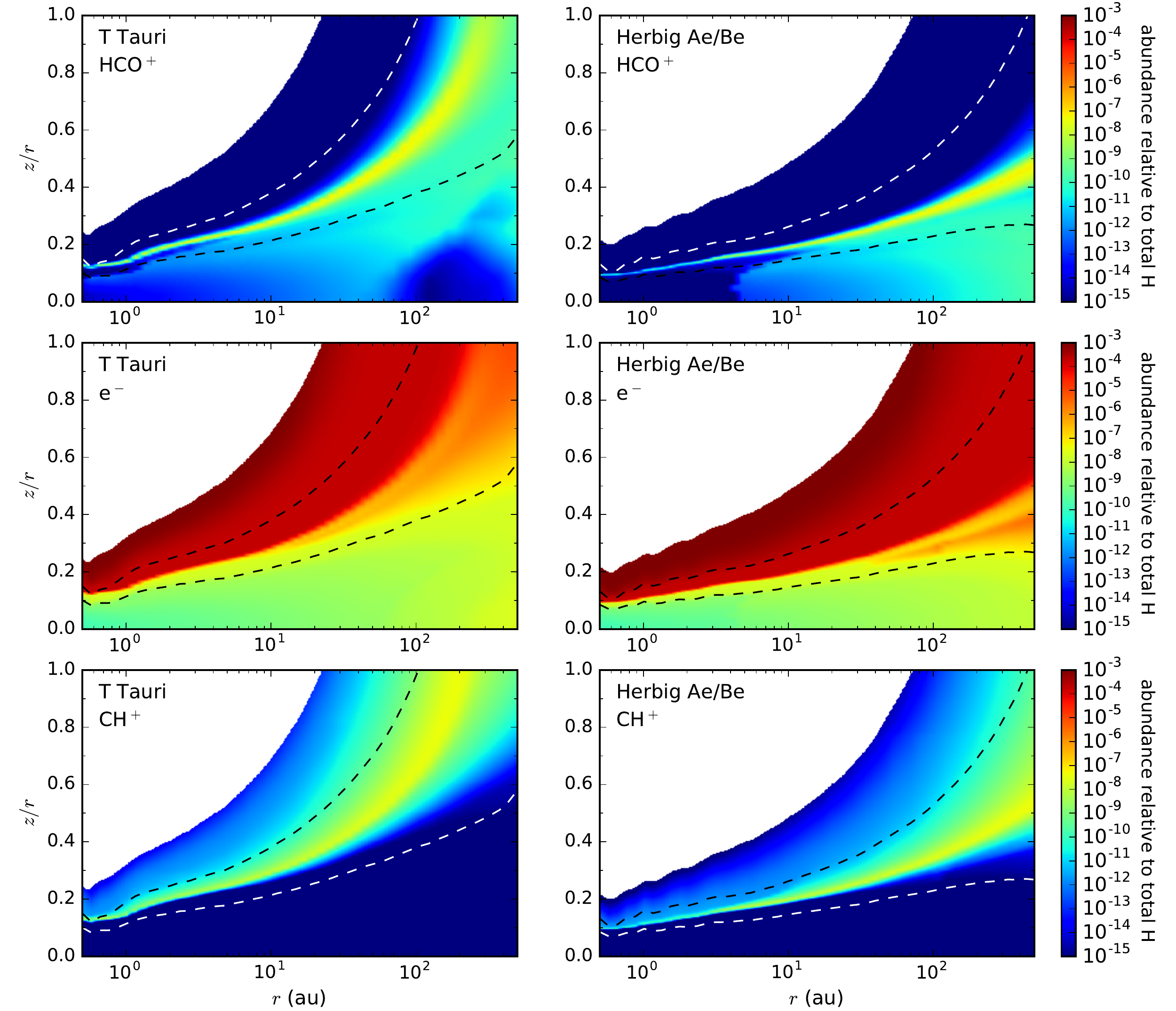}
\caption{Same as Fig.~\ref{fig:map_h2o} but for HCO$^+$, electrons, and CH$^+$.} \label{fig:map_ions}
\end{figure*}

\begin{figure*}
\centering
\includegraphics[angle=0,width=0.88\textwidth]{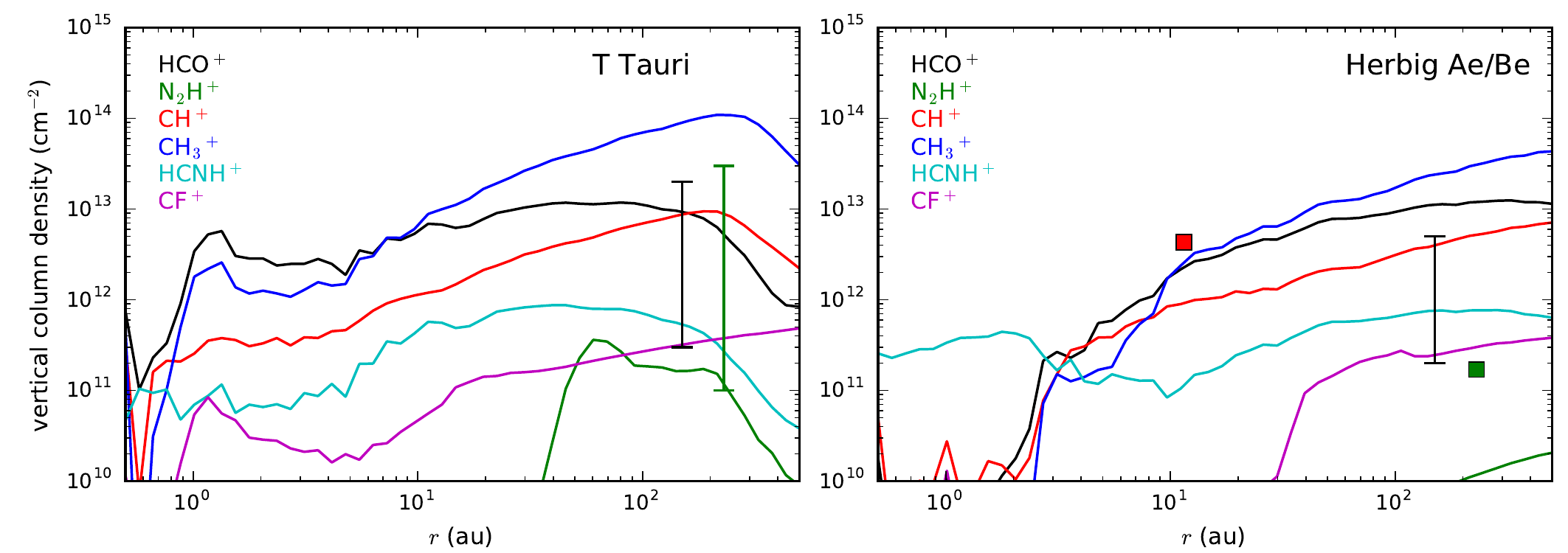}
\caption{Calculated vertical column densities (down to the midplane) of selected molecular ions as a function of radius $r$ for the T\,Tauri (left panel) and Herbig\,Ae/Be (right panel) disks. Column densities derived from observations are indicated by vertical lines or squares, with their radial locations corresponding to the approximate region probed by observations (see Table~\ref{table:observations}).} \label{fig:cdmol_ions}
\end{figure*}

The molecular ion HCO$^+$ is probably one of the most widely observed species in protoplanetary disks (see references in Table~\ref{table:observations}). In fact, it is remarkable that the detection rate of HCO$^+$ is 100 \% in all the T\,Tauri and Herbig\,Ae disks targeted by \cite{obe2010} (2010, 2011) and by \cite{gui2016} (2016). Observed column densities are of the order of 10$^{12}$ cm$^{-3}$, with no significant difference between T\,Tauri and Herbig\,Ae disks (see Table~\ref{table:observations}). In our model, HCO$^+$ is mainly present in the outer disk at intermediate heights (see Fig.~\ref{fig:map_ions}), where it is mainly formed by the reaction between H$_3^+$ and CO. The calculated vertical column density in the outer disk is in the range 10$^{12}$-10$^{13}$ cm$^{-2}$ in both the T\,Tauri and the Herbig\,Ae/Be disks, in good agreement with the values derived from observations (see Fig.~\ref{fig:cdmol_ions}) and in line with results from previous chemical models of T\,Tauri disks (e.g., \cite{aik1999b} 1999b, 2002; \cite{van2003} 2003; \cite{wal2010} 2010, 2012; \cite{sem2011} 2011). Unlike in the Herbig\,Ae/Be disk model, in the T\,Tauri disk model the column density of HCO$^+$ (and other molecular ions) experience a decline beyond $\sim$200 au. In these outer regions, the upper layers are more exposed to stellar FUV photons due to the flared shape of the disk, and atomic ions are favored at the expense of polyatomic ions such as HCO$^+$. This effect is probably a consequence of the particular geometry of the disk model adopted here and may not be a general characteristic of T\,Tauri disks. Therefore, according to the model, there is no reason to expect significantly different HCO$^+$ column densities in T\,Tauri and Herbig\,Ae disks.

Molecular ions other than HCO$^+$ are difficult to observe in protoplanetary disks. Nevertheless, sensitive observations with (sub-)mm interferometers have enabled the detection of N$_2$H$^+$ in a few T\,Tauri disks and one Herbig\,Ae disk, HD\,163296 (see references in Table~\ref{table:observations}). Derived column densities are highly dependent on each source and on each particular study. For example, in the disk around LkCa\,15, \cite{qi2003} (2003) derive $N$(N$_2$H$^+$) = $3.1\times10^{13}$ cm$^{-2}$ from observations with the OVRO array, while \cite{dut2007} (2007) find a column density around 100 times smaller using IRAM PdBI. The chemistry of N$_2$H$^+$ is relatively simple as it is formed by the reaction between H$_3^+$ and N$_2$, which is favored in the coldest regions where CO is mostly in the form of ice, that is, in the outer midplane region. In our particular disk models, CO ice is present beyond a few tens of au in the T\,Tauri disk, while in the Herbig\,Ae/Be disk, due to the much warmer dust temperatures, it is barely formed just in the outer edge (see Sec.~\ref{subsec:ices}). As a consequence, N$_2$H$^+$ is present in the outer T\,Tauri disk with a vertical column density of $\sim10^{11}$ cm$^{-2}$ while it is almost absent in the Herbig\,Ae/Be disk (see Fig.~\ref{fig:cdmol_ions}). The vertical column densities calculated for N$_2$H$^+$ are in line with the low range of observed values in the case of the T\,Tauri disk, while in the Herbig\,Ae/Be disk, $N$(N$_2$H$^+$) is lower than observed in HD\,163296 by more than one order of magnitude (see Fig.~\ref{fig:cdmol_ions}). The model also predicts that N$_2$H$^+$ must have a ring-like distribution with an inner gap, something that has been verified observationally in the TW\,Hya disk (\cite{qi2013c} 2013c). The inner radius and vertical column density calculated for N$_2$H$^+$ are highly dependent on which is the dust temperature structure across the disk, which in turn depends on parameters such as the stellar luminosity, the size and optical properties of dust grains, and the disk geometry (flared vs flat). In any case, since disks around Herbig\,Ae/Be stars are expected to be significantly warmer than around T\,Tauri stars, one should expect larger amounts of N$_2$H$^+$ in T\,Tauri disks than in Herbig\,Ae/Be disks.

Some deuterated ions, mostly DCO$^+$ but also H$_2$D$^+$ and N$_2$D$^+$, have been observed at (sub-)mm wavelengths in a few protoplanetary disks (\cite{vand2003} 2003; \cite{cec2004} 2004; \cite{gui2006} 2006; \cite{qi2008} 2008; \cite{obe2010} 2010, 2011, \cite{mat2013} 2013; \cite{hua2015} 2015; \cite{tea2015} 2015; \cite{hua2017} 2017). Here we do not specifically model deuterium chemistry (see, e.g., \cite{wil2007} 2007; \cite{tea2015} 2015), but we point out that these molecular ions are usually present in the cool outer disk and are useful to probe the fractional ionization. Estimates of the ionization fraction from observations of H$_2$D$^+$ (which is present in the midplane of the cold outer disk) are a few times 10$^{-10}$ (\cite{cec2004} 2004), while values inferred from observations of DCO$^+$ (which is expected in upper layers than H$_2$D$^+$) are as high as $\sim$10$^{-7}$ (\cite{qi2008} 2008; \cite{tea2015} 2015). In our T\,Tauri and Herbig\,Ae/Be disk models, the ionization fraction in the midplane ranges from $\sim10^{-10}$ in the denser inner regions to $\sim10^{-8}$ in the outer disk (see Fig.~\ref{fig:map_ions}). In these midplane regions, the fractional ionization is controlled by the cosmic-ray ionization rate and also depends on the gas density. As one moves to upper layers, where the gas is less dense, warmer, and less shielded against interstellar and stellar FUV photons, the ionization fraction increases gradually up to very high values in the surface of the disk (see Fig.~\ref{fig:map_ions}). According to the model, there are no substantive differences in the ionization degree of disks around T\,Tauri and Herbig\,Ae/Be stars.

The ion CH$^+$ has been detected in a couple of disks, around the Herbig\,Be star HD\,100546 and the Herbig\,Ae star HD\,97048, using \emph{Herschel} (\cite{thi2011} 2011; \cite{fed2013} 2013). The analysis of the emission lines in HD\,100546 indicates that CH$^+$ has an excitation temperature of 100-300 K, with most emission arising from regions inner to 100 au. In our model, CH$^+$ is distributed along a surface layer on top of HCO$^+$ (see Fig.~\ref{fig:map_ions}), where it is mainly formed by the reaction of C$^+$ with hot H$_2$ and FUV-pumped vibrationally excited H$_2$ (\cite{agu2010} 2010) and reaches maximum fractional abundances of $\sim10^{-7}$. The calculated distribution of CH$^+$ is very similar in the T\,Tauri and Herbig\,Ae/Be disks. The model predicts that the largest column densities are reached in the outer disk ($>$100 au; see Fig.~\ref{fig:cdmol_ions}) contrary to what observations of HD\,100546 indicate. We however note that CH$^+$ emission is probably very sensitive to the temperature structure and geometry of each particular disk. In fact, CH$^+$ has only been detected in a couple of disks from the sample of 22 Herbig\,Ae/Be and 8 T\,Tauri systems surveyed by \cite{fed2013} (2013).

Apart from HCO$^+$, the most abundant molecular ions in the disk are H$_3^+$, CH$_2^+$, CH$_3^+$ (see Fig.~\ref{fig:cdmol_ions}), and C$_2$H$_2^+$, species which are very difficult to detect. While H$_3^+$ is mostly present in the midplane of the outer disk, the hydrocarbon cations are essentially present in the disk surface. For example, the formation of CH$_2^+$ and CH$_3^+$ is strongly linked to that of CH$^+$ (\cite{agu2010} 2010). Molecular ions not yet observed in disks but potentially detectable are HCNH$^+$ (the precursor of HCN and HNC in the outer disk) and CF$^+$ (formed in the disk PDR). Both ions are polar and have column densities in excess of 10$^{11}$ cm$^{-2}$ (see Fig.~\ref{fig:cdmol_ions}), although detecting them may be challenging because their dipole moments are not very high. An additional interesting feature emerging from the model is that most of the positive charge is not carried out by molecular ions but by atomic ions (metals in the disk midplane and carbon in the disk surface).

\subsection{Ices and snowlines} \label{subsec:ices}

\begin{figure*}
\centering
\includegraphics[angle=0,width=0.88\textwidth]{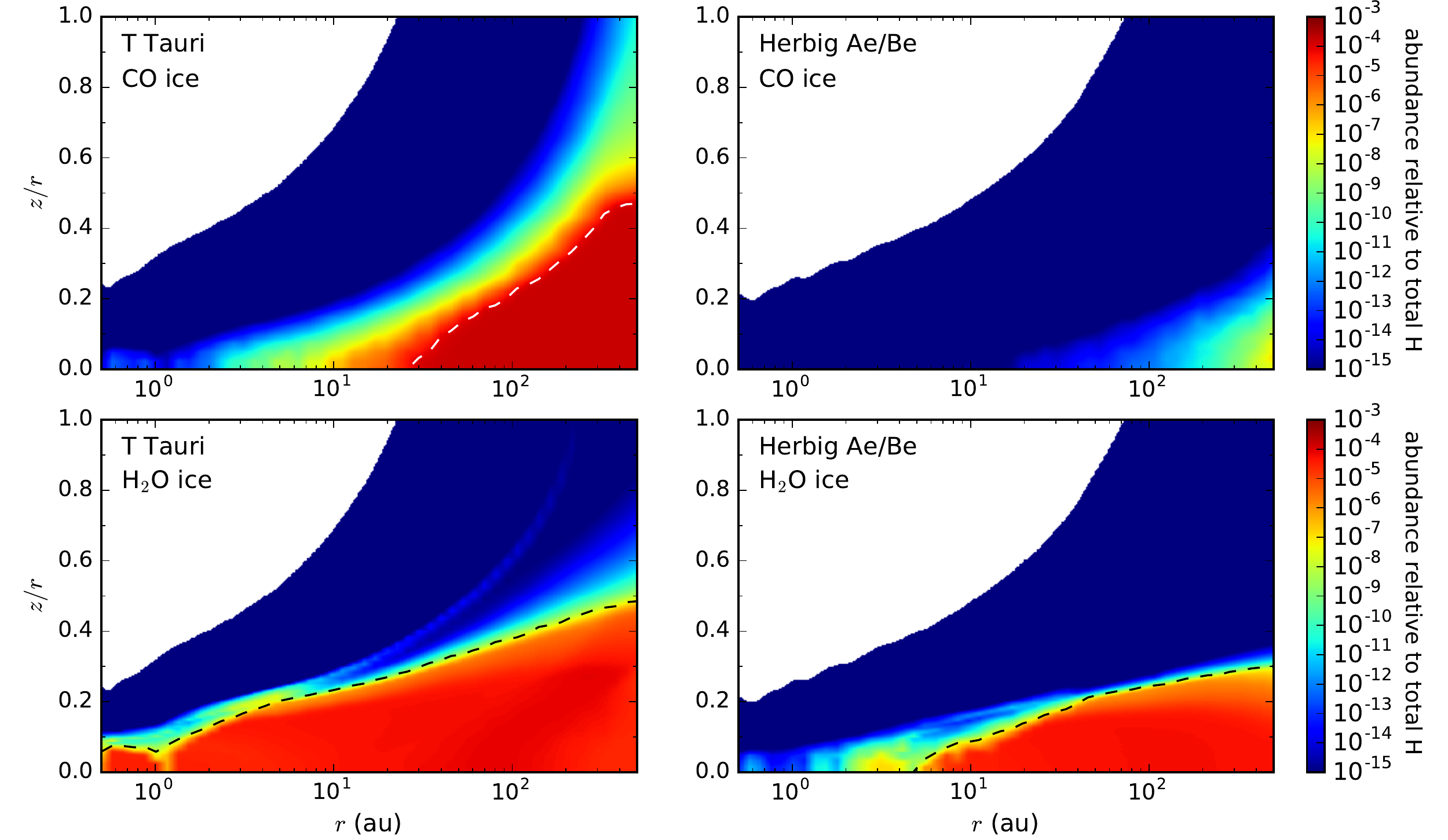}
\caption{Calculated distributions of CO and H$_2$O ices as a function of radius $r$ and height over radius $z/r$ for the T\,Tauri (left) and Herbig\,Ae/Be (right) disks. The dashed line in each panel indicates the location of the corresponding snowline.} \label{fig:map_ices}
\end{figure*}

\begin{figure}
\centering
\includegraphics[angle=0,width=\columnwidth]{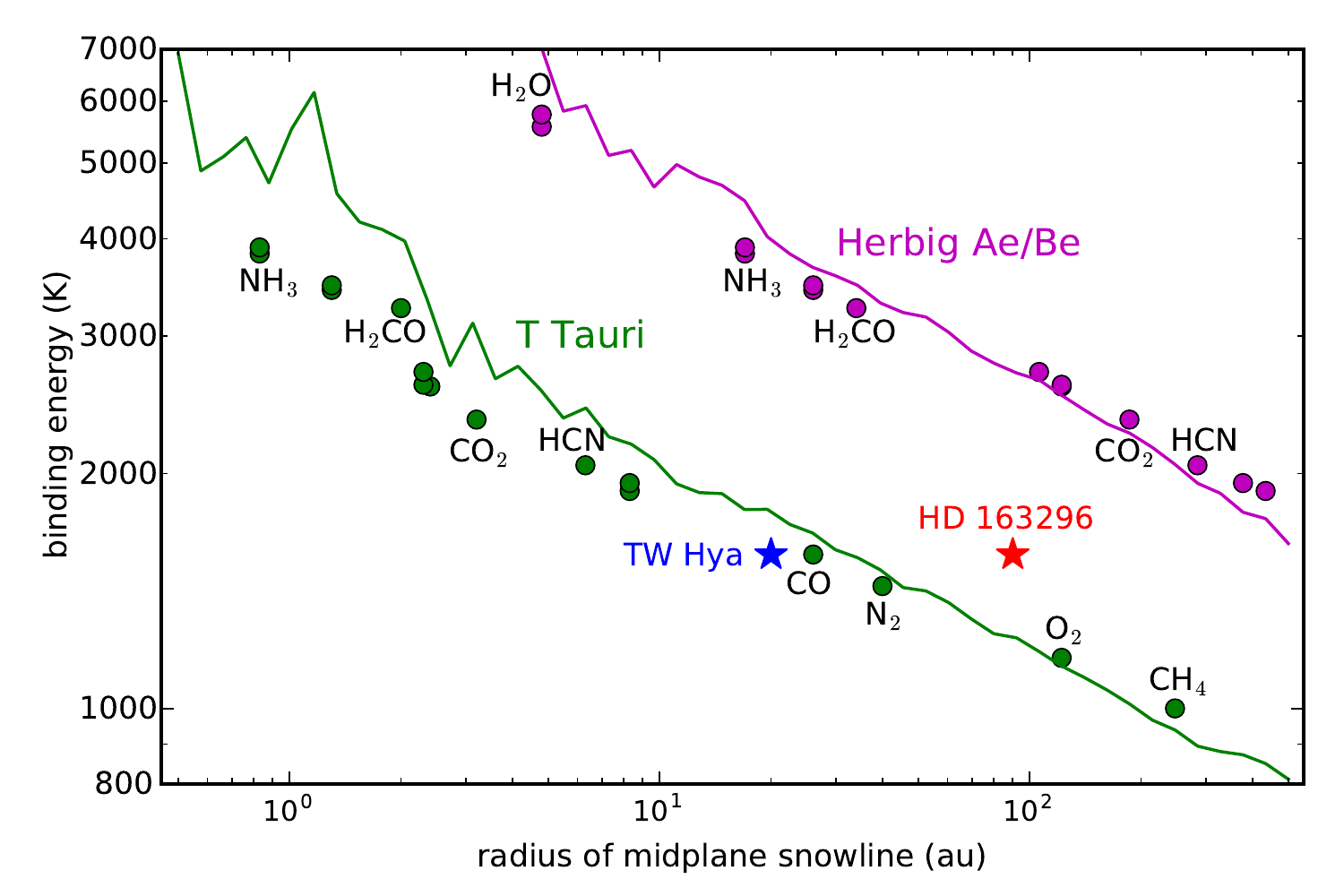}
\caption{Midplane snowline of various ices (along $x$-axis) as a function of the binding energy (along $y$-axis) in the T\,Tauri (green circles) and Herbig\,Ae/Be (magenta circles) disks. Observational CO snowlines in TW\,Hya (\cite{sch2016} 2016; \cite{zha2017} 2017) and HD\,163296 (\cite{qi2015} 2015) are also shown. We also plot the dust temperature in the midplane (scaled up by a factor of 45) as a function of radius in the T\,Tauri (green solid line) and Herbig\,Ae/Be (magenta solid line) disk models.} \label{fig:trend_snowlines}
\end{figure}

Ices account for an important percentage of the matter of protoplanetary disks. The high densities and cold temperatures prevailing in the midplane regions ensure a rapid and efficient adsorption of gas-phase molecules onto dust grains, where they settle and live long in the form of ice. The presence or absence of a particular type of ice in a certain disk region depends on the balance between adsorption and desorption, and, because adsorption rates are similar for most molecules (see Sec.~\ref{subsubsec:adsorption}), the extent of each particular type of ice is determined by its specific desorption rate. According to the model, among the various desorption mechanisms considered (see Sec.~\ref{subsubsec:desorption}), thermal desorption is clearly the most important and the one that shapes the bulk distribution of most ices. We however note that for those species with large binding energies, the vertical extent of ices in the outer disk is essentially controlled by photodesorption by interstellar and stellar FUV radiation. In these outer regions, dust temperatures may not be high enough to trigger thermal desorption of strongly bonded ices, while FUV photons can penetrate down to intermediate heights and provide an efficient means of desorbing ice molecules. The main effect of photodesorption is that it shifts down the snowline\footnote{The snowline of a particular species is defined here as the transition region where its gas and ice abundances become equal.} of highly polar molecules in the outer disk, enhancing their gas-phase abundance at intermediate heights. In summary, the distribution of ices with low binding energies, such as CO and N$_2$, is controlled by thermal desorption while for ices with large binding energies, like NH$_3$ and H$_2$O (see Table~\ref{table:ices}), their distribution is determined by thermal desorption in the inner regions and by photodesorption in the outer disk. Other desorption mechanisms, such as cosmic-ray induced desorption or photodesorption by FUV photons generated through the Prasad-Tarafdar mechanism, are much less important.

In protoplanetary disks, ices are mostly distributed around the midplane, from the very outer disk down to an inner edge, which is given by the location of the snowline in the midplane. Since the midplane snowline is essentially controlled by thermal desorption (and thus by the binding energy of the particular ice and by the dust temperature structure of the disk along the midplane), ices appear progressively in the radial direction according to their binding energies. In Fig.~\ref{fig:map_ices}, we show the calculated distributions of CO and H$_2$O ices in the T\,Tauri and Herbig\,Ae/Be disks. First focusing on the T\,Tauri disk model, we see that CO ice, for which the adopted binding energy is 1575 K, is only present beyond a few tens of au, while H$_2$O ice, which has a much higher binding energy (the adopted value is 5773 K), is present all over the disk midplane. The fact that water ice extends down to the inner disk edge is a particular outcome of the T\,Tauri disk model adopted here, which results in dust that is too cold to allow for efficient thermal desorption of water ice in the inner disk midplane. Since dust temperatures are strongly dependent on parameters such as the size of dust grains, other T\,Tauri disk models may find outer water snowlines compared to that obtained here (see, e.g., \cite{wal2015} 2015). An obvious difference between the T\,Tauri and the Herbig\,Ae/Be disk models is that in the later, snowlines shift to larger radii owing to the warmer dust temperatures (see Fig.~\ref{fig:map_ices}) and as a consequence the mass of ices in the disk becomes smaller.




The abundance distributions of CO and H$_2$O ices shown in Fig.~\ref{fig:map_ices} serve to illustrate two extreme cases of molecules with low and high, respectively, binding energies (see Table~\ref{table:ices}). Other ices with binding energies between those of CO and H$_2$O have intermediate distributions between those of these two molecules. In general, the smaller the binding energy, the snowline shifts to larger radii. This relationship is illustrated in Fig.~\ref{fig:trend_snowlines}, where we plot the radius at which the midplane snowline lies for various ices as a function of their binding energies. There is a clear trend which indicates that both quantities are related by a power law of the type $E_D \propto r_{\rm ms}^{-q}$, where $r_{\rm ms}$ stands for radius of midplane snowline. A fit to the data points in Fig.~\ref{fig:trend_snowlines} yields an exponent $q$ of $\sim$0.23 in both the T\,Tauri and the Herbig\,Ae/Be disks. This behavior is not surprising since the midplane snowline of a given ice is essentially located at the radius at which the dust temperature in the midplane becomes similar to the condensation temperature of the ice, and the latter is directly proportional to the binding energy of the ice. That is, the observed relation between binding energy and midplane snowline merely reflects how the midplane dust temperature varies with radius. In Fig.~\ref{fig:trend_snowlines} we have overplotted as a solid line the radial profile of the dust temperature in the midplane scaled up by a factor of 45. The comparison between solid line and data points in Fig.~\ref{fig:trend_snowlines} suggests that a factor of proportionality between binding energy and condensation temperature in the range 30-50 is adequate for the protoplanetary disks modeled here. For comparison, \cite{hol2009} (2009) calculate this factor of proportionality to be $\sim$50 while \cite{mar2014} (2014) quote a value of $\sim$30 (from \cite{att1998} 1998).


\begin{figure*}
\centering
\includegraphics[angle=0,width=\textwidth]{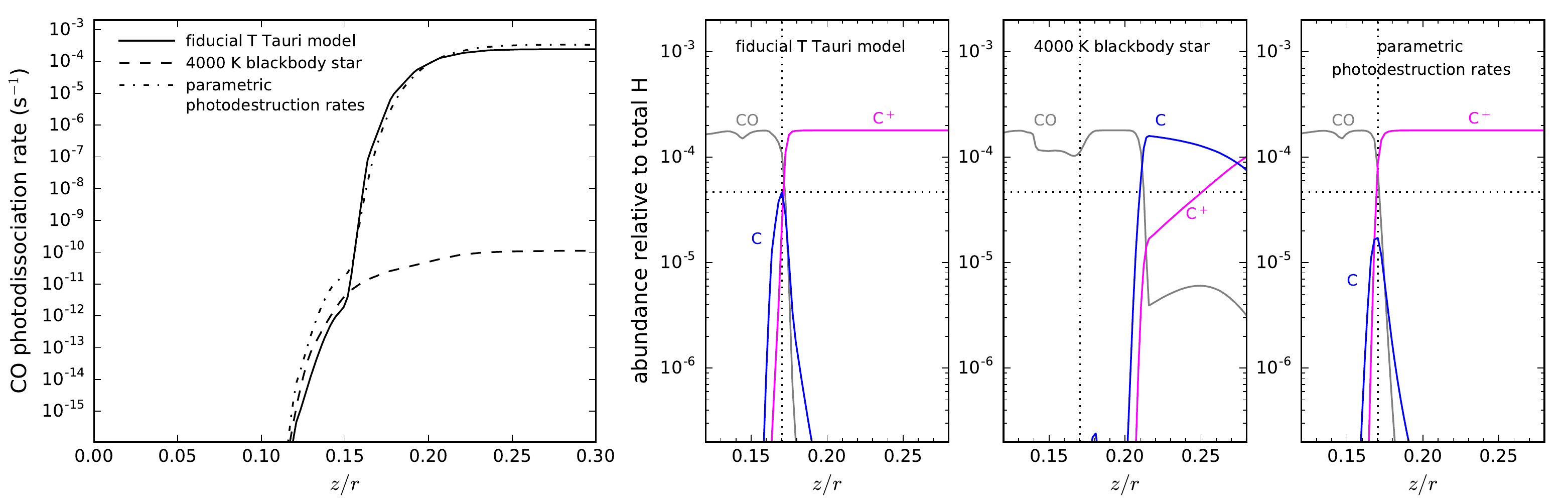}
\caption{Vertical structure of the CO photodissociation rate (left panel) and the CO/C/C$^+$ interface (right panels) at a radius of 1 au from the star for the T\,Tauri disk. We show results from our fiducial model, a model in which a 4000 K blackbody stellar spectrum is considered, and a model in which all photodestruction rates are computed using parametric expressions (see text). The dotted lines in the three right panels indicate what is the peak abundance of atomic carbon and at which height it is reached in the fiducial model. The small drop in the abundance of CO seen around $z/r\sim0.15$, which is more marked in the 4000 K blackbody star model, is due to a maximum in the abundance of water.} \label{fig:carbon_interface}
\end{figure*}

Using observations to reveal the location of snowlines in protoplanetary disks is still challenging, although ALMA observations are starting to put interesting constraints. For example, in the disk around the T\,Tauri star TW\,Hya the CO snowline has been found to lie at a radius of $\sim20$ au (\cite{sch2016} 2016; \cite{zha2017} 2017), while in the disk around the Herbig\,Ae star HD\,163296 it has been located at a radius of 90 au from the star (\cite{qi2015} 2015). These two observational points are included in Fig.~\ref{fig:trend_snowlines}. While the CO snowline derived for TW\,Hya is close to the value calculated in our T\,Tauri disk model, the radius derived for HD\,163296 is much farther in than indicated by our Herbig\,Ae/Be disk model, which puts the CO snowline even beyond the outer disk edge. We however note that the coincidence in the case of TW\,Hya is very likely accidental because our generic T\,Tauri disk model is not aimed to represent neither the TW\,Hya disk nor any other particular disk. The observational finding of an outer CO snowline in the HD\,163296 disk than in the TW\,Hya disk is in line with the general expectation that snowlines are shifted to larger radii in Herbig\,Ae/Be disks compared to T\,Tauri disks, although it may be also fortuitous taking into account the poor statistics, consisting of just one object of each type. Expanding the sample of disks seems mandatory to draw more definitive conclusions. It will be also very interesting to observationally locate the snowline of different molecules in the same disk, so that the sequence of snowlines of ices with different binding energies depicted in Fig.~\ref{fig:trend_snowlines} can be tracked.

\section{Influence of the stellar spectrum and the photodestruction rates} \label{sec:influence}

\begin{figure*}
\centering
\includegraphics[angle=0,width=\textwidth]{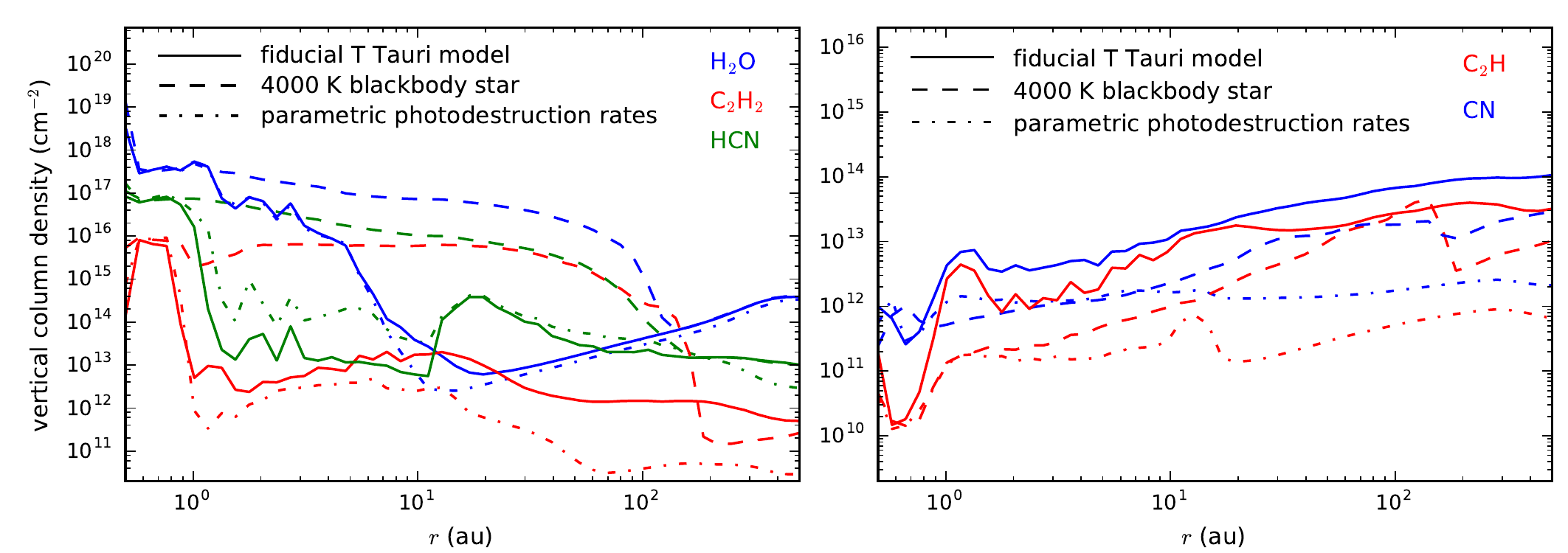}
\caption{Calculated vertical column densities down to the midplane of H$_2$O, C$_2$H$_2$, and HCN (left panel) and the radicals C$_2$H and CN (right panel) as a function of radius $r$ for the T\,Tauri disk. We show results from our fiducial model, a model in which a 4000 K blackbody stellar spectrum is considered, and a model in which all photodestruction rates are computed using parametric expressions (see text).} \label{fig:cdmol_comparison}
\end{figure*}

Chemical models of protoplanetary disks contain so many ingredients to deal with the multiple processes at work that the output abundance distributions can be sensitive to many of them. In this study we put the focus on the photochemistry and thus here we evaluate the influence of a couple of related aspects: the spectrum of the star and the way in which photodestruction rates are computed. We concentrate on the T\,Tauri disk model for this sensitivity analysis.

T\,Tauri stars are cool, with effective temperatures of the order of 4000 K, although they usually have an important FUV excess that can greatly affect the photochemistry of the disk. To evaluate the influence of this FUV excess we compare our fiducial T\,Tauri disk model, in which we consider the stellar spectrum of TW\,Hya, with a model in which we assume that the star emits as a blackbody at a temperature of 4000 K (both spectra are shown in Fig.~\ref{fig:uv_spectra}). The lack of FUV excess in this latter case results in much lower photodestruction rates (compare our values in Table~\ref{table:app_photorates} with those of \cite{van2006} 2006, or see \cite{hea2017} 2017). The net effect is that without FUV excess, most photodestruction rates are dominated by the ISRF rather than by stellar radiation. As as example, in the left panel of Fig.~\ref{fig:carbon_interface} we show the photodissociation rate of CO as a function of height at a radius of 1 au, where both the contributions of the ISRF and the star are taken into account. It is seen that in the upper disk, where photodestruction rates are dominated by stellar photons (see Fig.~\ref{fig:photorates}), the photodissociation rate of CO in the 4000 K blackbody star model is below that in the fiducial model by many orders of magnitude. A similar behavior occurs for the photodestruction rates of other species. As a consequence, the CO/C/C$^+$ interface shifts to upper heights and the atomic carbon layer becomes wider (see right panels in Fig.~\ref{fig:carbon_interface}). In general, the photochemically active layer, where for example the transition H$_2$O/OH is located and the radicals C$_2$H and CN are, is shifted to upper layers. This overall shift of the photochemistry to upper, less dense, and warmer layers, together with the lower strength of the FUV radiation field, induce important changes in the abundances of several molecules. For example, stable molecules like water and the simple organics  C$_2$H$_2$ and HCN become more abundant at intermediate disk radii compared to the fiducial model (see left panel in Fig.~\ref{fig:cdmol_comparison}), while the radicals C$_2$H and CN experience a decline in their abundances (see right panel in Fig.~\ref{fig:cdmol_comparison}).

The way in which photodestruction rates are computed may also have an influence on the chemical structure of the photochemically active layer of disks (e.g., \cite{wal2012} 2012). In our fiducial model, we have used the Meudon PDR code to compute the photodissociation rates of H$_2$ and CO by solving the excitation and the line-by-line FUV radiative transfer to properly account for self and mutual shielding effects and have computed photodestruction rates for a variety of species using the cross sections compiled in Appendix~\ref{sec:app_sections}. To investigate the effect of using a simpler approach, we have run a T\,Tauri disk model in which all photodestruction rates are computed through the parametric expression in Eq.~(\ref{eq:photorate}). The $\alpha$ and $\gamma$ parameters are taken from Table~\ref{table:app_photorates}, while for H$_2$ and CO we adopt as unattenuated rates and dust shielding factors the values given by \cite{hea2017} (2017). In the left panel of Fig.~\ref{fig:carbon_interface} we compare the resulting CO photodissociation rate as a function of height at a radius of 1 au given by this parametric photodestruction rates approach with that resulting from our fiducial model. It is seen that in the upper disk both approaches yield similar results although at lower heights, where photodestruction is dominated by ISRF photons (see Fig.~\ref{fig:photorates}), the parametric approach overestimates the photodissociation rate of CO. The resulting CO/C/C$^+$ interface is very similar in both scenarios, although it is worth noting that the peak abundance of atomic carbon is lower when the parametric photodestruction rates are used. This has some consequences for the abundances of carbon-bearing species typically formed under the action of photochemistry, such as the radicals C$_2$H and CN. In fact, these two radicals reach abundances significantly lower in the outer disk when parametric photodestruction rates are used compared to the fiducial model (see right panel in Fig.~\ref{fig:cdmol_comparison}). Other species are also affected to different degrees. For example, while the column density of water does not change much, acetylene becomes less abundant and hydrogen cyanide is slightly enhanced in the range of radii between 1 and 10 au (see left panel in Fig.~\ref{fig:cdmol_comparison}).

\section{Conclusions} \label{sec:conclusions}

We have developed a model aimed to compute the chemical composition of a generic protoplanetary disk around a young star. The model considers a passively irradiated disk in steady state and computes the physical and chemical structure of the disk with particular attention to the disk photochemistry. In particular we have compiled cross sections for 29 molecules and 8 atoms and computed the photodissociation and photoionization rates at each location in the disk by solving the FUV radiative transfer with the Meudon PDR code in a 1+1D approach.

We have applied the model to perform a comparative study of the chemistry of disks around low-mass (T\,Tauri) and intermediate-mass (Herbig\,Ae/Be) stars. Infrared and (sub-)mm observations of T\,Tauri and Herbig disks point to a lower detection rate of molecules in the latter type of disks and, for some species, somewhat lower abundances. Motivated by the observational studies, we have investigated the potential chemical differentiation between disks around these two types of stars, which have very different masses and spectra. We find that globally the chemical behavior of these two types of disks is quite similar, with some important differences driven by the higher stellar ultraviolet flux and the warmer temperatures of Herbig\,Ae/Be disks.

Water vapor and the simple organic molecules C$_2$H$_2$, HCN, and CH$_4$ are predicted to be very abundant ($\sim10^{-4}$ for H$_2$O and a few $\times10^{-5}$ for the organics) in the hot inner regions of disks around both T\,Tauri and Herbig\,Ae/Be stars. The main difference between the two types of disks is that these molecules extend over a larger region in Herbig\,Ae/Be disks due to the warmer temperatures attained, a finding that is in agreement with the models by \cite{wal2015} (2015) but in contrast with infrared observations that find a much lower detection rate of water and simple organics toward Herbig\,Ae/Be disks than toward T\,Tauri disks. This latter fact is probably caused by observational aspects rather than by substantive differences in the chemistry between these two types of disks. For example, \cite{ant2016} (2016) point out that Herbig stars are brighter than T\,Tauri stars resulting in a higher level of mid-IR continuum and smaller line/continuum ratios. If true, the higher sensitivity and spectral resolution of the James Webb Space Telescope with respect to \emph{Spitzer} could allow to increase the detection rate of water and simple organics in Herbig\,Ae/Be systems.

Concerning the outer regions of disks, observations point to a lower detection rate of molecules and to somewhat lower abundances in Herbig\,Ae disks compared to T\,Tauri disks, although the statistics of Herbig\,Ae objects observed is small. Our model indicates that in general there are not drastic differences between T\,Tauri and Herbig\,Ae disks concerning the abundances of molecules typically observed in disks at (sub-)mm wavelengths. More specifically, various molecules, such as H$_2$CO, CS, SO, and HCO$^+$, are observed with similar abundances in the outer regions of T\,Tauri and Herbig\,Ae disks and the model satisfactorily finds no clear differentiation between both types of disks for these species. For other species, such as HCN and the radicals C$_2$H and CN, observations suggest that T\,Tauri disks may retain somewhat larger abundances than Herbig\,Ae disks in the outer regions, while the model indeed predicts that these species are slightly less abundant in Herbig\,Ae disks than in T\,Tauri disks. In the case of the radicals C$_2$H and CN, which are produced by the action of photochemistry, the slightly lower abundances calculated in Herbig\,Ae disks are caused by the higher ultraviolet flux, which narrows the photochemically active layer and limits the ability of photochemistry to synthesize molecules. In any case, for HCN, C$_2$H, and CN, the observed and calculated abundance differences between both types of disks are small, of a factor of a few at most.

For other organic molecules with a certain complexity such as CH$_3$OH, HC$_3$N, CH$_3$CN, and $c$-C$_3$H$_2$, the match with observed values is not satisfactory. This fact, together with the ample dispersion of abundances calculated by different chemical models in the literature point to the chemistry of these molecules not being yet as robust as for simpler species, also because grain-surface chemistry, not included in our model, is likely playing an active role in the synthesis of some of these molecules.

A clear differentiation between T\,Tauri and Herbig\,Ae disks is however found concerning ices. The warmer temperatures of Herbig\,Ae disks shift snowlines to larger radii compared to T\,Tauri disks and as a consequence disks around intermediate-mass stars are expected to contain a substantially lower mass of ices compared to T\,Tauri disks.

\begin{acknowledgements}

We thank the anonymous referee for a detailed reading and for a constructive report which helped to improve this manuscript. M.A. acknowledges funding support from the European Community 7th Framework Programme through a \emph{Marie Curie Intra-European Individual Fellowship} (grant 235753), from the European Research Council through ERC Synergy grant 610256, and from Spanish MINECO through the Ram\'on y Cajal programme (RyC-2014-16277) and grant AYA2016-75066-C2-1-P. This work was partly supported by the CNRS program \emph{Physique et Chimie du Milieu Interstellaire} (PCMI) co-funded by the Centre National d'\'Etudes
Spatiales (CNES). We thank J. R. Goicoechea, E. Bron, S. Cazaux, G. M. Mu\~noz-Caro, R. Mart\'in-Dom\'enech, and A. Fuente for useful discussions, S. Miyake for help in the treatment of cross sections from the former version of the Leiden database, and E. A. Bergin, C. E. Brion, J. A. Nuth, D. M. P. Holland, K.-L. Han, and J.-H. Fillion for kindly providing useful data.

\end{acknowledgements}

\clearpage

\appendix

\section{Photodissociation and photoionization cross sections} \label{sec:app_sections}

Photodissociation and photoionization processes are treated using the relevant cross section as a function of wavelength. We adopt this approach for key species for which cross section data are available in the literature, excluding H$_2$ and CO, the photodissociation of which are treated solving the excitation and line-by-line radiative transfer and taking into account self-shielding effects. We are mainly interested in cross sections ranging from the Lyman cutoff, at 911.776 $\AA$, to wavelengths in the range 1500-4000 $\AA$, where most molecules have their photodissociation threshold. We note that in protoplanetary disks, stellar photons with wavelengths shorter than the Lyman cutoff may also have an impact on the photodissociation and photoionization rates in disk regions exposed to the star. It will be worth to take into account this in the future.

Cross sections of molecules with a closed electronic shell have been mostly taken from experimental studies carried out at room temperature. In the case of radicals, most of the cross section data comes from theoretical studies. We note that cross sections can show a great dependence with temperature, something that may have an important impact on the chemistry of protoplanetary disks taking into account that the gas kinetic temperature can take values from hundreds to thousands of degrees Kelvin in the photon-dominated region of the disk. Of course it would be desirable to use temperature-dependent cross sections, although such data is currently limited to just a few species (e.g., \cite{ven2013} 2013; \cite{mcm2016} 2016). This will be something to take into account in the future.

\begin{figure*}
\centering
\includegraphics[angle=0,width=\textwidth]{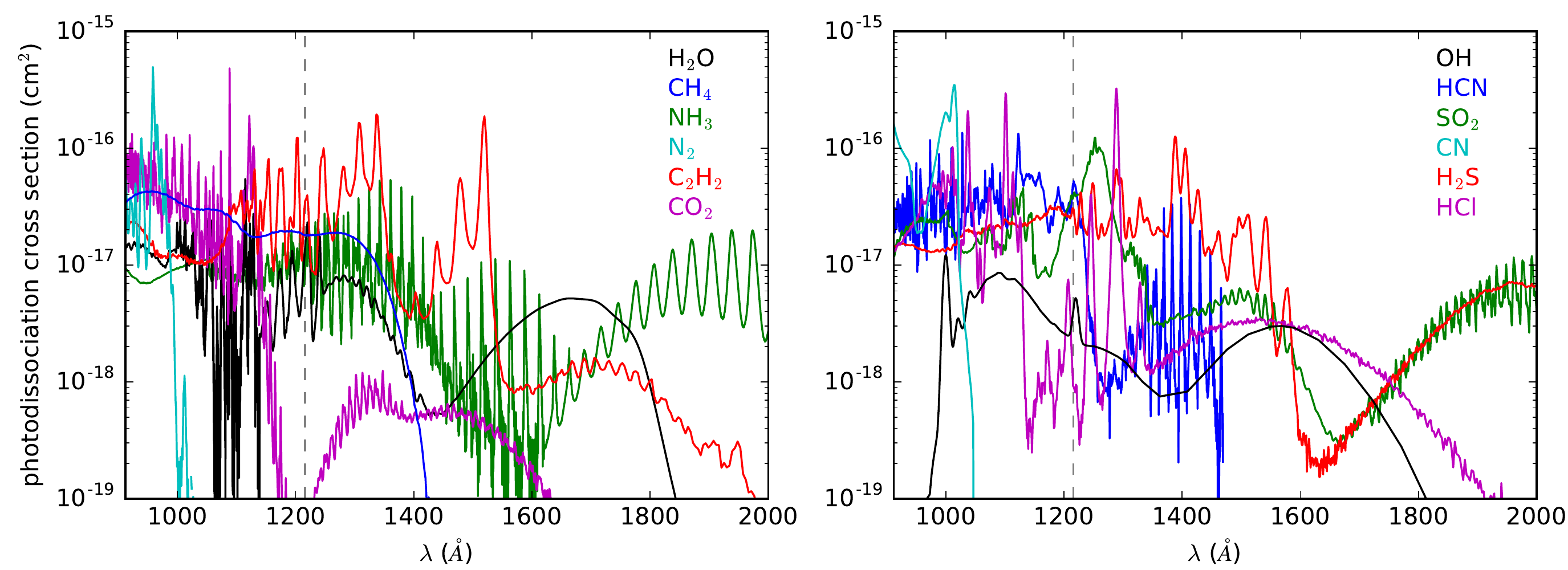}
\caption{Photodissociation cross section of various molecules as a function of wavelength. The position of the Ly-$\alpha$ line at 1216 $\AA$ is marked with a vertical dashed line.}
\label{fig:app_section}
\end{figure*}

We have compiled photo cross sections for 29 molecules and 8 atoms from original sources in the literature or from databases such as the \textit{MPI-Mainz UV/VIS Spectral Atlas of Gaseous Molecules of Atmospheric Interest}\footnote{\texttt{http://www.uv-vis-spectral-atlas-mainz.org}}, which contains extensive information on experimental cross sections of stable molecules, the \textit{Photo Rate Coefficient Database}\footnote{\texttt{http://phidrates.space.swri.edu/}} (\cite{hue1992} 1992), the \textit{Leiden database of photodissociation and photoionization of astrophysically relevant molecules}\footnote{\texttt{http://home.strw.leidenuniv.nl/$\sim$ewine/photo/}} (\cite{van2006} 2006; \cite{hea2017} 2017), which contains theoretical data of numerous radicals. It is common that experimental studies provide the photoabsorption cross section without disentangling whether the absorption leads to fluorescence, dissociation, or ionization. Unless otherwise stated, we have assumed that absorption of FUV photons not leading to ionization leads to dissociation of the molecule, that is, that the contribution of fluorescence to the total photoabsortion cross section is negligible. This may be a bad approximation at long wavelengths but is likely to be correct at short wavelengths, typically below 1500 $\AA$. In the case of photoionization of atoms we have made use of databases such as TOPbase, the Opacity Project atomic database\footnote{\texttt{http://cdsweb.u-strasbg.fr/topbase/topbase.html}} (\cite{cun1993} 1993), to retrieve cross sections for some of the atoms. In the next subsections we detail the cross section adopted for each of the 29 molecules and 8 atoms considered. In Fig.~\ref{fig:app_section} we show the photodissociation cross section as a function of wavelength for some key molecules.

\subsection{CH}

The photodissociation cross section of methylidyne has been taken from the calculations of \cite{van1987} (1987), which comprise dissociation by lines and by continuum. The photoionization cross section has been obtained from the theoretical study by \cite{bar1978} (1978). The photoionization threshold of CH is 1170 $\AA$ (\cite{hue1992} 1992).

\subsection{CH$^+$}

The photodissociation cross section of CH$^+$ is taken from the theoretical calculations of \cite{kir1980} (1980), where we have adopted a cross section of 10 Mb\footnote{1 Mb is equal to 10$^{-18}$ cm$^2$.} at the resonances lying at 1523, 1546, and 1579 $\AA$. Photodissociation through the three excited electronic states considered by \cite{kir1980} (1980)$, 2^1\Sigma^+$, $2^1\Pi$, and $3^1\Sigma^+$, yields as products C + H$^+$.

\subsection{CH$_2$}

The photodissociation cross section of the methylene radical has been taken from \cite{van1996} (1996), whose calculations comprise dissociation by lines and by continuum radiation. Methylene has two main dissociation channels:
\begin{eqnarray}
\rm CH_2 + \emph{h$\nu$} & \rightarrow & \rm CH + H \\
 & \rightarrow & \rm C + H_2,
\end{eqnarray}
but the latter is a minor one (\cite{kro1997} 1997), and has therefore been neglected here. The ionization threshold of CH$_2$ is 1193 $\AA$, although there are not cross section data in the literature, except for the relative data in the ionization threshold region by \cite{lit1998} (1998).

\subsection{CH$_3$}

The UV spectra of CH$_3$ has been recorded in photographic plates (\cite{her1961} 1961; \cite{cal1976} 1976), although no quantitative measurement of the intensity was carried out. More recently, the absolute photodissociation cross section has been measured in the 2000-2400 $\AA$ region by \cite{cam2002} (2002) and in the band at 2164 $\AA$ by \cite{kha2007} (2007), in good agreement with previous absolute measurements: 40.1-41.2 Mb at 2136.6 $\AA$ (\cite{mac1985} 1985) and 35.1 Mb at 2164 $\AA$ (\cite{art1986} 1986). We have adopted the photodissociation cross section measured by \cite{cam2002} (2002) and \cite{kha2007} (2007) at wavelengths longer than 2000 $\AA$, and have calibrated the band intensities classified in five categories, from very weak to very strong, by \cite{her1961} (1961) based on the absolute cross section measured for the 2164 $\AA$ band. The dissociation of the methyl radical has two possible channels:
\begin{eqnarray}
\rm CH_3 + \emph{h$\nu$} & \rightarrow & \rm CH_2 + H \\
 & \rightarrow & \rm CH + H_2,
\end{eqnarray}
whose quantum yields are not clear. Channel (A.3) is observed at 2163 $\AA$ by \cite{wil1994} (1994), at 1933 $\AA$ by
\cite{nor1995} (1995), and at 2125 $\AA$ by \cite{wu2004} (2004), while channel (A.4) is observed through fluorescence of CH at $\lambda$ $>$ 1050 $\AA$ by \cite{kas1994} (1994). In the absence of better constraints we have assumed a quantum yield of 50 \% for each channel. As concerns photoionization, we adopt the absolute cross section measured by \cite{gan2010} (2010) from the photoionization threshold (1260 $\AA$) to 1085 $\AA$, and consider a constant cross section shortward of 1085 $\AA$. The adopted cross section is, within uncertainties, similar to those reported in previous studies (\cite{taa2008} 2008; \cite{loi2010} 2010).

\subsection{CH$_4$}

In the case of methane the photoabsorption cross section has been taken from the experimental study by \cite{au1993} (1993) while the photoionization cross section (the ionization threshold of CH$_4$ is 983.2 $\AA$) has been measured by \cite{wan2008} (2008). Methane has various photodissociation channels leading to the radicals CH$_3$, CH$_2$, and CH:
\begin{eqnarray}
\rm CH_4 + \emph{h$\nu$} & \rightarrow & \rm CH_3 + H \\
 & \rightarrow & \rm CH_2(\emph{$X^3B_1$}) + H + H~~or~~CH_2(\emph{$a^1A_1$}) + H_2, \\
 & \rightarrow & \rm CH + H_2 + H~~or~~CH + H + H + H,
\end{eqnarray}
whose quantum yields have been measured by \cite{gan2011} (2011) at 1182 $\AA$ and 1216 $\AA$.

\subsection{C$_2$}

The photodissociation and photoionization cross sections of C$_2$ have been taken from the theoretical studies by \cite{pou1983} (1983) and \cite{tof2004} (2004), respectively. The photoionization threshold of C$_2$ is 1020 $\AA$ (\cite{tof2004}
2004).

\subsection{C$_2$H}

In the case of the ethynyl radical we adopt the photodissociation cross section by lines calculated by \cite{van2008} (2008). The main dissociation channel yields C$_2$ + H (\cite{duf1994} 1994; \cite{sor1997} 1997; \cite{meb2001} 2001; \cite{apa2004} 2004). The photoionization threshold of C$_2$H is 1068 $\AA$ (\cite{lid2009} 2009), although to our knowledge there are not cross section data available in the literature.

\subsection{C$_2$H$_2$}

For acetylene the photoionization cross section at wavelengths shorter than the ionization threshold of 1087.6 $\AA$ has been taken from the compilation by \cite{hud1971} (1971). We have substracted the cross section of photoionization to the
photoabsorption cross section measured by \cite{coo1995} (1995) to obtain the photodissociation cross section, which is assumed to yield the C$_2$H radical with a 100 \% efficiency.

\subsection{OH}

The photoabsorption cross section of the hydroxyl radical has been measured by \cite{nee1984} (1984) in the wavelength range 1150-1830 $\AA$. On the theoretical side, the photodissociation has also been studied by \cite{van1984} (1984). The experimental values are somewhat higher than the theoretical ones, although, on the other hand, calculations indicate that photodissociation by lines is also important at wavelengths shorter than 1150 $\AA$. We adopt the theoretical photodissociation cross section, mainly because it spans over a broader wavelength range. Relative photoionization cross section has been measured by \cite{deh1984} (1984) up to the threshold at 952.5 $\AA$ (\cite{lid2009} 2009). The absolute scale of the photoionization cross section is fixed by setting a value of 2.5 Mb at 946 $\AA$, according to the calculations of \cite{ste1988} (1988).

\subsection{H$_2$O}

The photoabsorption cross section of water has been taken from various experimental studies: \cite{fil2004} (2004) for the
wavelength range 999-1139 $\AA$, \cite{mot2005} (2005) in the wavelength range 1148-1939 $\AA$, and \cite{cha1993a} (1993a) elsewhere in the range from the Lyman cutoff to 2060.5 $\AA$. The photoionization cross section for wavelengths shorter than the ionization threshold of H$_2$O (982.4 $\AA$) has been taken from the experimental study of \cite{fil2003} (2003). Water has two main photodissociation channels, yielding OH radicals and O atoms:
\begin{eqnarray}
\rm H_2O + \emph{h$\nu$} & \rightarrow & \rm OH + H, \\
 & \rightarrow & \rm O(\emph{ $^3$P}) + H + H~~or~~O(\emph{$^1$D}) + H_2,
\end{eqnarray}
where channel (A.8) dominates over channel (A.9). Water can be photodissociated in the first band $\tilde{X}^1A_1-\tilde{A}^1B_1$ at wavelengths longer than 1360 $\AA$ and in the second
$\tilde{X}^1A_1-\tilde{B}^1A_1$ and higher bands at shorter wavelengths. Quantum yields of 78 \% and 22 \% for channels (A.8) and (A.9) have been measured by \cite{sla1982} (1982) at 1216 $\AA$ (Ly$\alpha$). We adopt these values for $\lambda$ $<$ 1360 $\AA$ while at longer wavelengths we take quantum yields of 99 \% and 1 \% (\cite{cro1989} 1989).

\subsection{O$_2$}

In the wavelength range of photoionization of molecular oxygen (up to the ionization threshold of 1027.8 $\AA$, e.g., \cite{hue1992} 1992), the photoabsorption and photoionization cross sections have been taken from the experimental study of \cite{hol1993} (1993). Up to 1750 $\AA$ the photoabsorption cross section has been taken from in the \cite{yos2005} (2005), and from \cite{cha1993b} (1993b) at longer wavelengths. The dissociation threshold of O$_2$
is 2423.7 $\AA$ (\cite{hue1992} 1992), although at wavelengths longer than $\sim$1850 $\AA$ the photodissociation cross section already becomes negligible.

\subsection{H$_2$CO}

The photoabsorption cross section of formaldehyde is taken from the measurements by \cite{coo1996} (1996). The cross section of photoionization, which occurs at wavelengths shorter than $\sim$1400 $\AA$, has been measured by \cite{men1971} (1971). The photodissociation channels of H$_2$CO are
\begin{eqnarray}
\rm H_2CO + \emph{h$\nu$} & \rightarrow & \rm CO + H_2, \\
 & \rightarrow & \rm CO + H + H, \\
 & \rightarrow & \rm HCO + H,
\end{eqnarray}
where channel (A.12) can be neglected as an important one because it occurs in the wavelength range 2000-3340 $\AA$, where the absorption cross section of H$_2$CO is three orders of magnitude lower than at $\lambda$ $<$ 2000 $\AA$. On the other hand, channels (A.10) and (A.11) occur with similar quantum yields (\cite{sti1972} 1972). However, since the implications of producing either molecular or atomic hydrogen in the photodissociation of H$_2$CO are small for the chemistry of protoplanetary disks, we have assumed that only channel (A.10) occurs.

\subsection{CO$_2$}

The ionization threshold of CO$_2$ is 900 $\AA$ and thus photoionization does not occur in our wavelength range of interest. The adopted photoabsorption cross section is based on the critical evaluation by \cite{hue2010} (2010). The only allowed channel in the photodissociation of carbon dioxide is that yielding CO + O (see, e.g., \cite{hue1992} 1992).

\subsection{NH}

In the case of the NH radical the photodissociation cross section has been calculated by \cite{kir1991} (1991). The photoionization cross section has been calculated by \cite{wan1990} (1990) to be around 8 Mb in the narrow wavelength range from the Lyman cutoff to the ionization threshold at 919.1 $\AA$.

\subsection{NH$_2$}

The photodissociation cross section of NH$_2$ is taken from the theoretical calculations of \cite{koc1997} (1997). There are two possible dissociation channels:
\begin{eqnarray}
\rm NH_2 + \emph{h$\nu$} & \rightarrow & \rm NH + H, \\
 & \rightarrow & \rm N + H_2,
\end{eqnarray}
where only channel (A.13) is considered here, based on experimental and theoretical evidence that the major channel in
the photodissociation of NH$_2$ leads to NH radicals (\cite{bie1994} 1994; \cite{vet1996} 1996). As concerns photoionization, we used the relative cross section measured by \cite{gib1985} (1985) in the wavelength range 745-1125 $\AA$ (the ionization threshold is 1130 $\AA$) and fixed the absolute scale by assuming a guess value of 10 Mb for the cross section at 912 $\AA$.

\subsection{NH$_3$}

The photoabsorption cross section of ammonia is taken from \cite{edv1999} (1999) up to the ionization threshold of 1231
$\AA$, from \cite{wu2007} (2007) in the wavelength range 1231-1440 $\AA$, and from \cite{che2006a} (2006a) at longer wavelengths. The data measured by \cite{bur1993} (1993) at a lower spectral resolution is consistent with the data described above. The photoionization cross section has been taken from \cite{edv1999} (1999). As concerns photodissociation, there are different channels for ammonia, leading to NH$_2$ and NH radicals,
\begin{eqnarray}
\rm NH_3 + \emph{h$\nu$} & \rightarrow & \rm NH_2 + H, \\
 & \rightarrow & \rm NH(\emph{$X^3\Sigma^-$}) + H + H~~or~~NH(\emph{$a^1\Delta$}) + H_2,
\end{eqnarray}
where channel (A.15) dominates over the production of NH with a yield $\ge$ 0.694. The quantum yields over the wavelength range of interest have been taken from various measurements carried out at different wavelengths (\cite{mcn1962} 1962; \cite{oka1967} 1967; \cite{gro1968} 1968; \cite{lil1973} 1973; \cite{sla1982} 1982).

\subsection{N$_2$}

Molecular nitrogen is ionized by photons with energies higher than 15.58 eV and thus only photodissociation, and not photoionization, occurs in our wavelength range of interest. The photoabsorption cross section has been taken from the experiments carried out by \cite{cha1993c} (1993c). The photodissociation threshold of N$_2$ is 1270.85 $\AA$, although in the practice only photons with $\lambda$ $<$ 1000 $\AA$ are efficient enough in dissociating molecular nitrogen.

\subsection{CN}

The photodissociation cross section of the CN radical has been calculated by \cite{lav1987} (1987). The photodissociation threshold is 1600 $\AA$ (\cite{hue1992} 1992), although in the practice the cross section becomes vanishingly small at wavelengths longer than $\sim$1100 $\AA$. Since the ionization threshold of CN is 911.756 $\AA$ (\cite{lid2009} 2009), slightly shorter than that of hydrogen, we do not consider photoionization here.

\subsection{HCN}

Hydrogen is ionized by photons with wavelengths shorter than the Lyman cutoff and thus photoionization of HCN is not considered here. We have adopted the photoabsorption cross section measured by \cite{nut1982} (1982) up to $\lambda$ = 1469.2 $\AA$. We assume that at longer wavelengths the contribution to photodissociation is small and that the main photodissociation channel leads to the CN radical.

\subsection{HNC}

The photoabsorption cross section of HNC has been recently calculated simultaneously with that of the most stable isomer HCN by \cite{che2016} (2016) in the 7-10 eV energy range. More recently, \cite{agu2017} (2017) have extended those calculations including higher electronic states to cover the 7-13.6 eV energy range. We adopted the cross section calculated by \cite{agu2017} (2017) and assumed that absorption in the studied wavelength range leads mainly to dissociation rather than fluorescence. The ionization threshold of HNC is 992 $\AA$ (\cite{lid2009} 2009), although the relevant cross section is not known.

\subsection{NO}

The photoabsorption cross section of nitric oxide was taken from \cite{wat1967} (1967) up to 1350 $\AA$ and from
\cite{gue1981} (1981) at longer wavelengths. The photoionization cross section has been measured by \cite{wat1967} (1967) up to the ionization threshold, at 1340 $\AA$. The photodissociation cross section has been obtained by substracting the cross sections due to photoionization (\cite{wat1967} 1967) and to fluorescence (as characterized by \cite{gue1981} 1981) to the total photoabsorption cross section.

\subsection{SH}

In the case of the mercapto radical, various studies have investigated theoretically the photodissociation dynamics involving selected electronic states, mostly the first excited state $A^2\Sigma^+$ but also higher ones such as $^2\Sigma^-$, $^2\Delta$, and $2^2\Pi$ (e.g., \cite{whe1997} 1997; \cite{che2006b} 2006b; \cite{jan2007} 2007). However, no quantitative measurement of the photodissociation cross section across the FUV range is available. We have therefore adopted the photodissociation cross section from the Leiden database, whose data are largely based on the calculations of \cite{bru1987} (1987). The photoionization threshold of SH is 1190 $\AA$, although there are no data on the relevant cross section.

\subsection{SH$^+$}

For the SH$^+$ ion, the photodissociation cross section was taken from the Leiden database, where the recent calculations by \cite{mcm2016} (2016) were used adopting the cross section of photodissociation from the ground $v=0$ and $J=0$ level (see \cite{hea2017} 2017). Photodissociation at FUV wavelengths is dominated by transitions involving the excited electronic states $3^3\Sigma^-$ and $3^3\Pi$, which yield S + H$^+$ as products (\cite{mcm2016} 2016).

\subsection{H$_2$S}

For hydrogen sulfide the photoabsorption cross section has been taken from \cite{fen1999a} (1999a) and the photoionization yield from \cite{fen1999b} (1999b). We just consider the dissociation channel leading to SH + H, which dominates over the others (see, e.g., \cite{coo2001} 2001). There are two possible ionization channels (\cite{hue1992} 1992):
\begin{eqnarray}
\rm H_2S + \emph{h$\nu$} & \rightarrow & \rm H_2S^+ + e^-, \\
 & \rightarrow & \rm S^+ + H_2 + e^-,
\end{eqnarray}
where channel (A.17) has a threshold of 1185.25 $\AA$ and dominates, while channel (A.18) has a threshold of 927 $\AA$ and accounts for just a 5 \% of the total ionization quantum yield. We therefore consider that ionization yields H$_2$S$^+$ in all cases.

\subsection{CS}

In the case of carbon monosulfide the photodissociation cross section was taken from the Leiden database, whose data is based on absorption lines measurements by \cite{sta1987} (1987), vertical excitation energies calculated by \cite{bru1975} (1975), plus some oscillator strength guesses. For the photoionization cross section we used the relative measurements carried out by \cite{nor1991} (1991) in the wavelength range 1000-1100 $\AA$ (the ionization threshold is 1095.5 $\AA$) and fixed the absolute scale by assuming a guess value of 10 Mb for the cross section at $\leq$ 1000 $\AA$.

\subsection{SO}

For sulfur monoxide the cross section data available in the literature cover just a limited spectral region. The photoabsorption cross section has been measured in the wavelength ranges 1150-1350 $\AA$ and 1900-2350 $\AA$ by \cite{nee1986} (1986) and by \cite{phi1981} (1981), respectively. In the gap between these two spectral regions and at wavelengths shorter than 1150 $\AA$ we have adopted an arbitrary photodissociation cross section of 5 Mb. The ionization threshold of SO is 1205 $\AA$ (\cite{hue1992} 1992). As photoionization cross section we used the photoelectron spectrum measured by \cite{nor1989} (1989) in the wavelength range 1025-1225 $\AA$ and scaled it assuming a cross section of 10 Mb at $\leq$ 1025 $\AA$.

\subsection{SO$_2$}

The photoabsorption cross section of sulfur dioxide is based on various experimental studies dealing with different wavelength ranges: \cite{hol1995} (1995) up to the ionization threshold of SO$_2$ at 1004 $\AA$, \cite{fen1999c} (1999c) in the range 1004-1061 $\AA$, \cite{man1993} (1993) in the range 1061-1717.7 $\AA$, and \cite{wu2000} (2000) at longer wavelengths. The photoionization cross section is taken from \cite{hol1995} (1995). The photodissociation of SO$_2$ has two possible channels:
\begin{eqnarray}
\rm SO_2 + \emph{h$\nu$} & \rightarrow & \rm SO + O, \\
 & \rightarrow & \rm S + O_2,
\end{eqnarray}
with quantum yields of 50 \% for each channel, as measured by \cite{dri1968} (1968) at 1849 $\AA$. We adopt these quantum yields from the Lyman cutoff to the photodissociation threshold of channel (A.20), lying at 2070 $\AA$ (\cite{hue1992} 1992), and assume that only channel (A.19) occurs from 2070 $\AA$ to 2179 $\AA$, this latter value being the photodissociation threshold of the channel leading to SO (\cite{hue1992} 1992).

\subsection{HF}

Hydrogen fluoride is ionized by photons with energies above 16 eV and thus only photodissociation is considered here. The photodissociation cross section of HF was obtained from the theoretical calculations by \cite{li2010} (2010), which are in good agreement with the experimental values measured in the wavelength range 1070-1450 $\AA$ by \cite{nee1985} (1985). The photodissociation cross section of HF has the shape of
a continuous band centered at $\sim$1230 $\AA$, which corresponds to the $X^1\Sigma^+-a^1\Pi$ transition.

\subsection{HCl}

The photoabsorption cross section of hydrogen chloride has been measured by \cite{bri2005} (2005). The photoionization yield up to the ionization threshold of HCl, at 972.5 $\AA$, has been taken from the study of \cite{dav1984} (1984).

\subsection{Photoionization of atoms}

We have also taken into account the photoionization cross sections of those atoms which can be ionized by photons with wavelengths longer than 911.776 $\AA$. Among the elements included in the chemical network those atoms are, in order of increasing photoionization threshold (taken from \cite{lid2009} 2009 and given in parentheses): Cl (956.11 $\AA$), C (1101.07 $\AA$), P (1182.30 $\AA$), S (1196.76 $\AA$), Si (1520.97 $\AA$), Fe (1568.9 $\AA$), Mg (1621.51 $\AA$), and Na (2412.58 $\AA$). Data for chlorine have been taken from the measurements by \cite{rus1983} (1983). Cross sections for sulfur and sodium were taken from the TOPbase database, while for phosphorus, carbon, and silicon we adopted the analytic fits by \cite{ver1996} (1996)\footnote{\texttt{http://www.pa.uky.edu/$\sim$verner/photo.html}}. For magnesium and iron we used the cross sections in the Leiden database.

\section{Photodissociation and photoionization rates} \label{sec:app_rates}

The photodissociation and photoionization rate of a given species depends on the relevant cross section and the strength and spectral shape of the FUV radiation field. In protoplanetary disks there are two main sources of FUV radiation, the interstellar radiation field (ISRF) and the star, each one having a different strength, spectral shape, and illumination geometry. Therefore, the relative contribution of each field to the various photoprocesses can be quite different depending on the particular process (via the spectral shape of the cross section) and the position in the disk (via the exposure to each radiation field).

The effect of the different spectral shapes of interstellar and stellar radiation fields on the photoprocesses occurring in protoplanetary disks has been investigated by \cite{van2006} (2006). These authors approximated the stellar emission of T\,Tauri and Herbig\,Ae stars as black bodies at temperatures of 4000 K and 10,000 K, respectively. One of the most dramatic effects found is that the 4000 K black body field is much less efficient in photodissociating and photoionizing molecules than the interstellar and 10,000 K black body fields. The reason is that the 4000 K black body emits little at short wavelengths, where dissociation and ionization take place. A similar study using an updated set of cross sections has been recently carried out by \cite{hea2017} (2017). Here we have adopted as proxies of the T\,Tauri and Herbig\,Ae/Be radiation fields the spectra of TW Hya and AB Aurigae described in section~\ref{subsubsec:uv_field} and shown in Fig.~\ref{fig:uv_spectra}. As a consequence of the strong FUV excess and Ly$\alpha$ emission usually present in T\,Tauri stars, which is accounted for by the TW Hya spectrum but not by a blackbody at 4000 K, the photodissociation and photoionization efficiency of the T\,Tauri radiation field is greatly enhanced with respect to the black body assumption.

\begin{figure}
\centering
\includegraphics[angle=0,width=\columnwidth]{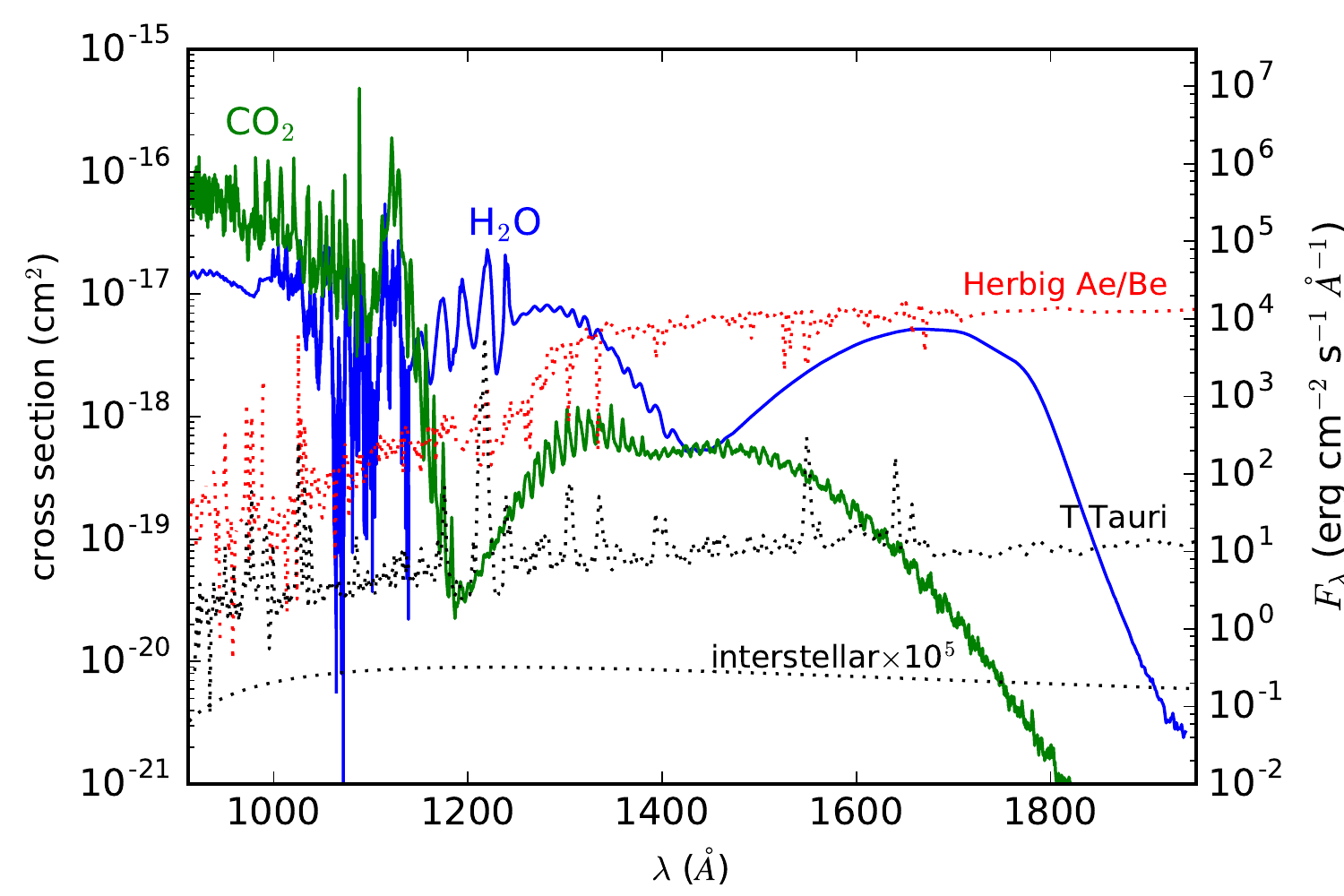}
\caption{FUV photodissociation cross sections of H$_2$O and CO$_2$ superimposed on the FUV spectrum of the T\,Tauri and Herbig\,Ae/Be stars (flux at 1 au) and on the spectral shape of the interstellar radiation field. Note for example how Ly$\alpha$ radiation becomes very important to photodissociate H$_2$O but not CO$_2$, which is photodissociated more efficiently at wavelengths shorter than $\sim$1150 $\AA$.} \label{fig:section}
\end{figure}

Following the idea of \cite{van2006} (2006), here we are interested in evaluating the effect of the different spectral shapes of interstellar and stellar radiation fields (see, e.g., Fig.~\ref{fig:section}) on the photodissociation and photoionization rates of assorted species. To that purpose we have scaled the T\,Tauri and Herbig\,Ae/Be radiation fields to get the same energy density of the ISRF over the wavelength range 912-2400 $\AA$, which amounts to 1.07 $\times$ 10$^{-13}$ erg cm$^{-3}$ adopting the ISRF of \cite{dra1978} (1978) and the expressions given in section~\ref{subsubsec:uv_field}. This is almost twice the value given by \cite{hab1968} (1968), 5.6 $\times$ 10$^{-14}$ erg cm$^{-3}$. For a given radiation field characterized by a specific intensity $I_\lambda$ and a photoprocess characterized by a cross section $\sigma_\lambda$, the rate $\Gamma$ can be computed as
\begin{equation}
\Gamma = \int_{912}^{\lambda_{max}} \big( 4 \pi \frac{\lambda}{h c} I_\lambda \big) \sigma_\lambda d\lambda,
\end{equation}
where the term in parentheses is the photon flux per unit time, area, and wavelength, and the integral extends from the Lyman cutoff (912 $\AA$) to a maximum wavelength $\lambda_{max}$, which depends on each process.

In order to evaluate how the rates of the various photoprocesses vary with the depth into the disk, we have made use of the Meudon PDR code (\cite{lep2006} 2006) to compute the photodissociation and photoionization rates as a function of $A_V$. We consider a plane-parallel cloud illuminated on one side by an external radiation field corresponding to either the ISRF, the T\,Tauri star, or the Herbig\,Ae/Be star, where the stellar fields have been scaled to the FUV energy density of the ISRF. The cloud has uniform density of H nuclei and gas kinetic temperature. We have verified that densities in the range 10$^4$-10$^8$ cm$^{-3}$ and temperatures ranging from 100 to 1000 K, typical values in the photon dominated region of protoplanetary disks, yield identical results. The resulting photodissociation and photoionization rates as a function of the visual extinction have been fitted in the range $A_V$=0-5 according to the standard expression given by Eq.~(\ref{eq:photorate}). We use this expression for simplicity, although we note that more accurate fits can be obtained with more elaborate expressions, e.g., considering two coefficients instead of one in the exponential term (\cite{rob1991} 1991) or involving the 2$^{nd}$-order exponential integral $E_2$ (\cite{neu2009} 2009; \cite{rou2014} 2014; \cite{hea2017} 2017).

\begin{table*}
\caption{Photodissociation and photoionization rates for various radiation fields.} \label{table:app_photorates} \centering
\begin{tabular}{l@{\hspace{1.0cm}}lcclcclc}
\hline \hline
& \multicolumn{2}{c}{interstellar$^a$} & & \multicolumn{2}{c}{T\,Tauri$^b$} & & \multicolumn{2}{c}{Herbig\,Ae/Be$^c$} \\
\cline{2-3} \cline{5-6} \cline{8-9}
Reaction & \multicolumn{1}{c}{$\alpha$ (s$^{-1}$)} & $\gamma$ & & \multicolumn{1}{c}{$\alpha$ (s$^{-1}$)} & $\gamma$ & &  \multicolumn{1}{c}{$\alpha$ (s$^{-1}$)} & $\gamma$ \\
\hline
 CH + $h \nu$ $\rightarrow$ C + H                                     &   9.8 $\times$ 10$^{-10}$ & 1.58 &  &   9.9 $\times$ 10$^{-10}$ & 1.02 &  &   2.5 $\times$ 10$^{-9}$ & 0.89 \\
 CH + $h \nu$ $\rightarrow$ CH$^+$ + e$^-$                    &   9.3 $\times$ 10$^{-10}$ & 3.20 &  &   1.6 $\times$ 10$^{-10}$ & 2.71 &  &   1.1 $\times$ 10$^{-11}$ & 2.64 \\
 CH$^+$ + $h \nu$ $\rightarrow$ C + H$^+$                     &   3.3 $\times$ 10$^{-10}$ & 3.02 &  &   1.0 $\times$ 10$^{-10}$ & 2.39 &  &   2.2 $\times$ 10$^{-11}$ & 1.89 \\
 CH$_2$ + $h \nu$ $\rightarrow$ CH + H                          &   6.5 $\times$ 10$^{-10}$ & 1.89 &  &   3.3 $\times$ 10$^{-10}$ & 1.45 &  &   9.4 $\times$ 10$^{-10}$ & 1.42 \\
 CH$_3$ + $h \nu$ $\rightarrow$ CH$_2$ + H                  &   3.6 $\times$ 10$^{-10}$ & 2.06 &  &   2.0 $\times$ 10$^{-10}$ & 1.55 &  &   4.8 $\times$ 10$^{-10}$ & 1.54 \\
 CH$_3$ + $h \nu$ $\rightarrow$ CH + H$_2$                  &   3.6 $\times$ 10$^{-10}$ & 2.06 &  &   2.0 $\times$ 10$^{-10}$ & 1.55 &  &   4.8 $\times$ 10$^{-10}$ & 1.54 \\
 CH$_3$ + $h \nu$ $\rightarrow$ CH$_3^+$ + e$^-$              &   3.3 $\times$ 10$^{-10}$ & 2.91 &  &   5.4 $\times$ 10$^{-10}$ & 2.23 &  &   8.1 $\times$ 10$^{-12}$ & 2.32 \\
 CH$_4$ + $h \nu$ $\rightarrow$ CH$_3$ + H                    &   4.7 $\times$ 10$^{-10}$ & 2.75 &  &   7.8 $\times$ 10$^{-10}$ & 2.22 &  &   5.3 $\times$ 10$^{-11}$ & 2.07 \\
 CH$_4$ + $h \nu$ $\rightarrow$ CH$_2$ + H$_2$                &   8.8 $\times$ 10$^{-10}$ & 2.83 &  &   1.0 $\times$ 10$^{-9}$ & 2.23 &  &   7.0 $\times$ 10$^{-11}$ & 2.09 \\
 CH$_4$ + $h \nu$ $\rightarrow$ CH + H$_2$ + H                &   1.2 $\times$ 10$^{-10}$ & 2.83 &  &   1.4 $\times$ 10$^{-10}$ & 2.23 &  &   9.7 $\times$ 10$^{-12}$ & 2.09 \\
 CH$_4$ + $h \nu$ $\rightarrow$ CH$_4^+$ + e$^-$              &   1.4 $\times$ 10$^{-11}$ & 3.95 &  &   1.8 $\times$ 10$^{-12}$ & 3.36 &  &   1.4 $\times$ 10$^{-13}$ & 3.37 \\
 C$_2$ + $h \nu$ $\rightarrow$ C + C                          &   1.3 $\times$ 10$^{-10}$ & 2.97 &  &   3.9 $\times$ 10$^{-11}$ & 2.52 &  &   3.4 $\times$ 10$^{-12}$ & 2.45 \\
 C$_2$ + $h \nu$ $\rightarrow$ C$_2^+$ + e$^-$                &   2.1 $\times$ 10$^{-10}$ & 3.82 &  &   3.9 $\times$ 10$^{-11}$ & 3.23 &  &   2.3 $\times$ 10$^{-12}$ & 3.22 \\
 C$_2$H + $h \nu$ $\rightarrow$ C$_2$ + H                     &   1.6 $\times$ 10$^{-9}$ & 2.48 &  &   1.8 $\times$ 10$^{-9}$ & 2.13 &  &   5.6 $\times$ 10$^{-10}$ & 1.83 \\
 C$_2$H$_2$ + $h \nu$ $\rightarrow$ C$_2$H + H                &   4.4 $\times$ 10$^{-9}$ & 2.46 &  &   4.2 $\times$ 10$^{-9}$ & 2.08 &  &   1.9 $\times$ 10$^{-9}$ & 1.81 \\
 C$_2$H$_2$ + $h \nu$ $\rightarrow$ C$_2$H$_2^+$ + e$^-$      &   3.3 $\times$ 10$^{-10}$ & 3.37 &  &   9.0 $\times$ 10$^{-11}$ & 2.87 &  &   6.0 $\times$ 10$^{-12}$ & 2.87 \\
 OH + $h \nu$ $\rightarrow$ O + H                             &   3.8 $\times$ 10$^{-10}$ & 2.33 &  &   5.1 $\times$ 10$^{-10}$ & 1.99 &  &   1.7 $\times$ 10$^{-10}$ & 1.62 \\
 OH + $h \nu$ $\rightarrow$ OH$^+$ + e$^-$                    &   5.2 $\times$ 10$^{-12}$ & 3.95 &  &   6.5 $\times$ 10$^{-13}$ & 3.35 &  &   5.5 $\times$ 10$^{-14}$ & 3.37 \\
 H$_2$O + $h \nu$ $\rightarrow$ OH + H                        &   6.8 $\times$ 10$^{-10}$ & 2.22 &  &   1.4 $\times$ 10$^{-9}$ & 2.01 &  &   3.4 $\times$ 10$^{-10}$ & 1.52 \\
 H$_2$O + $h \nu$ $\rightarrow$ O + H$_2$                     &   1.0 $\times$ 10$^{-10}$ & 2.70 &  &   3.5 $\times$ 10$^{-10}$ & 2.21 &  &   9.5 $\times$ 10$^{-12}$ & 1.74 \\
 H$_2$O + $h \nu$ $\rightarrow$ H$_2$O$^+$ + e$^-$            &   2.7 $\times$ 10$^{-11}$ & 3.84 &  &   6.2 $\times$ 10$^{-12}$ & 3.22 &  &   3.6 $\times$ 10$^{-13}$ & 3.23 \\
 O$_2$ + $h \nu$ $\rightarrow$ O + O                          &   7.3 $\times$ 10$^{-10}$ & 2.31 &  &   3.5 $\times$ 10$^{-10}$ & 1.81 &  &   5.6 $\times$ 10$^{-10}$ & 1.77 \\
 O$_2$ + $h \nu$ $\rightarrow$ O$_2^+$ + e$^-$                &   5.1 $\times$ 10$^{-11}$ & 3.72 &  &   1.2 $\times$ 10$^{-11}$ & 3.08 &  &   8.6 $\times$ 10$^{-13}$ & 3.08 \\
 H$_2$CO + $h \nu$ $\rightarrow$ CO + H$_2$                   &   1.6 $\times$ 10$^{-9}$ & 2.16 &  &   2.5 $\times$ 10$^{-9}$ & 1.80 &  &   1.0 $\times$ 10$^{-9}$ & 1.37 \\
 H$_2$CO + $h \nu$ $\rightarrow$ H$_2$CO$^+$ + e$^-$          &   4.3 $\times$ 10$^{-10}$ & 3.22 &  &   1.0 $\times$ 10$^{-10}$ & 2.76 &  &   7.6 $\times$ 10$^{-12}$ & 2.69 \\
 CO$_2$ + $h \nu$ $\rightarrow$ CO + O                        &   1.1 $\times$ 10$^{-9}$ & 3.01 &  &   2.4 $\times$ 10$^{-10}$ & 2.41 &  &   4.3 $\times$ 10$^{-11}$ & 1.93 \\
 NH + $h \nu$ $\rightarrow$ N + H                             &   4.8 $\times$ 10$^{-10}$ & 2.46 &  &   2.0 $\times$ 10$^{-10}$ & 2.00 &  &   2.7 $\times$ 10$^{-10}$ & 1.94 \\
 NH + $h \nu$ $\rightarrow$ NH$^+$ + e$^-$                    &   1.8 $\times$ 10$^{-12}$ & 4.00 &  &   7.4 $\times$ 10$^{-15}$ & 3.41 &  &   1.7 $\times$ 10$^{-14}$ & 3.42 \\
 NH$_2$ + $h \nu$ $\rightarrow$ NH + H                        &   8.9 $\times$ 10$^{-10}$ & 1.92 &  &   4.2 $\times$ 10$^{-10}$ & 1.50 &  &   1.1 $\times$ 10$^{-9}$ & 1.47 \\
 NH$_2$ + $h \nu$ $\rightarrow$ NH$_2^+$ + e$^-$              &   1.1 $\times$ 10$^{-10}$ & 3.44 &  &   3.1 $\times$ 10$^{-11}$ & 2.92 &  &   2.1 $\times$ 10$^{-12}$ & 2.91 \\
 NH$_3$ + $h \nu$ $\rightarrow$ NH$_2$ + H                    &   1.2 $\times$ 10$^{-9}$ & 1.99 &  &   6.9 $\times$ 10$^{-10}$ & 1.58 &  &   1.2 $\times$ 10$^{-9}$ & 1.41 \\
 NH$_3$ + $h \nu$ $\rightarrow$ NH + H$_2$                    &   3.1 $\times$ 10$^{-10}$ & 2.72 &  &   1.1 $\times$ 10$^{-9}$ & 2.21 &  &   3.6 $\times$ 10$^{-11}$ & 2.03 \\
 NH$_3$ + $h \nu$ $\rightarrow$ NH$_3^+$ + e$^-$              &   4.2 $\times$ 10$^{-10}$ & 3.04 &  &   1.8 $\times$ 10$^{-10}$ & 2.34 &  &   8.2 $\times$ 10$^{-12}$ & 2.50 \\
 N$_2$ + $h \nu$ $\rightarrow$ N + N                          &   3.6 $\times$ 10$^{-10}$ & 3.81 &  &   7.6 $\times$ 10$^{-11}$ & 3.21 &  &   3.9 $\times$ 10$^{-12}$ & 3.19 \\
 CN + $h \nu$ $\rightarrow$ C + N                             &   1.0 $\times$ 10$^{-9}$ & 3.55 &  &   1.8 $\times$ 10$^{-10}$ & 3.02 &  &   1.2 $\times$ 10$^{-11}$ & 3.04 \\
 HCN + $h \nu$ $\rightarrow$ CN + H                           &   1.9 $\times$ 10$^{-9}$ & 2.82 &  &   4.5 $\times$ 10$^{-9}$ & 2.22 &  &   1.6 $\times$ 10$^{-10}$ & 2.01 \\
 HNC + $h \nu$ $\rightarrow$ CN + H                           &   9.4 $\times$ 10$^{-10}$ & 2.45 &  &   3.2 $\times$ 10$^{-9}$ & 2.16 &  &   3.9 $\times$ 10$^{-10}$ & 1.80 \\
 NO + $h \nu$ $\rightarrow$ N + O                             &   4.7 $\times$ 10$^{-10}$ & 2.01 &  &   2.3 $\times$ 10$^{-10}$ & 1.60 &  &   4.1 $\times$ 10$^{-10}$ & 1.41 \\
 NO + $h \nu$ $\rightarrow$ NO$^+$ + e$^-$                    &   2.6 $\times$ 10$^{-10}$ & 3.00 &  &   2.4 $\times$ 10$^{-10}$ & 2.27 &  &   8.7 $\times$ 10$^{-12}$ & 2.24 \\
 SH + $h \nu$ $\rightarrow$ S + H                             &   1.3 $\times$ 10$^{-9}$ & 1.89 &  &   2.0 $\times$ 10$^{-9}$ & 1.54 &  &   1.8 $\times$ 10$^{-9}$ & 1.30 \\
 SH$^+$ + $h \nu$ $\rightarrow$ S + H$^+$                     &   6.9 $\times$ 10$^{-10}$ & 1.89 &  &   3.2 $\times$ 10$^{-10}$ & 1.05 &  &   3.9 $\times$ 10$^{-10}$ & 0.80 \\
 H$_2$S + $h \nu$ $\rightarrow$ SH + H                        &   3.2 $\times$ 10$^{-9}$ & 2.26 &  &   3.4 $\times$ 10$^{-9}$ & 1.91 &  &   1.8 $\times$ 10$^{-9}$ & 1.58 \\
 H$_2$S + $h \nu$ $\rightarrow$ H$_2$S$^+$ + e$^-$            &   7.2 $\times$ 10$^{-10}$ & 3.15 &  &   1.7 $\times$ 10$^{-10}$ & 2.68 &  &   1.3 $\times$ 10$^{-11}$ & 2.61 \\
 CS + $h \nu$ $\rightarrow$ C + S                             &   9.5 $\times$ 10$^{-10}$ & 2.60 &  &   4.2 $\times$ 10$^{-9}$ & 2.20 &  &   1.7 $\times$ 10$^{-10}$ & 1.98 \\
 CS + $h \nu$ $\rightarrow$ CS$^+$ + e$^-$                    &   1.7 $\times$ 10$^{-10}$ & 3.30 &  &   4.1 $\times$ 10$^{-11}$ & 2.81 &  &   2.8 $\times$ 10$^{-12}$ & 2.81 \\
 SO + $h \nu$ $\rightarrow$ S + O                             &   4.8 $\times$ 10$^{-9}$ & 2.24 &  &   1.1 $\times$ 10$^{-8}$ & 2.01 &  &   2.1 $\times$ 10$^{-9}$ & 1.47 \\
 SO + $h \nu$ $\rightarrow$ SO$^+$ + e$^-$                    &   2.1 $\times$ 10$^{-10}$ & 3.18 &  &   5.3 $\times$ 10$^{-11}$ & 2.72 &  &   4.0 $\times$ 10$^{-12}$ & 2.64 \\
 SO$_2$ + $h \nu$ $\rightarrow$ SO + O                        &   1.2 $\times$ 10$^{-9}$ & 2.25 &  &   2.1 $\times$ 10$^{-9}$ & 1.97 &  &   5.5 $\times$ 10$^{-10}$ & 1.47 \\
 SO$_2$ + $h \nu$ $\rightarrow$ S + O$_2$                     &   1.1 $\times$ 10$^{-9}$ & 2.29 &  &   2.1 $\times$ 10$^{-9}$ & 2.02 &  &   4.1 $\times$ 10$^{-10}$ & 1.47 \\
 SO$_2$ + $h \nu$ $\rightarrow$ SO$_2^+$ + e$^-$              &   1.3 $\times$ 10$^{-10}$ & 3.77 &  &   3.0 $\times$ 10$^{-11}$ & 3.17 &  &   2.1 $\times$ 10$^{-12}$ & 3.16 \\
 HF + $h \nu$ $\rightarrow$ H + F                             &   1.3 $\times$ 10$^{-10}$ & 2.57 &  &   3.1 $\times$ 10$^{-10}$ & 2.18 &  &   2.7 $\times$ 10$^{-11}$ & 1.84 \\
 HCl + $h \nu$ $\rightarrow$ H + Cl                           &   2.0 $\times$ 10$^{-9}$ & 2.53 &  &   7.7 $\times$ 10$^{-10}$ & 1.99 &  &   4.2 $\times$ 10$^{-10}$ & 1.68 \\
 HCl + $h \nu$ $\rightarrow$ HCl$^+$ + e$^-$                  &   1.9 $\times$ 10$^{-11}$ & 3.91 &  &   2.7 $\times$ 10$^{-12}$ & 3.33 &  &   1.7 $\times$ 10$^{-13}$ & 3.35 \\
 & & & & & & & & \\
 C + $h \nu$ $\rightarrow$ C$^+$ + e$^-$                      &   3.3 $\times$ 10$^{-10}$ & 3.28 &  &   8.4 $\times$ 10$^{-11}$ & 2.82 &  &   5.8 $\times$ 10$^{-12}$ & 2.80 \\
 Si + $h \nu$ $\rightarrow$ Si$^+$ + e$^-$                    &   4.2 $\times$ 10$^{-9}$ & 2.49 &  &   4.4 $\times$ 10$^{-9}$ & 2.13 &  &   2.1 $\times$ 10$^{-9}$ & 1.91 \\
 P + $h \nu$ $\rightarrow$ P$^+$ + e$^-$                      &   1.1 $\times$ 10$^{-9}$ & 2.99 &  &   2.5 $\times$ 10$^{-10}$ & 2.48 &  &   2.3 $\times$ 10$^{-11}$ & 2.45 \\
 S + $h \nu$ $\rightarrow$ S$^+$ + e$^-$                      &   9.7 $\times$ 10$^{-10}$ & 3.09 &  &   2.1 $\times$ 10$^{-10}$ & 2.62 &  &   1.8 $\times$ 10$^{-11}$ & 2.55 \\
 Cl + $h \nu$ $\rightarrow$ Cl$^+$ + e$^-$                    &   4.7 $\times$ 10$^{-11}$ & 3.94 &  &   6.1 $\times$ 10$^{-12}$ & 3.35 &  &   4.4 $\times$ 10$^{-13}$ & 3.36 \\
 Na + $h \nu$ $\rightarrow$ Na$^+$ + e$^-$                    &   1.3 $\times$ 10$^{-11}$ & 2.12 &  &   1.6 $\times$ 10$^{-11}$ & 1.68 &  &   8.4 $\times$ 10$^{-12}$ & 1.47 \\
 Mg + $h \nu$ $\rightarrow$ Mg$^+$ + e$^-$                    &   6.7 $\times$ 10$^{-11}$ & 2.29 &  &   4.4 $\times$ 10$^{-11}$ & 1.80 &  &   5.4 $\times$ 10$^{-11}$ & 1.73 \\
 Fe + $h \nu$ $\rightarrow$ Fe$^+$ + e$^-$                    &   4.7 $\times$ 10$^{-10}$ & 2.45 &  &   6.9 $\times$ 10$^{-10}$ & 2.02 &  &   2.8 $\times$ 10$^{-10}$ & 1.88 \\
\hline
\end{tabular}
\tablenoted{\\
$^a$ Interstellar radiation field (ISRF) of \cite{dra1978} (1978); see expressions in section~\ref{subsubsec:uv_field}. \\
 $^b$ Radiation field of TW\,Hya scaled to the FUV energy density of the ISRF. \\
$^c$ Radiation field of AB\,Aurigae scaled to the FUV energy density of the ISRF.}
\end{table*}

In Table~\ref{table:app_photorates} we list unattenuated rates $\alpha$ and attenuation factors $\gamma$ under different radiation fields for the photodissociation and photoionization processes involving the 29 molecules and 8 atoms discussed in Appendix~\ref{sec:app_sections}\footnote{The interstellar unatennuated rates $\alpha$ in Table~\ref{table:app_photorates} have been scaled up by a factor of two with respect to the output values from the one-side illuminated clouds modeled with the Meudon PDR code. This way, our tabulated values are in line with those in the literature (\cite{rob1991} 1991; \cite{van2006} 2006; \cite{hea2017} 2017) and in the astrochemical databases {\footnotesize UMIST} and {\footnotesize KIDA}. Note, however, that when modeling protoplanetary disks or clouds that are illuminated on just one side, the other being optically thick, one must use half the unatennuated rates $\alpha$ listed in Table~\ref{table:app_photorates}.}. If we focus on the unattenuated rates $\alpha$, we see that the rates calculated under the T\,Tauri radiation field are not very different from those computed under the ISRF (within one order of magnitude), while in the case of the Herbig\,Ae/Be radiation field some photoprocesses have rates similar to those computed under the ISRF (within one order of magnitude) while for others the rates are lower by more than a factor of ten. By looking to the spectral shape of the different FUV radiation fields (see Fig.~\ref{fig:section}) we note that, except for the fact that stellar spectra have spectral features while the ISRF is a continuum, the spectra of the T\,Tauri star and the ISRF are more flat than that of the Herbig\,Ae/Be star, which shows a depletion of flux at short wavelengths (below $\sim$1300 $\AA$). Therefore, those photoprocesses which occur more effectively at short wavelengths, as, e.g., the photodissociation of N$_2$ (see Fig.\ref{fig:app_section}), have higher rates under the ISRF and the T\,Tauri radiation field than under the Herbig\,Ae/Be field.

It is interesting to compare our results with previous studies. In general, the unattenuated rates $\alpha$ and the attenuation parameters $\gamma$ calculated here under the ISRF are similar to those calculated by \cite{rob1991} (1991), \cite{van2006} (2006), and \cite{hea2017} (2017). As concerns the rates under the Herbig\,Ae/Be radiation field, our parameters $\alpha$ and $\gamma$ are also not very different from those calculated by \cite{van2006} (2006) and \cite{hea2017} (2017) for a 10,000 K black body radiation field, because the FUV spectrum of AB\,Aurigae used by us is not drastically different from that of a 10,000 K black body. The same is not true in the case of the radiation field of the T\,Tauri star. The FUV spectrum of a 4000 K black body is very different from that of TW\,Hya, which translates into very different photodissociation and photoionization rates. In general, the rates calculated by us for a T\,Tauri star are orders of magnitude higher than those computed by \cite{van2006} (2006) and \cite{hea2017} (2017).

The values provided in Table~\ref{table:app_photorates} can be useful in chemical models of protoplananetary disks around T\,Tauri and Herbig\,Ae/Be stars, as long as they allow to compute the photodissociation and photoionization rates in a simple way, avoiding the more expensive approach, in terms of computing time, of solving the FUV radiative transfer.

\end{document}